\documentclass[12pt]{article}
\usepackage{graphicx}
\usepackage{newtxtext}
\usepackage{newtxmath}
\usepackage{hyperref}
\hypersetup{
    colorlinks = true,
    urlcolor   = blue,
    citecolor  = black,
}

\newcommand{\RomanNumeralCaps}[1]
\linenumbers
\usepackage{tikz}
\usetikzlibrary{decorations.markings}
\usepackage{mathrsfs}
\usepackage[all,cmtip]{xy}
\usepackage{mleftright}
\usepackage{booktabs}
\usepackage{enumitem}
\usepackage[outdir=./]{epstopdf}
\usepackage{caption}
\usepackage{subcaption}
\usepackage{placeins}
\usepackage{comment}
\usepackage{chemmacros}
\usepackage{subcaption}
\usepackage{setspace}
\usepackage[ruled]{algorithm2e}

\graphicspath{ {./figures/} }
\def\bC{{\bf C}}

\def\PP{{\mathbb P}}
\def\RR{{\mathbb R}}

\def\I{{\mathcal I}}
\def\L{{\mathcal L}}

\def\bn{{\bf n}}
\def\bN{{\bf N}}

\def\bA{{\bf A}}
\def\bB{{\bf B}}
\def\bW{{\bf W}}

\def\JJ{{\mathcal{J}}}

\def\br{{\bf r}}

\def\bF{{\bf F}}

\def\bW{{\bf W}}
\def\bw{{\bf w}}
\def\bI{{\bf I}}
\def\bD{{\bf D}}

\def\brho{{\boldsymbol{\rho}}}
\def\bbeta{{\boldsymbol{\beta}}}
\def\bnab{{\boldsymbol{\nabla}}}

\numberwithin{equation}{section}

\renewcommand{\L}{\mathcal L}

\title{\bfseries\Large\vspace{-1em} Data-Driven Approach to Learning Optimal Forms of Constitutive Relations in Models Describing Lithium Plating in Battery Cells}

\author{\normalsize Avesta Ahmadi$^1$, Kevin J. Sanders$^2$, Gillian R. Goward$^2$ \ and Bartosz Protas$^{3,}$\thanks{Email address for correspondence: bprotas@mcmaster.ca}
	\\ 
	\footnotesize$^1$School of Computational Science \& Engineering, McMaster University,\\ \footnotesize Hamilton, Ontario, Canada, L8S 4K1\\
	\footnotesize$^2$Department of Chemistry \& Chemical Biology, McMaster University,\\ \footnotesize Hamilton, Ontario, Canada, L8S 4M1\\
	\footnotesize$^3$ Department of Mathematics \& Statistics, McMaster University,\\ \footnotesize Hamilton, Ontario, Canada, L8S 4K1 
}

\date{\normalsize \today}

\begin{document}
\maketitle

\begin{abstract}
In this study we construct a data-driven model
describing Lithium plating in a battery cell, which is a key process
contributing to degradation of such cells. Starting from the
fundamental Doyle-Fuller-Newman (DFN) model, we use asymptotic
reduction and spatial averaging techniques to derive a simplified
representation to track the temporal evolution of two key
concentrations in the system, namely, the total intercalated Lithium
on the negative electrode particles and total plated Lithium.  This
model depends on an a priori unknown constitutive relation
representing the plating dynamics of the cell as a function of the
state variables. An optimal form of this constitutive relation is
then deduced from experimental measurements of the time-dependent
concentrations of different Lithium phases acquired through Nuclear
Magnetic Resonance spectroscopy. This is done by solving an inverse
problem in which this constitutive relation is found subject to
minimum assumptions as a minimizer of a suitable constrained
optimization problem where the discrepancy between the model
predictions and experimental data is minimized. This optimization
problem is solved using a state-of-the-art adjoint-based technique.
In contrast to some of the earlier approaches to modelling Lithium
plating, the proposed model is able to predict non-trivial evolution
of the concentrations in the relaxation regime when no current is
applied to the cell. When equipped with an optimal constitutive
relation, the model provides accurate predictions of the time
evolution of both intercalated and plated Lithium across a wide
range of charging/discharging rates. It can therefore serve as a
useful tool for prediction and control of degradation mechanism in
battery cells.
\end{abstract}

\begin{flushleft}
\end{flushleft}


\section{Introduction}\label{sec:intro}
In recent years, due to the growing demand for green energy 
and the phasing out of fossil fuels in pursuit of a more sustainable 
future, rechargeable batteries have assumed a prominent role in 
the transition to green technologies. Lithium ion (Li-ion) batteries, 
among the most promising energy storage solutions, have found 
extensive applications in portable electronic devices, electric 
vehicles, and grid storage. With the increasing need for clean 
energy storage technologies, addressing challenges related to 
the performance and reliability of Li-ion batteries has become 
crucial. Aging and inefficiency mechanisms in cells contribute 
to their degradation. Battery degradation involves complex 
processes, both physical and chemical, within a cell. To comprehend, 
analyze, mitigate, and control the impact of these mechanisms, 
sophisticated experimental and computational techniques are 
essential. Current research aims to contribute to the understanding, 
prediction, and management of one of the primary degradation 
mechanisms in Li-ion batteries, commonly known as Lithium plating 
(Li-plating).

A Li-ion cell is composed of a pair of porous electrodes: the negative
electrode (anode) and the positive electrode (cathode), separated by a
porous separator, and immersed in a liquid electrolyte. These
components are enclosed between two current collectors, each connected
to an external circuit. The porous nature of the electrodes
facilitates the movement of Lithium ions within the material.
Typically, graphite is used as the material for the negative electrode
due to its layered crystalline structure. In recent years, silicon has
received significant attention as an electrode material alternative to
graphite due to its high capacity and abundance \cite{su2014silicon}.
The cathode material is typically a Lithium-metal-oxide, with Nickel,
Manganese, and Cobalt being common metal components. The primary
constituent in the electrolyte solution are $\ch{Li+}$ ions which
migrate between the electrodes during the cell operation. During the
charging process, Li ions are deintercalated from the cathode layers,
freeing up electrons.  Electrons then travel through the solid phase
of the cathode to the current collector, through an external circuit,
and into the solid phase of the anode. Simultaneously, Li ions
dissolve into the electrolyte and diffuse through the separator pores
to the anode layers, where they undergo intercalation. The charging
process continues as more Li ions intercalate into the anode.
Depending on the capacity of the anode to host Li ions, the charging
process continues until most available sites on the anode surface are
occupied by Li elements.  During cell discharge, a reverse process
occurs, with Li deintercalating from the anode surface, prompting the
migration of Li ions from the anode to the cathode. Intercalation of
Li ions on the anode solid phase during the charge process, and their
subsequent deintercalation during the discharge process, are the
desired mechanisms in the operation of the cell. However, these
processes are typically impaired by various degradation mechanisms.

Several degradation mechanisms contribute to the inefficiencies
observed in Li-ion cells, most importantly Solid-Electrolyte
Interphase (SEI) growth, Li-plating, and binder decomposition. These
degradation mechanisms can result in three distinct degradation modes,
namely, the loss of cycleable Lithium, loss of active materials, and
loss of electrolyte, as noted in different studies
\cite{birkl2017degradation,edge2021lithium,lin2021lithium}.  The loss
of cycleable Lithium, which leads to significant capacity fade in the
cell, is primarily caused by the consumption of Li ions through
undesirable side reactions such as irreversible Li-plating and SEI
growth.  Conversely, the loss of active material is linked to
structural changes in the anode, potentially leading to a reduction in
active sites available for Li intercalation. On the cathode side, the
loss of active material can occur due to structural changes in the
cathode, transition metal dissolution, and particle cracking.
Moreover, the consumption of electrolyte can also contribute to cell
degradation. This is driven by interactions with deposited Lithium at
the anode interface, ultimately leading to the depletion of cycleable
Lithium. The SEI growth is categorized as primary or secondary. The
primary SEI growth process is related to the creation of a SEI layer
on the anode surface during the initial cycle of the cell. Although it
consumes some cycleable Lithium, its presence is vital for the
performance and stability of the battery. The secondary SEI growth, on
the other hand, pertains to the creation of SEI layer during the
subsequent cycles of the cell, which could be another potential
mechanism contributing to the degradation of the cell. Also, inactive
particles within the negative electrode play a crucial role in
providing structural stability to the cell. However, binder
decomposition can lead to changes in the cell morphology, ultimately
also contributing to its degradation. Additionally, a primary
challenge associated with silicon anodes is their substantial volume
change during charge/discharge cycles, a characteristic that enhances
their energy density due to the presence of free sites for Lithium
ions to intercalate. However, the continuous volume fluctuations might
lead to the formation of secondary films on the anode surface,
increasing the chance of Li-plating, and thereby depleting cycleable
Lithium and contributing to capacity fade of the cell over time. Each
of these degradation mechanisms can become more prominent in certain
circumstances of cell operation such as extreme temperatures, high
charge/discharge rates, and overvoltage of the cell due to overcharge
and overdischarge.

Li-plating is a critical degradation mechanism that becomes more
pronounced under harsh charging conditions, as discussed by Zhang et
al.~\cite{zhang2022investigation}. It is primarily accelerated when
metallic Lithium forms during the charging process under conditions
such as high charging rates, overcharging at high states-of-charge,
and charging at low temperatures. At lower temperatures, the energy
density of the cell decreases due to several factors, including
reduced ionic conductivity and diffusivity of the electrolyte, lower
solid-state diffusivity of Li ions in the electrodes, and slower
intercalation rates. Higher charge rates introduce greater kinetic and
transport overpotentials, contributing to the Li-plating phenomenon,
as highlighted by Lin et al.~\cite{lin2021lithium}. Additionally, when
the state-of-charge of the cell is high, continued charging can lead
to an excess of Lithium ions saturating on the anode surface,
surpassing the maximum allowable Lithium levels, further accelerating
Li-plating. In essence, under low-temperature and high state-of-charge
conditions, the diffusion rate of Li ions within the electrolyte
toward the anode exceeds the rate of Li ions diffusing into the SEI
layer and graphite interlayer. This results in an accumulation of Li
ions on the surface of the SEI layer, which subsequently absorbs
electrons and forms metallic Lithium. This metallic Lithium is
deposited onto the surface of the SEI layer. The Li-plating process
can be either reversible or irreversible. The reverse process,
known as Lithium stripping, occurs when metallic Lithium maintains
electrical contact with the anode, allowing for the release of an
electron and the deposition of Li ions back into the electrolyte.
Conversely, if the plated Lithium loses electrical contact with the
anode, the process becomes irreversible, leading to the loss of
cycleable Lithium and the growth of dendrites on the anode surface.
This form of metallic Lithium is often referred to as ``dead
Lithium''. The growth of metallic Lithium dendrites on the anode
surface can potentially rupture the separator, creating an electrical
pathway between the anode and cathode, resulting in a cell short
circuit \cite{santhanagopalan2009analysis}. Furthermore, the high
surface area of dead Lithium can contribute to secondary SEI growth on
its surface, further reducing the available cycleable Lithium
\cite{fang2022quantifying}.  Parasitic reactions related to Li-plating
can be exacerbated during fast-charging operating conditions
\cite{bugga2010lithium}.

Quantifying plated Lithium in Li-ion batteries has been a
long-standing challenge in battery studies, with the task of
distinguishing between the SEI and metallic Lithium being especially
complicated.  Various techniques, categorized as \textit{ex-situ},
\textit{in-situ}, and \textit{operando}, have been proposed for
determining the dead Lithium content within the cell, as discussed by
Lin et al.~\cite{lin2021lithium}. Different experimental techniques
could be used for detection of metallic Lithium in the cell including
Scanning Electron Microscopy, Nuclear Magnetic Resonance Spectroscopy
(NMR), X-ray Photoelectron Spectroscopy, and Electrochemical Impedance
Spectroscopy.  A detailed discussion of these experimental techniques
can be found in references such as
\cite{lin2021lithium,paul2021review, tian2021detecting}.  In the
current study, in order to better understand the Li-plating phenomena,
we leverage experimental data obtained from a novel Li-NMR spectroscopy
technique introduced and developed by Sanders et
al.~\cite{sanders2023quantitative}. Fang et
al.~\cite{fang2022quantifying} have also used a similar approach for
quantification of metallic Lithium using Li-NMR technique as an
\textit{operando} approach.

In an effort to quantify Lithium plating in the cell and eliminate the
need for experimentation in an online application of cells or battery
packs, we aim to model the growth and decay of plated Lithium using
mathematical and computational tools.  We seek to track the evolution
of different phases of Lithium in operando under diverse
charge/discharge protocols and techniques of asymptotic analysis will
first be used to develop simplified models based on the physical
principles governing cell behavior. Then, state-of-the-art
computational tools will be employed to calibrate these models,
optimizing their alignment with experimental data. In particular, the
technique of inverse modeling will be utilized for this purpose
\cite{sethurajan2015accurate,sethurajan2019dendrites,
	escalante2020discerning, daniels2023learning}, where optimal forms
of electrochemical parameters and constitutive relations in the model
are inferred from experimental data by solving suitable optimization
problems.  The resulting calibrated model holds promise for online
applications, enabling real-time monitoring, recommending optimal
charge/discharge protocols, and ultimately enhancing cell performance
while mitigating degradation in the long run.

The paper is organized as follows: details of the experimental data
are presented in Section \ref{sec:experiment}; then in Section
\ref{sec:modeling} we introduce the mathematical modeling framework
for the this problem and develop a dynamical system governing the
evolution of lumped concentrations of different phases of Lithium in
the cell; in Section \ref{sec:inverse} we introduce the inverse
modeling framework and the computational tools used for calibrating
the dynamical system; Section \ref{sec:results} presents the results
of this analysis and compares them to the experimental data; finally,
the summary of the work and the conclusions are deferred to Section
\ref{sec:discussion}. Some more technical material is collected in two
appendices.

\section{Experimental Data}\label{sec:experiment}
\begin{figure}[!t]
	\centering
	\mbox{
		\begin{subfigure}[b]{0.3\textwidth}
			\centering
			\includegraphics[width=1\textwidth]{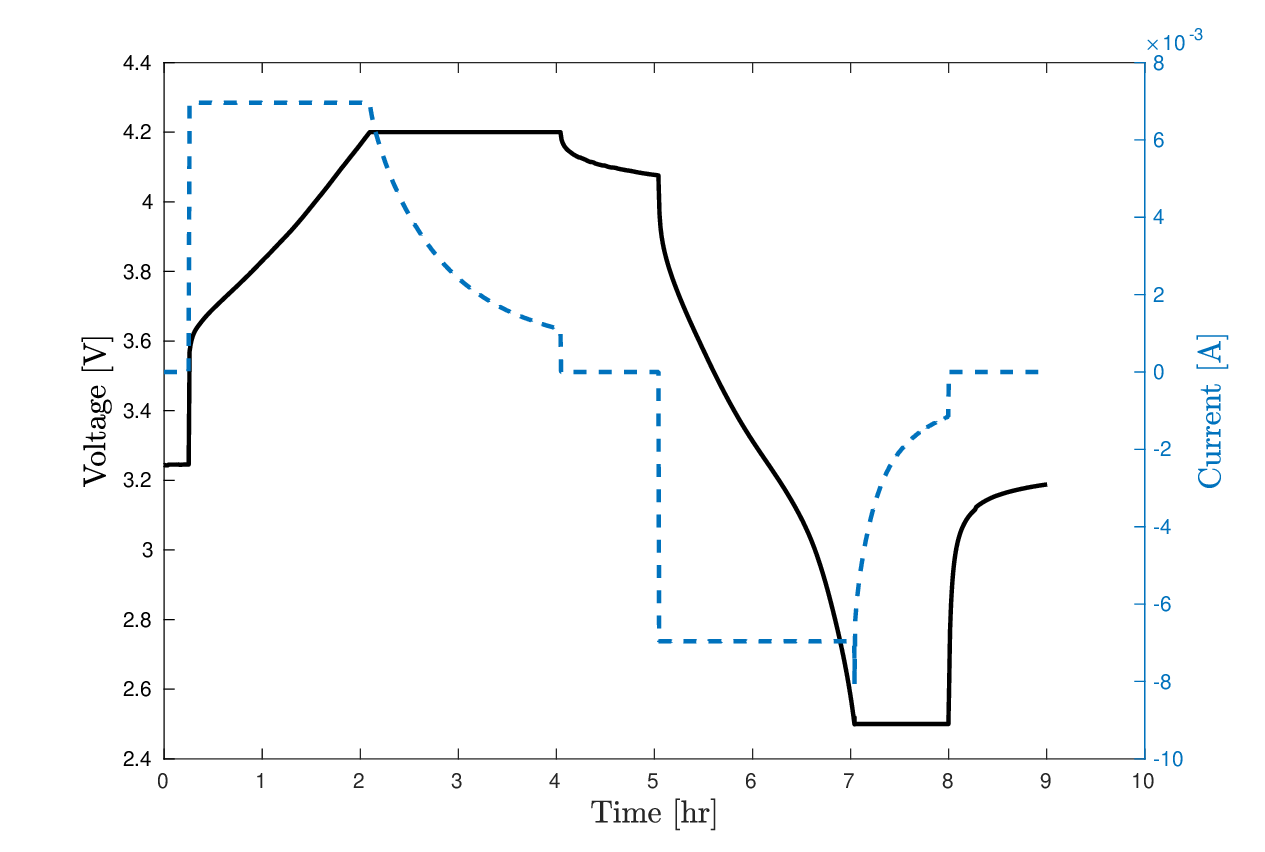}
			\subcaption{C3 cycle}
		\end{subfigure}
		\begin{subfigure}[b]{0.3\textwidth}
			\centering
			\includegraphics[width=1\textwidth]{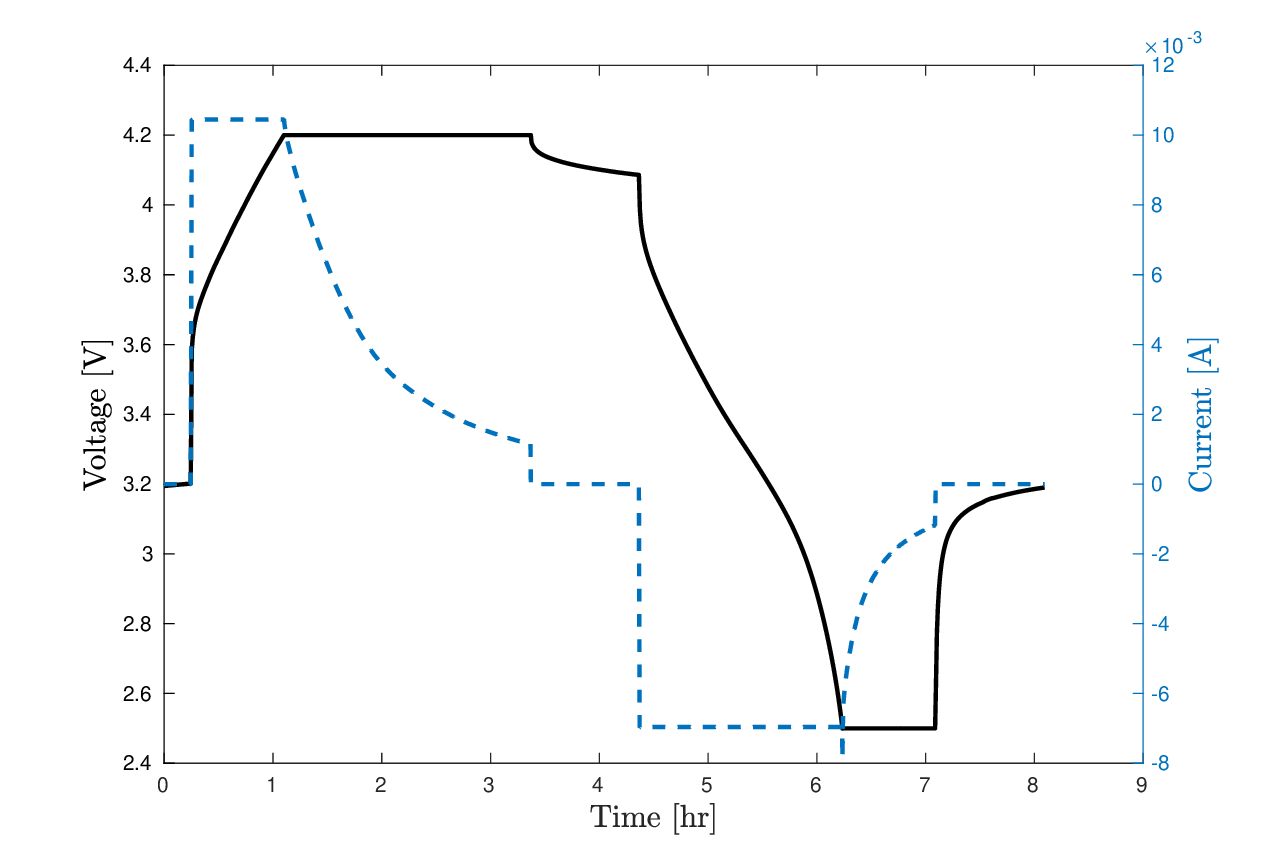}
			\subcaption{C2 cycle}
		\end{subfigure}
		\begin{subfigure}[b]{0.3\textwidth}
			\centering
			\includegraphics[width=1\textwidth]{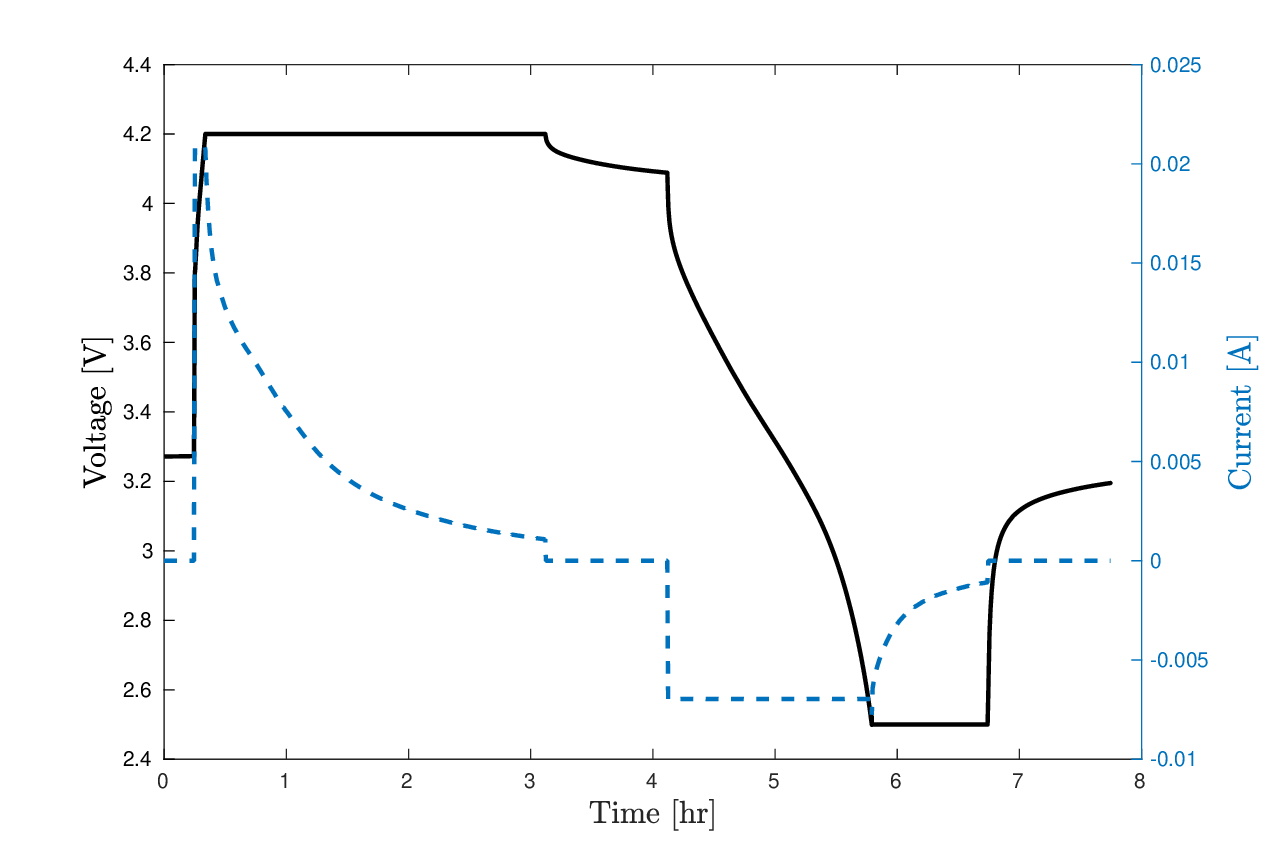}
			\subcaption{1C cycle}
	\end{subfigure}}
	\mbox{
		\begin{subfigure}[b]{0.3\textwidth}
			\centering
			\includegraphics[width=1\textwidth]{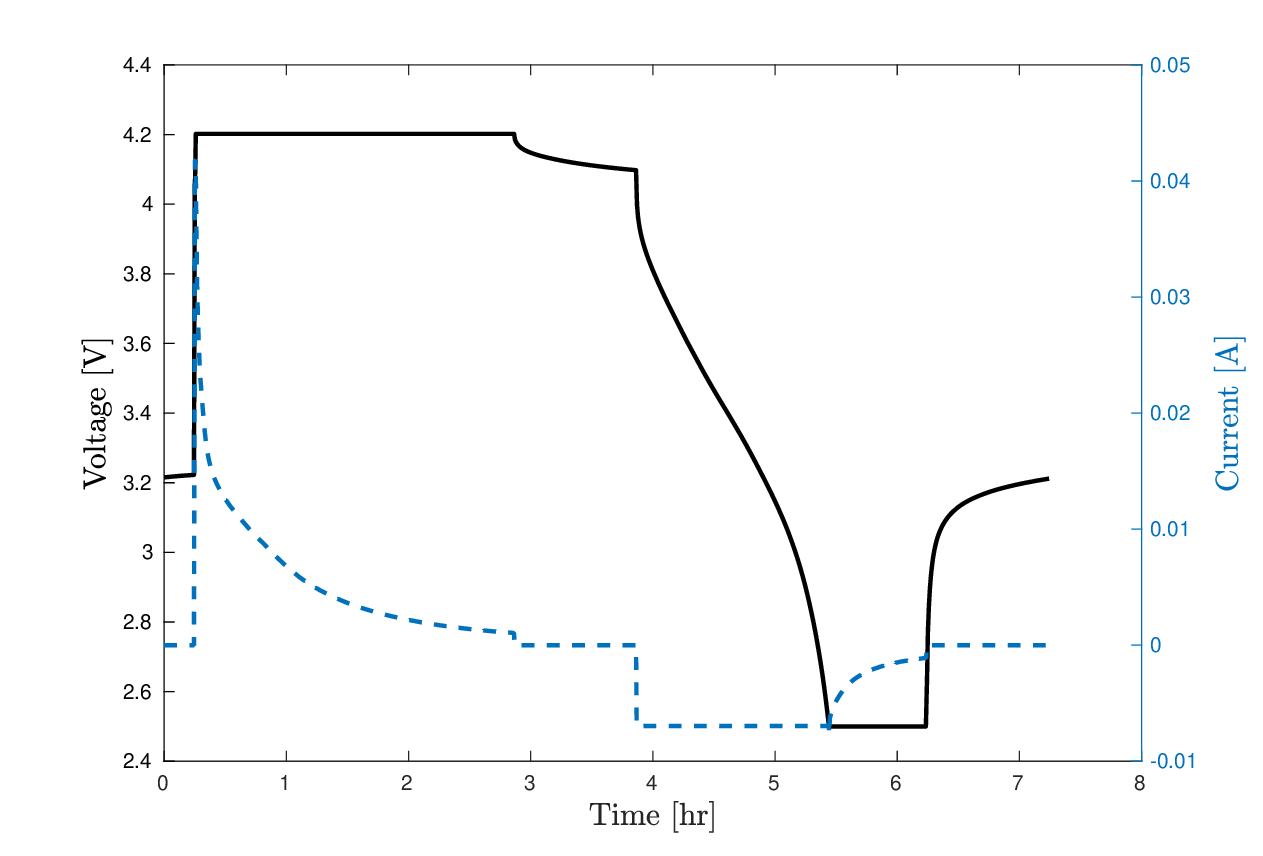}
			\subcaption{2C cycle}
		\end{subfigure}
		\begin{subfigure}[b]{0.3\textwidth}
			\centering
			\includegraphics[width=1\textwidth]{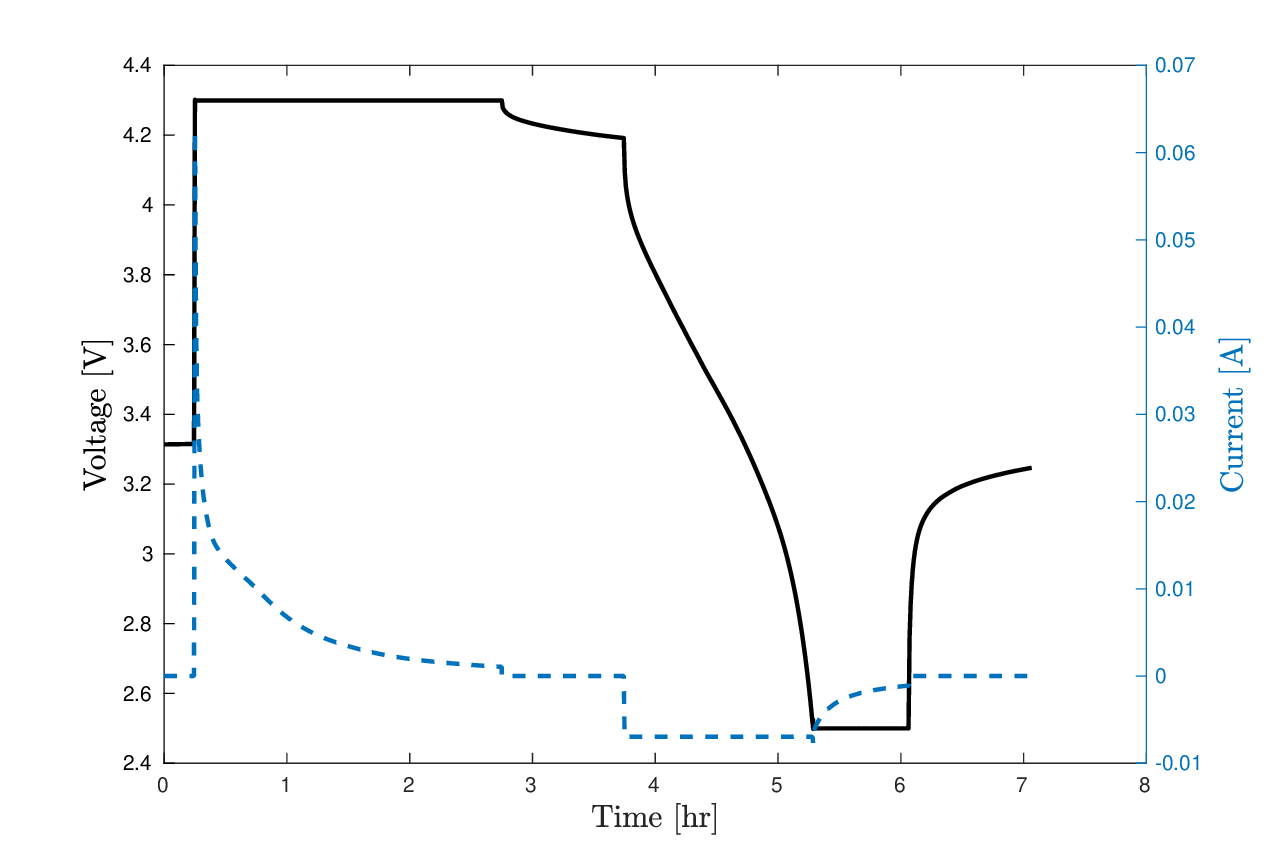}
			\subcaption{3C cycle}
	\end{subfigure}} \\ \medskip
	\mbox{
		\begin{subfigure}[b]{0.3\textwidth}
			\centering
			\includegraphics[width=1\textwidth]{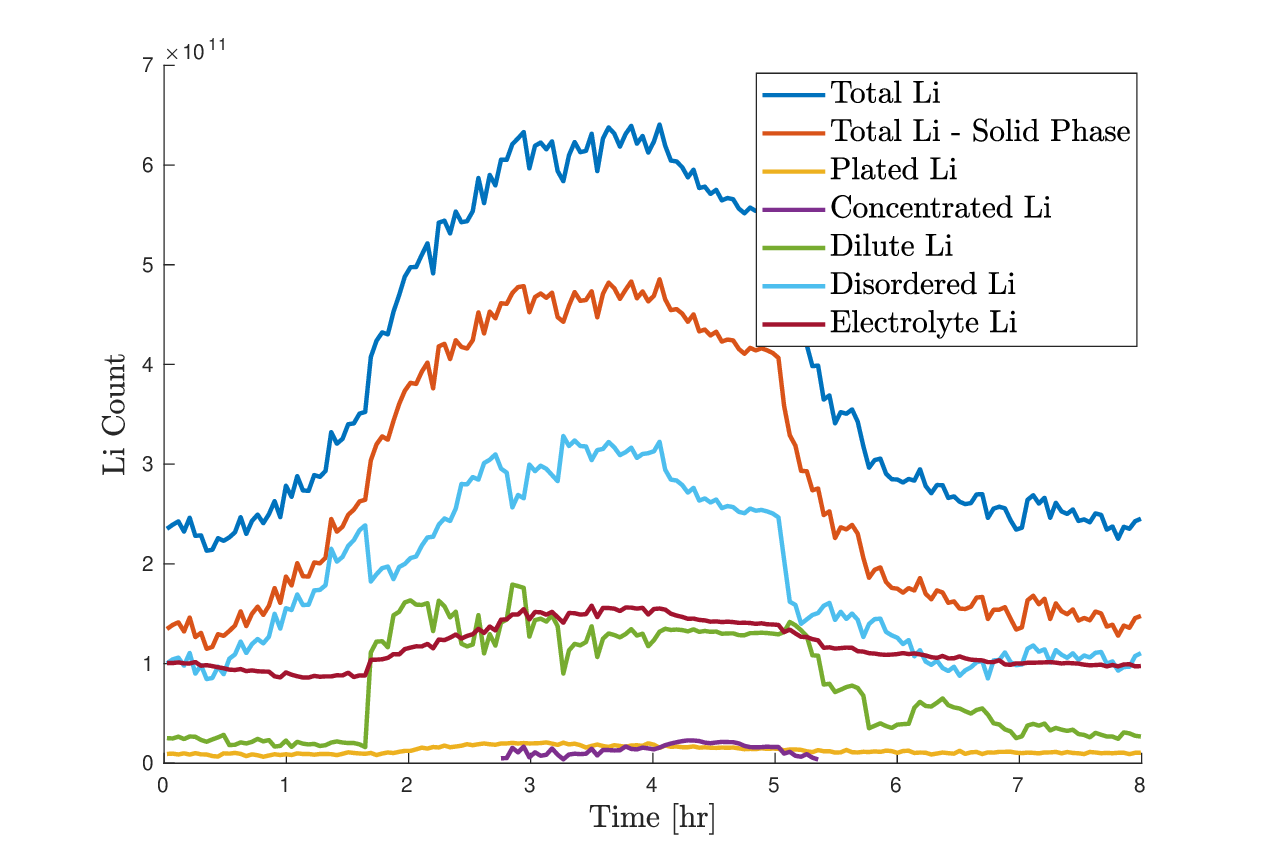}
			\subcaption{C3 cycle}
		\end{subfigure} 
		\begin{subfigure}[b]{0.3\textwidth}
			\centering
			\includegraphics[width=1\textwidth]{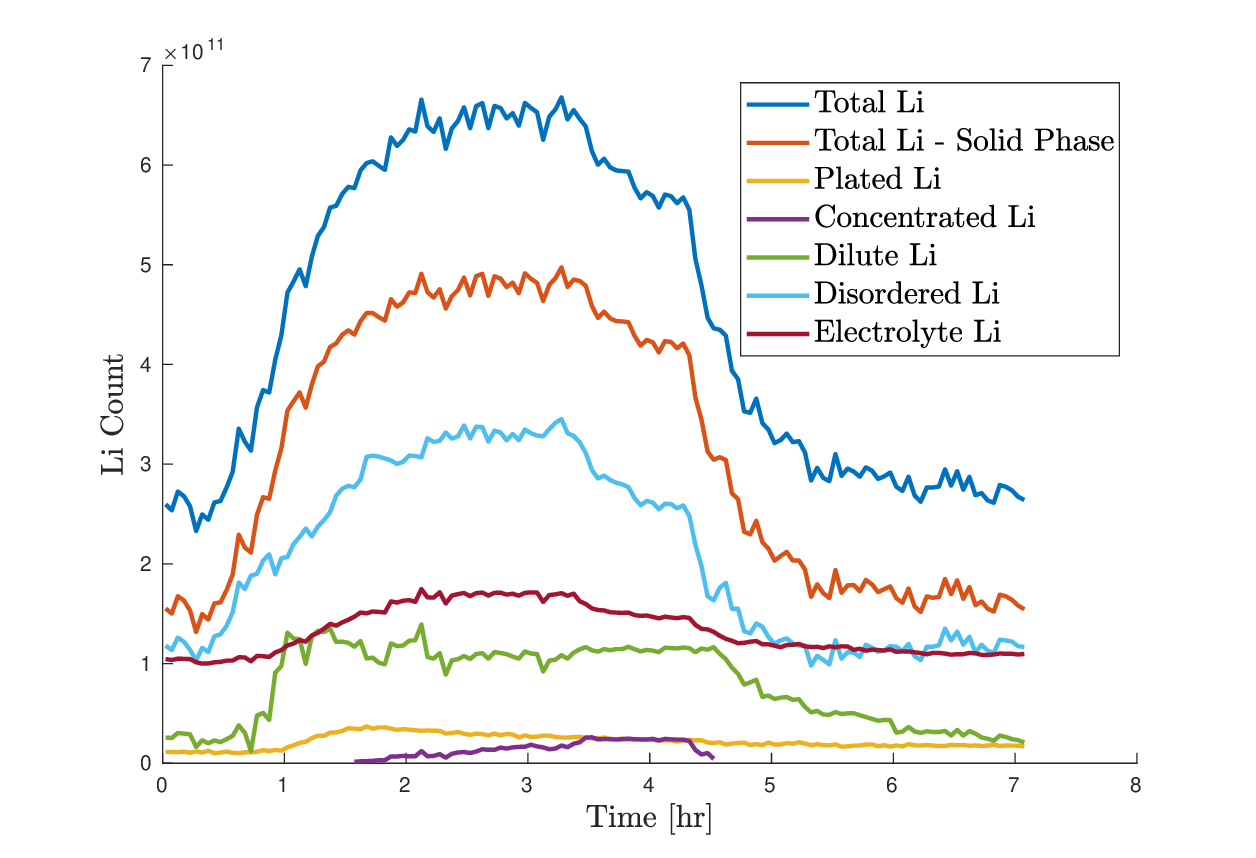}
			\subcaption{C2 cycle}
		\end{subfigure}
		\begin{subfigure}[b]{0.3\textwidth}
			\centering
			\includegraphics[width=1\textwidth]{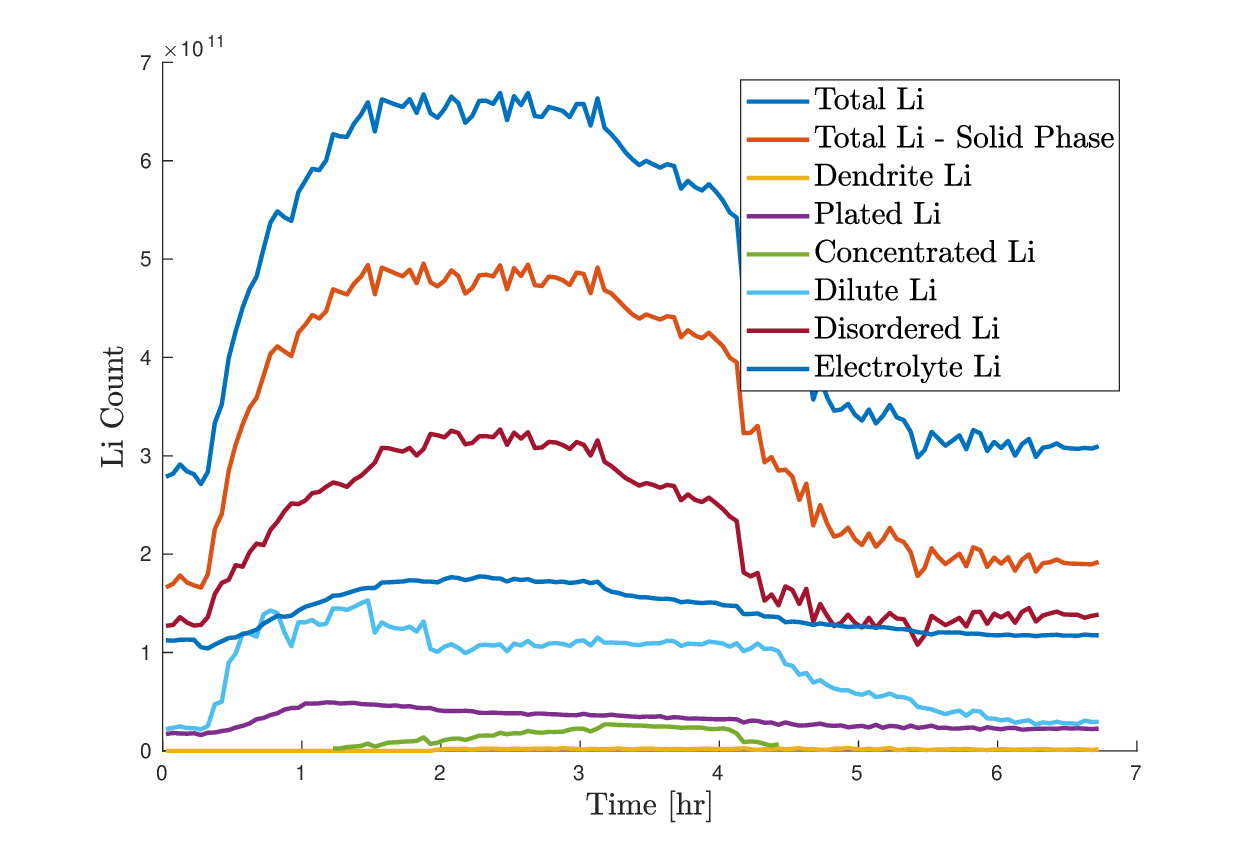}
			\subcaption{1C cycle}
	\end{subfigure}}
	\mbox{
		\begin{subfigure}[b]{0.3\textwidth}
			\centering
			\includegraphics[width=1\textwidth]{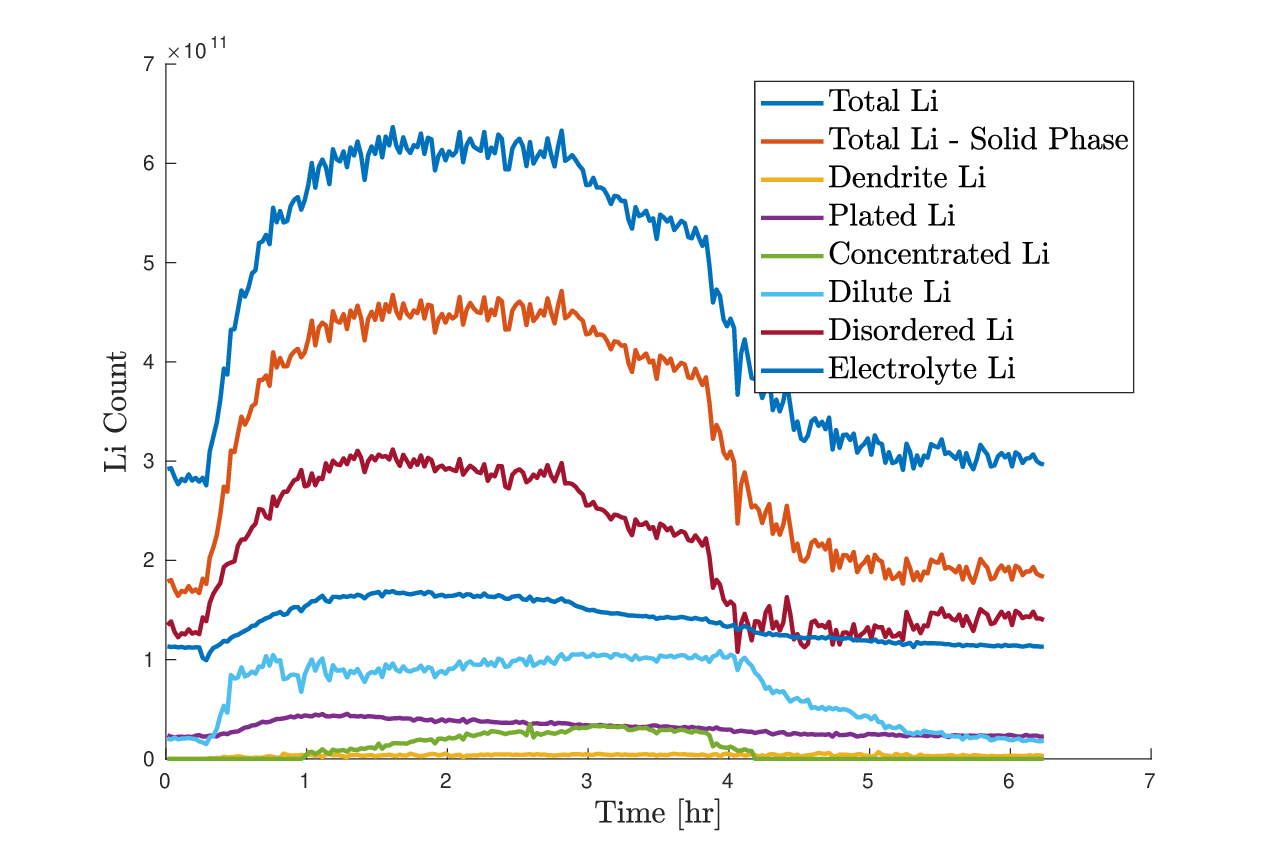}
			\subcaption{2C cycle}
		\end{subfigure}
		\begin{subfigure}[b]{0.3\textwidth}
			\centering
			\includegraphics[width=1\textwidth]{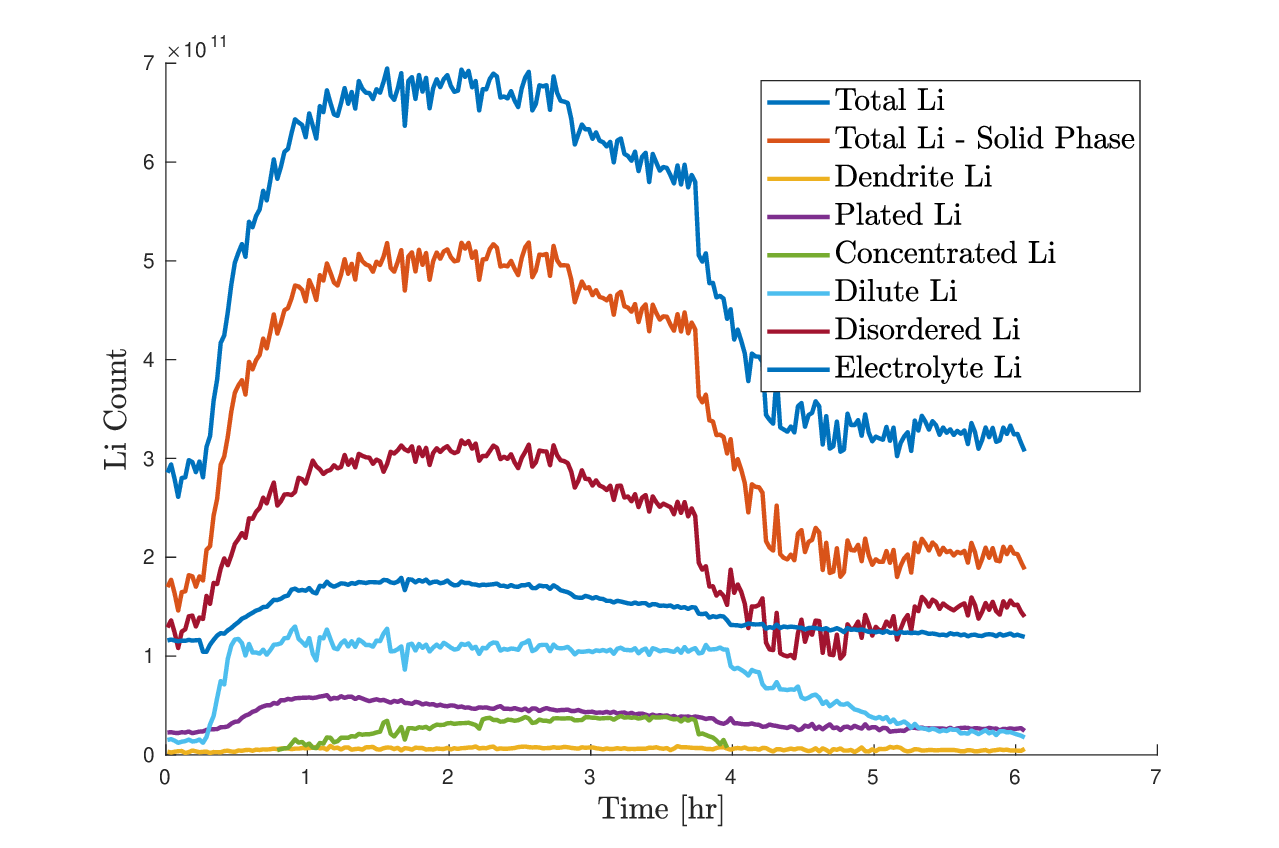}
			\subcaption{3C cycle}
	\end{subfigure}}
	\caption{Terminal voltage and current applied to the cell
		(a,b,c,d,e), and evolution of the Li content in time in
		different phases obtained via the Li-NMR spectroscopy method
		(f,g,h,i,j) using different test protocols of the cell.}
	\label{fig:raw_data}
\end{figure}

To calibrate the mathematical models for subsequent prediction and control, 
one requires experimental data tailored for this purpose. The experimental 
data utilized in this study was collected using the \textit{operando} Li-NMR 
spectroscopy technique, as introduced in a prior publication by Sanders 
et al.~\cite{sanders2023quantitative}. This technique enables the identification 
and quantification of various Lithium phases within the anode while the cell 
is in operation, as depicted in Figure \ref{fig:raw_data}.

The anode material used for these experiments is silicon, recognized
as one of the most promising materials due to its high energy density.
The cathode material employed is NMC622
($\ch{LiNi_{0.6}Mn_{0.2}Co_{0.2}O_2}$). The test protocol of each
experiment comprises a constant-current (CC) charge followed by
constant-voltage (CV) discharge, and open-circuit resting (OCV)
phases. The charge rates for the CC phase are C/3 ($3$-hour charge),
C/2 ($2$-hour charge), 1C ($1$-hour charge), 2C ($30$-minute charge),
and 3C ($20$-minute charge); the discharge rate for the CC phase
remains constant at C/3 ($3$-hour discharge) for all cycles. Here, 'C'
denotes the capacity of the cell. Note that, for simplicity of
notation, the cycles C/3 and C/2 are hereafter denoted C3 and C2,
respectively. The voltage of the cell ranges from $2.5V$ to $4.2V$,
with the lower value representing the full discharge of the cell,
while the higher value corresponds to the fully charged state. The OCV
segment after charge and discharge is set for the duration of one
hour. \textit{Operando} NMR measurements were conducted at intervals
of $5$ minutes for the C3 cycle, $3$ minutes for the C2 and 1C cycles,
and $1.5$ minutes for the 2C and 3C cycles. The evolution of various
Lithium phases from Li-NMR experiments, alongside their operational
current profile and the cell's terminal voltage, is depicted in Figure
\ref{fig:raw_data}.

Several peaks are modelled when fitting NMR spectra to quantify different phases 
of Lithium, including 
\begin{enumerate}
	\item Lithium in the electrolyte or the SEI,
	\item Lithium in dilute $\ch{Li_xSi}$ where $x<2.0$ 
	in a locally-ordered environment (referred to as dilute Li),
	\item Lithium in concentrated $\ch{Li_xSi}$ where $x>2.0$ in a locally-ordered environment 
	(referred to as concentrated Li),
	\item $\ch{Li_xSi}$ in a disordered environment (referred to as disordered Li), and
	\item dendritic and plated Lithium.
\end{enumerate}
These different phases of Lithium are manifested through distinct
chemical shifts in the \textit{operando} NMR data. It is worth noting
that all dendritic Lithium formed is irreversible, while plated
Lithium may exhibit reversible or irreversible behavior. As depicted
in Figure \ref{fig:raw_data}, the evolution of different phases at
constant rates demonstrates a nonlinear behavior during cell
operation. Also, the dendritic Lithium content does not appear in all
cycles, but only in the ones with higher C-rates. In other words, in
the cycles with lower C-rates, the formation of dendritic Lithium is
smaller than the sensitivity of the measurement device.

\begin{figure}[!ht]
	\centering
	\begin{subfigure}[b]{0.44\textwidth}
		\centering
		\includegraphics[width=1\textwidth]{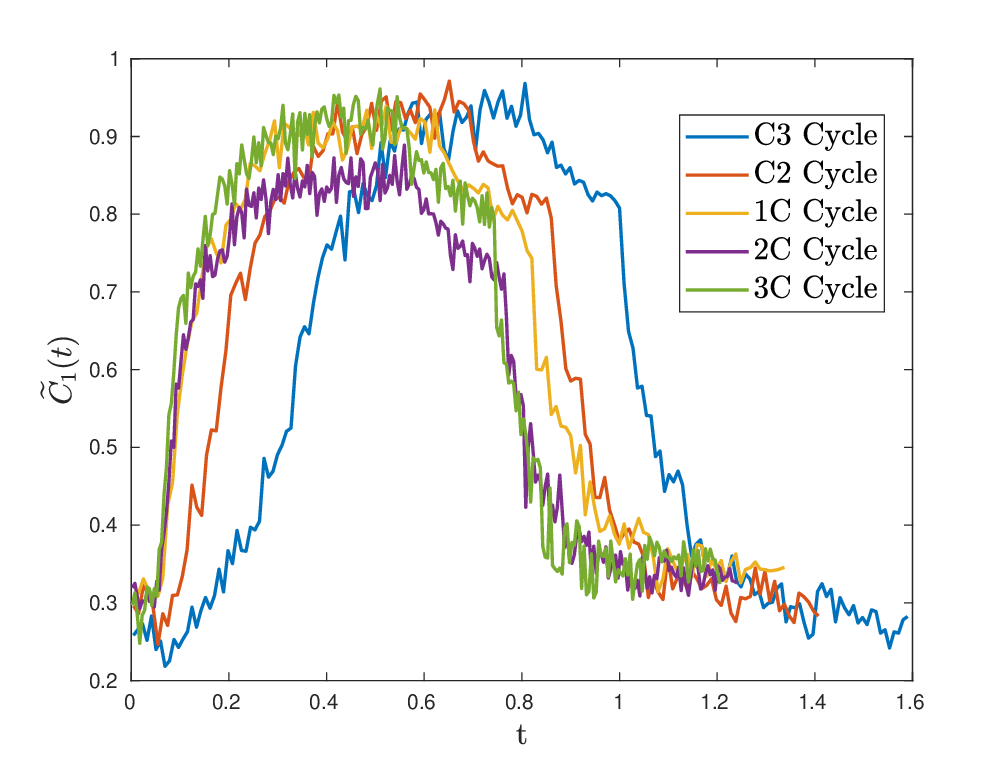}
		\subcaption{}
	\end{subfigure}
	\begin{subfigure}[b]{0.44\textwidth}
		\centering
		\includegraphics[width=1\textwidth]{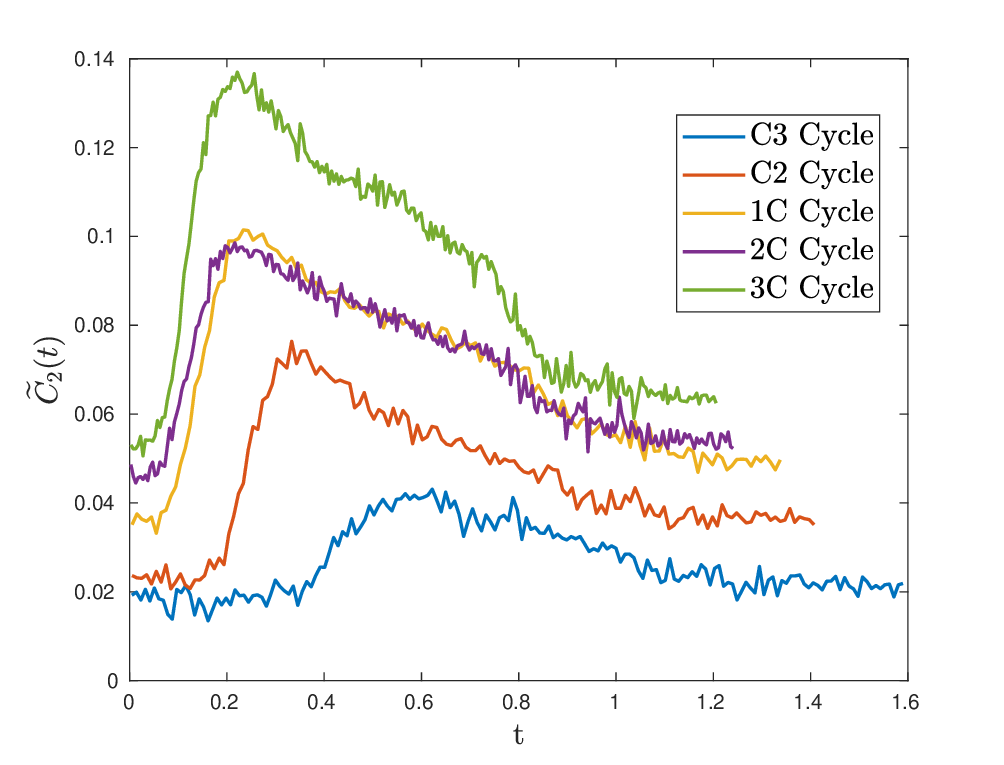}
		\subcaption{}
	\end{subfigure}
	\caption{Evolution of Li content in time in negative electrode solid phase corresponding to intercalated Li (a), and plated Li (b) for different C-rates. Note that the variables are normalized.}
	\label{fig:experimental}
\end{figure}

In order to pre-process the experimental data depicted in Figure
\ref{fig:raw_data} and make format it for the mathematical model and
further analysis, the Lithium in the anode solid phase (dilute Li,
concentrated Li, and disordered Li) is combined to form the solid
phase concentration denoted $\widetilde{C}_1(t)$. Similarly, addition
of plated and dendritic Li content in the cell forms the Li phase
$\widetilde{C}_2(t)$ corresponding to side reactions.  Note that the
subscripts $1$ and $2$ refer to the intercalated Lithium and Lithium
involved in side-reactions, respectively; a notation that is
consistent with the mathematical model in Section \ref{sec:forward}.
These concentrations are normalized and their evolution for each cycle
are shown in Figure \ref{fig:experimental}.  As can be observed in
Figure \ref{fig:raw_data}, the total Li content in the cell does not
add up to a constant and is changing with the cell operation, due to
several factors. First, the Li content in the positive electrode of
the cell is not accounted for in the Li-NMR measurements. The
complement of the Li content in the cell could be stored in the
positive electrode which is not modelled in this case.  Second, the
presence of noise in Li-NMR measurements is another source of
deviation from the conservation of Lithium, cf.~Section
\ref{sec:asymptotic}.  It is also notable that two forms of dynamics
are evident in the cell: the excitation dynamics and the relaxation
dynamics. The excitation dynamics is the response of the system to an
external current and is the dominant regime in the dynamics in the
cell.  The relaxation dynamics represents the evolution of the system
while the cell is at rest in the absence of an external current
source. We note that the dynamics of the system are primarily
determined by excitation, and hence the change in the dynamics due to
excitation is larger than the change due to relaxation.  This fact
will be used in mathematical modeling, cf.~Section \ref{sec:modeling}.

The data for each experiment is split into three regimes: the charging
regime, the OCV regime, and the discharge regime. We will use the
following notation for the amalgamated data $\mathcal{D}_t =
\bigoplus_i \mathcal{D}_i, i\in
\{\text{C3},\text{C2},\text{1C},\text{2C},\text{3C}\}$, where
$\mathcal{D}_i$ refers to the total concentration data available for
the cycle with rate $i$, and $\bigoplus$ denotes the concatenation
operator. Each cycle consists of three regimes: $\mathcal{D}_i =
\bigoplus_{j} \mathcal{D}_i^{j}, j\in\{ch,ocv,dch\}$, where
$\{ch,ocv,dch\}$ refer to charge, OCV, and discharge regimes of the
cell testing protocols, respectively. Different segments of
$\mathcal{D}_t$ will be used for analysis.

\section{Physical Modeling}\label{sec:modeling}
Mathematical modeling plays a pivotal role in comprehending the intricate physical 
processes within a battery cell, exploring degradation mechanisms and their influencing 
factors, and crafting effective mitigation and control strategies for the challenges 
encountered in large-scale applications of Li-ion batteries. Various approaches can be 
employed for mathematical modeling, with two primary paradigms being physics-based modeling 
and data-driven machine-learning modeling. In this study, we aim to navigate the border 
between these two paradigms, with more emphasis placed on physics-based modeling of cells. 
This approach involves utilizing fundamental physical principles to construct a mathematical 
framework that represents the behavior of the cell, augmenting the model, and leveraging 
data-driven strategies for calibration. The ultimate objective is to predict the Li-plating 
dynamics of the cell using experimental data obtained from Li-NMR spectroscopy. The use of 
Li-NMR spectroscopy data in the modeling process highlights the integration of experimental 
data into the physics-based framework. This coupling of experimental observations with 
theoretical modeling can yield highly informative and predictive models for understanding 
and mitigating the critical issue of Li-plating in Li-ion batteries.

The operation of a Li-ion cell involves a multitude of physical and chemical processes, 
each occurring at different spatial and temporal scales. This complex multi-physics 
multi-scale nature of the phenomenon makes it challenging to develop suitable models 
for specific applications aimed at investigating various aspects of the cell's behavior. 
One of the most widely accepted physics-based modeling approaches for Li-ion cells is 
rooted in the porous electrode theory initially introduced by 
Newman et al.~\cite{newman2021electrochemical}. In these models, the cell is treated 
as a continuum medium, and it operates on larger temporal and spatial scales compared to 
discrete particle-level models that necessitate fine-scale resolution. Continuum models, 
which are widely regarded as the fundamental approach, serve as the basis for modeling 
Li-ion cells. Depending on the specific application and research objectives, models of 
varying levels of complexity can be adapted. These models allow researchers to delve 
into the intricacies of the cell's behavior, considering the multitude of physical 
and chemical processes occurring within it. The review by Brosa Planella 
et al.~\cite{planella2022continuum} explores various modeling approaches for Li-ion cells
and introduces a systematic reductive framework, called asymptotic reduction, to simplify 
complex mathematical models using physical assumptions. Among the different modeling 
approaches, the most intricate is the microscale model, which operates at the finest 
temporal and spatial scales to capture the detailed physical phenomena within the cell. 
In microscale modeling, the framework is built upon the conservation laws for Lithium ions 
and counter-ions in the electrolyte, as well as the conservation of Lithium ions and 
electrons in the solid phase of the electrodes. Because electrons serve as charge carriers 
in the solid phase, the conservation of electrons and Lithium ions is treated separately. 
In contrast, ions act as the charge carriers in the electrolyte phase, resulting in more 
intricate conservation equations in the electrolyte phase. To close these conservation 
laws and make them mathematically complete, suitable constitutive relations are employed. 
These relations establish the connection between the flux of a species (e.g. Lithium ions 
or electrons) and the thermodynamic forces acting upon them, allowing for a comprehensive 
representation of the physical processes occurring at the microscale within the Li-ion 
cell. The process of Lithium intercalation and deintercalation primarily takes place on 
the surface of the anode particles and is considered as an interfacial phenomenon. The 
rate of intercalation reactions on the surface of the anode active material depends 
on the surface overpotential. This overpotential is defined as the difference between 
the electrochemical potential of Li ions on the surface of the solid phase and the Li 
ions in the adjacent electrolyte, and is typically represented by the well-known 
Butler-Volmer (BV) relation. Solid-phase diffusion of Lithium ions within the anode is 
a complex process that may involve phase transition phenomena. While simplifying 
assumptions are often applied to this diffusion process, conventional diffusion equations 
are commonly used to model it. However, recent modeling techniques have relaxed the 
assumption of linear diffusion and instead incorporate nonlinear diffusion within the 
solid phase. This nonlinear diffusion accounts for concentration-dependent diffusion 
coefficients, making it capable of capturing phase-transition behavior 
\cite{escalante2020discerning, o2022lithium}. Furthermore, an alternative approach based 
on the Cahn-Hilliard modeling framework has gained attention due to its ability to 
naturally capture the dynamics of phase transitions during solid-state diffusion
\cite{guo2016li,escalante2020discerning}. In the electrolyte, charge transport is described 
using various theories. Two common theories include (i) Dilute Electrolyte Theory, 
which is based on Nernst-Planck equations and is applicable to dilute electrolytes where 
there is limited interaction between species, and (ii) Concentrated Electrolyte Theory, 
which is based on Stefan-Maxwell type equations and is more suitable for concentrated 
electrolyte solutions. These models govern charge transport in the electrolyte phase and 
are essential components of comprehensive Li-ion cell models. Microscale models offer a 
highly detailed representation of Li-ion cells, but they come with the drawback of 
demanding significant computational resources and requiring extensive knowledge about 
the microstructure of various cell components. They also depend on more parameters which 
makes them harder to calibrate. As a result, they are not well-suited for online 
estimation and control, where real-time decision-making is essential.

To address these challenges, one could rely 
on some simplifying assumptions to transform complex microscale models into
more manageable homogenized models \cite{planella2022continuum}. In homogenized 
models, the porous media within the cell is treated as a continuum, and the equations 
are modified to incorporate the influence of the microstructure. This approach allows 
for the resolution of electrolyte flow at the macroscale, while retaining the microscale 
representation of solid-state diffusion, as this process is typically slow and 
involves significant concentration gradients in fine spatial scales. In this type of 
modeling, microstructural information is still required, but the model simplifies this 
by generalizing a small subdomain to represent the entire domain using periodic boundary 
conditions. These homogenized models can be reduced to the well-known Doyle-Fuller-Newman (DFN) 
model by assuming a simpler geometry for all electrode particles, a model that is also 
referred to as the pseudo-two-dimensional (P2D) or Newman model, firstly introduced by 
Fuller, Doyle, and Newman \cite{fullerdoylenewman1994}. The DFN model simplifies the 
representation of electrode particles by assuming them to be spherical. Consequently, 
it solves the solid-state diffusion equations in a 1D radial coordinate, rather than 
attempting to capture the intricate 3D microstructure of the electrode particles. 
Similarly, the electrolyte equations are solved in a 1D planar geometry. This 
simplification results in a model that can be conceptually described as 1D+1D, giving 
rise to the term ``pseudo-two-dimensional''. The P2D model is renowned for its 
computational efficiency while retaining the capability to capture the internal 
dynamics and behavior of Li-ion cells.

In pursuit of enhanced computational efficiency and suitability for online estimation 
and control, reduced-order models have been introduced as alternatives to the comprehensive 
DFN models. Two notable reduced-order models are the Single-Particle Model (SPM), originally 
introduced by Atlung et al.~\cite{Atlung1979dynamic}, and the Single-Particle Model with 
Electrolyte (SPMe), developed by Prada et al.~\cite{prada2012simplified}. The fundamental 
assumption in these models is that the spherical electrode particles, as considered in the
DFN model, are sufficiently similar in nature. This similarity allows these particles 
to be effectively represented by a single averaged or representative particle. It is 
assumed that the intercalation and deintercalation processes occur almost uniformly 
across all electrode particles, making it feasible to describe these processes using 
a single representative particle. In this setting, the partial differential equations 
(PDEs) governing the Li-ion cell behavior can be effectively decoupled into micro and 
macro scales. It is worth noting that the SPMe model, unlike the simpler SPM model, accounts 
for the electrolyte dynamics, offering a more comprehensive representation of cell dynamics 
by considering the behavior of the electrolyte phase. The SPM and SPMe models could be 
achieved by asymptotic reduction of DFN model as developed by different authors 
\cite{planella2023single, marquis2019asymptotic, richardson2020generalised}. 
Certain physical assumptions are 
used by Marquis et al.~\cite{marquis2019asymptotic} to systematically reduce the 
DFN model to a much simpler SPM model which will affect the range of validity of these 
models. The physical assumptions include high electrical conductivity in the electrodes 
and electrolyte, and fast Li ion migration in electrolyte in comparison to the discharge 
timescale. The range of validity of the SPM model according to 
Brosa Planella et al.~\cite{planella2023single} is small overpotentials from 
open-circuit-voltage, and weak side reactions. The two assumptions hold for low 
to moderate charge rates and will break at high rates. In summary, starting from the most 
complex microscale model and utilizing a systematic asymptotic reduction framework, 
the order of complexity can be progressively reduced.

The objective of this research is to adopt a simple model that can effectively capture
the internal dynamics of a Li-ion cell, focusing on the interactions among various particles 
within the cell. The physical model is developed in a manner 
to match the experimental data obtained from Li-NMR experiments. The physical modeling
framework of this study is inspired by the SP model with side reactions, recently introduced by
Brosa Planella et al.~\cite{planella2023single}. This study also finds close connections
to a recent study by Sahu et al.~\cite{sahu2023continuum}.
The model used in this study could be 
seen as a simpler version of the SP model with side reactions, where certain parameters 
and functions are to be calibrated using experimental data. The final model takes the 
form of a system of ordinary differential equations (ODE). It involves employing a SP
model in the form of partial differential equations (PDE) and applying reduction and 
averaging techniques to derive a suitable ODE model that describes the evolution of key 
space-averaged concentrations within the cell. Some aspects of the model are shown to 
increase its flexibility in fitting the experimental data. We begin by introducing 
the DFN model in Section \ref{sec:DFN}, developing the dimensionless model in Section 
\ref{sec:dimensionless}, applying the asymptotic reduction technique in Section 
\ref{sec:asymptotic}, and finally introducing our dynamical system as forward model 
in Section \ref{sec:forward}. The key differences of our model with similar studies are highlighted 
in Section \ref{sec:comparison}.

\subsection{DFN Model}\label{sec:DFN}
In this study, we begin by presenting the 1D DFN model. The SP model is 
derived from an asymptotic reduction analysis. This model is then further simplified 
using averaging techniques to yield a mathematical representation suitable for 
modeling our experimental data, cf.~Section \ref{sec:experiment}. 
It is noteworthy that our modeling approach is inspired by the SPMe+SR (Single-Particle Model with 
Side Reactions) framework of Brosa Planella et al.~\cite{planella2023single}, albeit with 
some modifications to the underlying assumptions, which serve to mitigate certain limitations 
associated with the Brosa Planella's model. The differences in modeling assumptions are highlighted 
in Section \ref{sec:asymptotic}. The current study also finds close connections to 
Li plating modeling efforts of Sahu et al.~\cite{sahu2023continuum}.
The SP model assumes the presence of a representative 
(averaged) particle to describe the transport of species within the solid state of the 
electrode. The key assumption is that all solid particles within the electrode are 
indistinguishable, allowing a single particle to serve as a representative for the entire 
solid phase. It is important to note that the cathode component of the cell is 
also considered in the modeling effort, however, the final model (presented in 
Section \ref{sec:forward}) eliminates the need for solving for the positive electrode 
components, as the experimental data does not contain information from the positive 
electrode domain. The model is composed of five distinct components, namely, charge 
conservation in the solid phase of positive and negative electrodes, Li Transport in the 
solid phase of positive and negative electrodes, Li transport in the electrolyte phase, 
charge conservation in the electrolyte phase, and models of side reactions through interfacial 
dynamics. Each of these components is explained in more detail below. 
Note that in our model we only take into account the Li-plating side reaction and we disregard other
side reactions in the cell (e.g., SEI growth). We also disregard the film resistance formed on the surface of 
the anode particle due to side reactions, and porosity change of the anode particles in time is not 
modelled. Also, the volume change of anode particles (which could be significant in silicon anodes) 
is not explicitly considered in this model, however, the concentration-dependent constitutive 
relations can implicitly take this effect into account, as described in Section \ref{sec:forward}. 

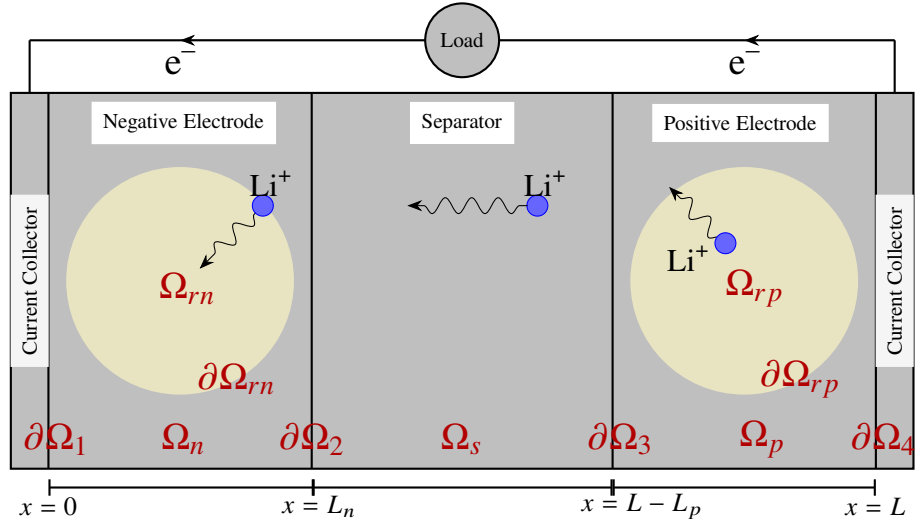
\begin{figure}
	\centering
	\scalebox{1}{
		\begin{tikzpicture}
		\node[rectangle, thick, draw = black,text = olive, fill = black!25, minimum width = 12cm, minimum height = 5cm, font=\sffamily\Large\bfseries] (r) at (0,0) {};
		
		\draw[black,thick,solid] (-5.5,-2.5) -- (-5.5,2.5);
		\draw[black,thick,solid] (5.5,-2.5) -- (5.5,2.5);
		\draw[black,thick,solid] (-2,-2.5) -- (-2,2.5);
		\draw[black,thick,solid] (2,-2.5) -- (2,2.5);
		
		\draw[black,thick,solid] (-5.75,2.5) -- (-5.75,3.2);
		\draw[black,thick,solid] (5.75,2.5) -- (5.75,3.2);
		\draw[black,thick,solid] (-5.75,3.2) -- (5.75,3.2);
		\draw[-{Stealth[length=2mm, width=1.5mm]}] (5.75,3.2) -> (3.75,3.2);
		\draw[-{Stealth[length=2mm, width=1.5mm]}] (3.75,3.2) -> (-3.75,3.2);
		\node [thick, text=black, inner sep=2pt, font=\large] at (3.75,2.9) {$\ch{e-}$};
		\node [thick, text=black, inner sep=2pt, font=\large] at (-3.75,2.9) {$\ch{e-}$};
		\node[circle, thick, draw = black,text = black, minimum size=0.8cm, fill = black!25, font=\scriptsize] () at (0,3.2) {Load};

		\node[rectangle, thick, draw = black!0,text = black, fill = black!0, minimum width = 0.2cm, minimum height = 0.2cm,font=\scriptsize] (r) at (-3.7,2.1) {Negative Electrode};
		\node[rectangle, thick, draw = black!0,text = black, fill = black!0, minimum width = 0.2cm, minimum height = 0.2cm,font=\scriptsize] (r) at (0,2.1) {Separator};
		\node[rectangle, thick, draw = black!0,text = black, fill = black!0, minimum width = 0.2cm, minimum height = 0.2cm,font=\scriptsize] (r) at (3.7,2.1) {Positive Electrode};
		\node[rectangle, thick, draw = black!4,text = black, fill = black!4, minimum width = 0.2cm, minimum height = 0.2cm, rotate=90,font=\scriptsize] (r) at (-5.75,0) {Current Collector};
		\node[rectangle, thick, draw = black!4,text = black, fill = black!4, minimum width = 0.2cm, minimum height = 0.2cm, rotate=90,font=\scriptsize] (r) at (5.75,0) {Current Collector};

		\node[circle, thick, draw = olive!20,text = olive, minimum size=3cm, fill = olive!20, font=\sffamily\Large\bfseries] () at (-3.75,0) {};
		\node[circle, thick, draw = olive!20,text = olive, minimum size=3cm, fill = olive!20, font=\sffamily\Large\bfseries] () at (3.75,0) {};
		
		\node[circle, draw = blue!100,text = , minimum size=0.1cm, fill = blue!60, font=\scriptsize, scale=0.7] () at (1,1) {};
		\draw[-{Stealth[length=2mm, width=1.5mm]},decorate,decoration={snake,amplitude=3pt,pre length=3pt,post length=3pt}] (0.87,1) -- ++(-1.6	,0);
		\node [thick, inner sep=1pt] at (1.1,1.3) {$\ch{Li+}$};
		\draw[-{Stealth[length=2mm, width=1.5mm]},decorate,decoration={snake,amplitude=3pt,pre length=3pt,post length=3pt}] (-2.69,0.96) -- ++(-0.8,-0.8);
		\node[circle, draw = blue!100,text = , minimum size=0.1cm, fill = blue!60, font=\scriptsize, scale=0.7] () at (-2.65,1) {};
		\node [thick, inner sep=1pt] at (-2.55,1.3) {$\ch{Li+}$};
		\draw[-{Stealth[length=2mm, width=1.5mm]},decorate,decoration={snake,amplitude=3pt,pre length=3pt,post length=3pt}] (3.5,0.5) -- ++(-0.75,+0.75);
		\node[circle, draw = blue!100,text = , minimum size=0.1cm, fill = blue!60, font=\scriptsize, scale=0.7] () at (3.5,0.5) {};
		\node [thick, inner sep=1pt] at (3,0.3) {$\ch{Li+}$};
		
		\node [thick, text=red!70!black, inner sep=2pt, font=\large] at (0,-2.1) {$\Omega_s$};
		\node [thick, text=red!70!black, inner sep=2pt, font=\large] at (-3.7,-2.1) {$\Omega_n$};
		\node [thick, text=red!70!black, inner sep=2pt, font=\large] at (-5.4,-2.1) {$\partial\Omega_1$};
		\node [thick, text=red!70!black, inner sep=2pt, font=\large] at (-2,-2.1) {$\partial\Omega_2$};
		\node [thick, text=red!70!black, inner sep=2pt, font=\large] at (2.1,-2.1) {$\partial\Omega_3$};
		\node [thick, text=red!70!black, inner sep=2pt, font=\large] at (4,-2.1) {$\Omega_p$};
		\node [thick, text=red!70!black, inner sep=2pt, font=\large] at (5.6,-2.1) {$\partial\Omega_4$};
		\node [thick, text=red!70!black, inner sep=2pt, font=\large] at (-3.65,-0.1) {$\Omega_{rn}$};
		\node [thick, text=red!70!black, inner sep=2pt, font=\large] at (-3,-1.3) {$\partial\Omega_{rn}$};
		\node [thick, text=red!70!black, inner sep=2pt, font=\large] at (3.9,-0.1) {$\Omega_{rp}$};
		\node [thick, text=red!70!black, inner sep=2pt, font=\large] at (4.5,-1.3) {$\partial\Omega_{rp}$};
		
		\node[thick, text=black, inner sep=2pt, font=\footnotesize] at (-5.5, -3)   (a) {$x=0$};
		\node[thick, text=black, inner sep=2pt, font=\footnotesize] at (-1.9, -3)   (b) {$x=L_n$};
		\node[thick, text=black, inner sep=2pt, font=\footnotesize] at (2.4, -3)   (c) {$x=L-L_p$};
		\node[thick, text=black, inner sep=2pt, font=\footnotesize] at (5.5, -3)   (c) {$x=L$};
		\draw[{Bar}-{Bar},decorate, thick] (-5.5,-2.75) -- ++(3.54,0);
		\draw[{Bar}-{Bar},decorate, thick] (-2,-2.75) -- ++(4,0);
		\draw[{Bar}-{Bar},decorate, thick] (2,-2.75) -- ++(3.5,0);
		
		\end{tikzpicture}}
	\caption{Schematic of a Li-ion cell in the charging state. Li ions deintercalate from the positive
		electrode surface, migrate toward the negative electrode through the electrolyte, and intercalate 
		int negative particles. Electrons will migrate through the external circuit toward the negative 
		electrode. The domain of the system is highlighted in red.}
	\label{fig:schematic}
\end{figure}

The model geometry consists of the negative electrode ($\Omega_n$), the separator ($\Omega_s$), 
and the positive electrode ($\Omega_p$) where $\Omega = \Omega_n \cup \Omega_s \cup \Omega_p$. 
The model's geometry is depicted in Figure \ref{fig:schematic}, where the 1D macroscale 
coordinate is indicated on the horizontal axis with $\Omega_n = \left[0,L_n\right]$,  
$\Omega_s = \left[L_n,L - L_p\right]$ and $\Omega_p = \left[L-L_p,L\right]$, where $L_n,L_p>0$ 
are the widths of the negative electrode and positive electrode, respectively. In contrast, 
the microscale dimension is described using 
spherical coordinates with $r \in \Omega_{rn} = \left[0,R_n\right]$ for the negative particle 
and $r \in \Omega_{rp} = \left[0,R_p\right]$ for the positive particle, where $R_n$ and $R_p$ 
represent the radii of the spherical negative and positive particles, respectively. In our study, each of 
these sub-models is averaged over its respective spatial domain to eliminate the spatial 
dependence of the model to match to the experimental data.

Note that in the following sections, where the mathematical model is presented, variables 
with a hat are dimensionless, variables in bold are vector quantities, and variables with 
a bar refer to quantities averaged over their spatial domain. Subscripts $n$, $e$, and $p$ 
refer to the negative electrode solid phase, the electrolyte phase, and the positive 
electrode solid phase, respectively. In each of the following subsections, different 
components of the DFN model are presented. 

\subsubsection{Charge Conservation in the Solid Phase} 
This sub-model describes charge conservation within the solid phase of the negative electrode. 
The charge conservation equation is stated in macroscale in $\Omega_n$. 
The potential profile in the solid phase is obtained by solving the following equation in 1D as
\begin{equation}
\begin{alignedat}{3}
\frac{\partial J_n }{\partial x} &= - a_n J_{n,tot},\\
J_n &= - \sigma_n \frac{\partial \phi_n}{\partial x},\\
J_n &= J_{app}, \qquad &&\qquad\text{at}\qquad x=0,\\
J_n &= 0, \qquad &&\qquad\text{at} \qquad x=L_n,
\end{alignedat}
\label{eq:charge_conservation_solid_1D}
\end{equation}
where $\phi_n [V]$ is the electrostatic potential in the solid phase, $J_n [\frac{A}{m^2}]$ is 
the current density in the solid phase,  $\sigma_n [S/m]$ is the effective conductivity of 
the solid particles, $J_{n,tot} [\frac{A}{m^2}]$ is the source/sink term representing the total 
current density flux at the solid-electrolyte interface of the negative electrode
due to intercalation and side reactions, 
$J_{app} [\frac{A}{m^2}]$ denotes the current density applied to the cell, and 
$a_n = \frac{3}{R_n} [\frac{1}{m}]$ is the effective surface area of the anode particles per unit 
volume. Similarly, the charge conservation in the solid phase for the positive electrode in 1D becomes
\begin{equation}
\begin{alignedat}{3}
\frac{\partial J_p }{\partial x} &= - a_p J_{p,tot},\\
J_p &= - \sigma_p \frac{\partial \phi_p}{\partial x},\\
J_p &= 0, \qquad &&\qquad\text{at}\qquad x=L-L_p,\\
J_p &= J_{app}, \qquad &&\qquad\text{at} \qquad x=L.
\end{alignedat}
\label{eq:charge_conservation_solid_1D_p}
\end{equation}
Note that the total current density is a source/sink term that is present in negative electrode 
and positive electrode only and vanishes in the separator, as
\begin{equation}
\begin{alignedat}{3}
J_{tot} &= 
\begin{cases}
J_{n,tot} = J_{n,int}+J_{n,sr} & 0\le x \le L_n,\\
0 & L_n \le x \le L-L_p,\\
J_{p,tot} = J_{p,int} & L - L_p \le x \le L,
\end{cases}
\end{alignedat}
\label{eq:Jtot}
\end{equation}
where $J_{n,int}$, $J_{n,sr}$ and $J_{p,int}$ represent intercalation and side reaction current 
densities at the solid-electrolyte interface of the negative electrode and intercalation current
density at the solid-electrolyte interface of the positive electrode, respectively. Note that 
no side reaction is assumed on the positive electrode.

\subsubsection{Li Ion Transport in the Solid Phase} 
This sub-model describes the slow diffusion of Li ions inside the solid phase. The diffusion 
equation for Li transport in the solid phase is stated in the microscale in the spherical 
coordinates for a representative particle (assuming uniformity along all particles).
In the 1D spherical coordinates, the system is
\begin{equation}
\begin{alignedat}{3}
\frac{\partial C_n}{\partial t} &= \frac{1}{r^2} \frac{\partial}{\partial r} \left(r^2 D_n \frac{\partial C_n}{\partial r}\right),\qquad && \qquad r\in (0,R_n),\\
\frac{\partial C_n}{\partial r} &= 0, \qquad &&\text{at} \qquad r=0,\\
-D_n \frac{\partial C_n}{\partial r} &= \frac{J_{n,tot}}{F}, \qquad &&\text{at} \qquad r=R_n,\\
C_n &= C_{n_{i}}(r), \qquad &&\text{at} \qquad t=0,
\end{alignedat}
\label{eq:Li_transport_solid_1D}
\end{equation}
where $C_n = C_n(r,t) [\frac{mol}{m^3}]$, $D_n [\frac{m^2}{s}]$, $C_{n_{i}}(r)$ 
are the Li concentration, the diffusion coefficient, 
and initial concentration profile, respectively. $N_{tot}$ is the total molar flux at 
the solid-electrolyte interface as $N_{tot} = N_{int} + N_{sr}$, where $N_{int}$ is the 
molar flux of Li corresponding to the intercalation process, whereas $N_{sr}$ is the molar 
flux of Li resulting from side reactions. The Li flux on the surface of the anode particle is 
obtained from $\bn\cdot\bN_n\big\rvert_{r=R_n} = -\frac{J_{n,tot}}{F}$, 
where $J_{n,tot} [\frac{A}{m^2}]$ is the current density flux at the interface (obtained 
from the Butler-Volmer relation), and $F [\frac{A.s}{mol}]$ is Faraday's constant. 
Note that the interfacial current density $J_{n,tot}$ will be replaced with the intercalation current 
density $J_{n,int}$ in Section \ref{sec:asymptotic}, to account for side reaction as well.
Similarly, Li transport in the solid phase of the positive particles in 1D spherical 
coordinates is governed by 
\begin{equation}
\begin{alignedat}{3}
\frac{\partial C_p}{\partial t} &= \frac{1}{r^2} \frac{\partial}{\partial r} \left(r^2 D_p \frac{\partial C_p}{\partial r}\right),\qquad && \qquad r\in (0,R_p),\\
\frac{\partial C_p}{\partial r} &= 0, \qquad &&\text{at} \qquad r=0,\\
-D_p \frac{\partial C_p}{\partial r} &= -\frac{J_{p,tot}}{F}, \qquad &&\text{at} \qquad r=R_p,\\
C_p &= C_{p_{i}}(r), \qquad &&\text{at} \qquad t=0.
\end{alignedat}
\label{eq:Li_transport_solid_1D_p}
\end{equation}

\subsubsection{Charge Conservation in the Electrolyte Phase}
This sub-model describes charge conservation within the electrolyte phase. The continuity equation of 
charge conservation in the electrolyte phase is defined in the macroscale in $\Omega$ in
terms of the potential profile and has the form
\begin{equation}
\begin{alignedat}{3}
\frac{\partial}{ \partial x} J_e &=  a J_{tot},\\
J_e &= -\sigma_e B(x)\left[\frac{\partial}{ \partial x}\phi_e - 2(1-t^+) \frac{RT}{F}\frac{\partial}{ \partial x}\log C_e\right],\\
\frac{\partial}{ \partial x}\phi_e  &= 2(1-t^+) \frac{RT}{F}\frac{\partial}{ \partial x}\log C_e, \qquad &&\qquad\text{at} \qquad x=0, L,\\
\end{alignedat}
\label{eq:charge_consservation_electrolyte_1D}
\end{equation}
where $J_e$ is the current density in the electrolyte phase, $\phi_e$ is the potential in 
electrolyte phase, $\sigma_e$ is the electrolyte conductivity, $B=B(x,t)$ is the permeability, 
and $t^+=t^+(C_e)$ is the transference number.

\subsubsection{Li Ion Transport in the Electrolyte Phase} 
This sub-model deals with the transport of Li ions within the electrolyte phase at macroscale. 
The continuity equation for Li ion conservation in the electrolyte phase is stated on $\Omega$ 
in terms of the concentration profile of Li ions and has the following form
\begin{equation}
\begin{alignedat}{3}
\frac{\partial}{ \partial t} (\epsilon C_e) &= -\frac{\partial}{\partial x} N_e + \frac{a}{F}J_{tot},\\
N_e &=  - D_eB(x) \frac{\partial}{\partial x} C_e + \frac{t^+}{F} J_e,\\
\frac{\partial}{\partial x} C_e &= \frac{1}{D_eB(x)}\frac{t^+}{F} J_e , \qquad&&\qquad\text{at}\qquad x=0, L,\\
C_e &= C_{e_{i}}, \qquad&&\qquad\text{at}\qquad t=0.
\end{alignedat}
\label{eq:Li_tranport_electrolyte_1D}
\end{equation}
where $\epsilon = \epsilon(x,t)$ is the porosity of the domain, $C_e=C_e(x,t)$ is the Li ion concentration 
in the electrolyte, $D_e=D_e(C_e)$ is the diffusion coefficient in the electrolyte phase, and $J_e$ is 
the current density vector in the electrolyte phase.

\subsubsection{Interfacial Dynamics}\label{sec:interfacial}
Interfacial processes in the cell are normally modelled through the well-known Butler-Volmer (BV) 
relation for electrochemical kinetics. It describes how the current density at the solid-electrolyte 
interface depends on the potential difference between the electrode surface and the neighbouring electrolyte.
Several variants of this semi-empirical relation exist in the literature as surveyed by 
Dickinson et al.~\cite{dickinson2020butler}. Depending on the nature of the problem and the level of 
complexity required for a given application, a suitable BV relation could be used. Lithium plating and 
stripping as interfacial processes can also be modelled by adding an extra BV equation for the side 
reaction. Essentially, one BV relation can be added for each of the side reactions, representing the 
intensity of each side reaction as a function of overpotential, as first introduced by 
Arora et. al.~\cite{arora1999mathematical} and later expanded by Yang et. al.~\cite{yang2018look}. 
The BV relations representing interfacial phenomena at the solid-electrolyte interface are represented 
by \cite{arora1999mathematical, yang2018look}
\begin{equation}
\begin{alignedat}{1}
J_{n,int} &= j_{int}\left[ \exp(\alpha_{a,int}f\eta_{int}) - \exp(-\alpha_{c,int}f\eta_{int}) \right],\\
J_{n,sr} &= j_{sr}\left[ \exp(\alpha_{a,sr}f\eta_{sr}) - \exp(-\alpha_{c,sr}f\eta_{sr}) \right],\\
\eta_{int} &= \phi_n - \phi_e - U_{n},\\
\eta_{sr} &= \phi_n - \phi_e - U_{sr},\\
j_{int} &= k_{a,int}^{\alpha_{c,int}} k_{c,int}^{\alpha_{a,int}} C_n^{\alpha_{c,int}} C_e^{\alpha_{a,int}} (C_{max}-C_n)^{\alpha_{a,int}},\\
j_{sr} &= k_{a,sr}^{\alpha_{c,sr}} k_{c,sr}^{\alpha_{a,sr}} C_e^{\alpha_{a,sr}},
\end{alignedat}
\label{eq:BVs}
\end{equation}
where $j_{int}$ and $j_{sr}$ are the exchange current densities for intercalation and side reaction, 
respectively, $\eta_{int}$ and $\eta_{sr}$ are the overpotentials at the solid-electrolyte 
interface for intercalation and side reactions, respectively, $U_n$ and $U_{sr}$ 
are equilibrium potentials, $k_{c,int}$, $k_{a,int}$, $k_{c,sr}$, $k_{a,sr}$ are the reaction 
rates for cathodic and anodic currents of intercalation and cathodic and anodic currents of side reaction, 
respectively, $C_{max}$ is the saturation concentration of Li in solid phase, and $C_n = C_n\big\rvert_{r=R_n}$ 
refers to solid phase concentration on the surface of the particle. On the other hand, 
Escalante et al.~\cite{escalante2020discerning} use a more sophisticated technique for modeling 
side reaction in the cell, where one BV relation is used to represent both Li intercalation/deintercalation 
and plating/stripping as
\begin{equation*}
\begin{alignedat}{3}
j_{int} &= k_0 C_n^{0.5} C_e^{0.5} (C_{max}-C_n)^{0.5} \tanh\left(\gamma \frac{C_{max - C_n}}{C_{max}} \right),\\
\end{alignedat}
\end{equation*}
where the intercalation exchange current density is multiplied by a factor describing how the total current 
is divided between the intercalation and side reactions, hence, reducing the formulations for 
exchange current densities. Thus, the two equations for exchange current density of intercalation and 
side reaction are replaced with one, with a significantly lower number of parameters. 
As can be observed,
the exchange current densities are concentration dependent, and most authors have used similar functional
forms to model this dependency. Daniels et al.~\cite{daniels2023learning} leverage a data-driven approach 
in order to optimally construct the exchange current density as a function of concentration, without assuming
any {\em a priori} functional form for this function. In our modeling approach, we use a combination
of aforementioned techniques. We use separate BV relations as in first approach in order to account for the side reaction, 
cf.~\eqref{eq:BVs}. We also introduce a variable as the ratio of exchange current densities for intercalation
and plating processes that represents the competition between the intercalation versus plating, 
as outlined in Section \ref{sec:asymptotic}. This key variable is concentration dependent, and 
the optimal functional form will be 
constructed using data-driven inverse modeling techniques.
Before we start the analysis, we present a dimensionless version of our model in the next section.

\subsection{Dimensionless Model}\label{sec:dimensionless}
The model introduced in \eqref{eq:charge_conservation_solid_1D}, \eqref{eq:charge_conservation_solid_1D_p}
\eqref{eq:Li_transport_solid_1D}, \eqref{eq:Li_transport_solid_1D_p}, 
\eqref{eq:charge_consservation_electrolyte_1D}, and \eqref{eq:Li_tranport_electrolyte_1D} could be 
rescaled to a dimensionless form that will facilitate its asymptotic reduction. We focus on the 1D
version of the model where the independent and dependent variables are rescaled as follows
\begin{equation*}
\begin{alignedat}{10}
& x &&= L \widehat{x},  ~~
& L_n &= L l_n, ~~
& L_p &= L l_p, ~~
& r &= R_n \widehat{r}_n, ~~
& r &= R_p \widehat{r}_p,\\
& J_k &&= J_t \widehat{J}_k, ~~
& J_{K} &= J_t \widehat{J}_{K}, ~~
& j_{int} &= J_t \widehat{j}_{int}, ~~
& j_{sr} &= J_t \widehat{j}_{sr},~~
& \phi_n &= \phi_t \widehat{\phi}_n,\\
& \phi_e &&= \frac{1}{f} \widehat{\phi}_e, ~~
& U_n &= \phi_t \widehat{U}_n,~~
& \eta_{int} &= \frac{1}{f}\widehat{\eta}_{int},~~
& \eta_{sr} &= \frac{1}{f}\widehat{\eta}_{sr},~~
&\sigma_e &= \sigma_e^{\text{typ}} \widehat{\sigma}_e,\\
& C_n &&= C_n^{\text{max}} \widehat{C}_n,~~
& C_p &= C_p^{\text{max}} \widehat{C}_p,~~
& C_e &= C_e^{\text{max}} \widehat{C}_e,~~
& C_{sr} &= C_{n}^{\text{max}} \widehat{C}_{sr}, 
& t &= \tau \widehat{t}, \\
& N_e &&= \frac{D_e^{\text{typ}} C_e^{\text{max}}}{L} \widehat{N}_e, ~~
& D_n &= D_n^{\text{typ}} \widehat{D}_n,~~
& D_e &= D_e^{\text{typ}} \widehat{D}_e,~~
& D_p &= D_p^{\text{typ}} \widehat{D}_p,~~
\end{alignedat}
\end{equation*}
where $\widehat{x}\in\left[0,l\right]$, $l=1$, $k\in\{n,e,p\}$, $K\in\{tot,int,sr,app\}$, $f=\frac{F}{RT}$, 
$J_t$ is the typical current density in the cell, $\phi_t$ is the typical potential in the cell 
components, $D_k^{\text{typ}}$ is the typical diffusion coefficient, $\sigma_e^{\text{typ}}$ is 
the typical conductivity, $C_k^{\text{max}}$ is the maximum concentration of Lithium 
in the corresponding domain, and $\tau = \frac{F C_n^{\text{max}} L}{J_t}$ is the discharge time scale. 
Note that $\frac{1}{f} = \frac{RT}{F}$ has the unit of Volts and is defined as the thermal voltage of 
the cell. The dimensionless parameters are then defined as 
\begin{equation*}
\begin{alignedat}{10}
&\lambda &&= \phi_t f,~~
&\Xi_n &= \frac{\sigma_n}{f L J_t},~~
&\Xi_e  &= \frac{\sigma_e^{\text{typ}}}{f L J_t},~~
&\Xi_p  &= \frac{\sigma_p^{\text{typ}}}{f L J_t},~~
&\mathcal{K}_n &= \frac{R_n^2}{\tau D_n^{\text{typ}}},\\
&\mathcal{K}_e &&= \frac{L^2}{\tau D_e^{\text{typ}}},~~
&\mathcal{K}_p &= \frac{R_p^2}{\tau D_p^{\text{typ}}},~~
&\gamma_n &= \frac{\tau J_t}{R_n F C_n^{\text{max}}},~~
&\gamma_e &= \frac{\tau J_t}{L F C_e^{\text{max}}}~~
&\gamma_p &= \frac{\tau J_t}{R_p F C_p^{\text{max}}}.
\end{alignedat}
\end{equation*}
The dimensionless system of equations will become
\begin{subequations}
	\begin{alignat}{3}
	\frac{\partial \widehat{J}_n }{\partial \widehat{x}}  &= - a_n L \widehat{J}_{n,tot}, \qquad&&\text{on}\qquad 0\le \widehat{x} \le l_n, \label{eq:dimensionlessa} \\
	\widehat{J}_n &= - \lambda \Xi_n\frac{\partial \widehat{\phi}_n}{\partial \widehat{x}}, \qquad&&\text{on}\qquad 0\le \widehat{x} \le l_n, \label{eq:dimensionlessb} \\
	\mathcal{K}_n \frac{\partial \widehat{C}_n}{\partial \widehat{t}} &= \frac{1}{\widehat{r}^2} \frac{\partial}{\partial \widehat{r}} \left(\widehat{r}^2 \widehat{D}_n \frac{\partial \widehat{C}_n}{\partial \widehat{r}}\right), \qquad&&\text{on}\qquad 0\le \widehat{r} \le \widehat{r}_n, \label{eq:dimensionlessc} \\
	\frac{\partial \widehat{J}_p }{\partial \widehat{x}}  &= - a_p L \widehat{J}_{p,tot}, \qquad&&\text{on}\qquad 1-l_p\le \widehat{x} \le 1, \label{eq:dimensionlessl} \\ 
	\widehat{J}_p &= - \lambda \Xi_p\frac{\partial \widehat{\phi}_p}{\partial \widehat{x}}, \qquad&&\text{on}\qquad 1-l_p\le \widehat{x} \le 1, \label{eq:dimensionlessm} \\ 
	\mathcal{K}_p \frac{\partial \widehat{C}_p}{\partial \widehat{t}} &= \frac{1}{\widehat{r}^2} \frac{\partial}{\partial \widehat{r}} \left(\widehat{r}^2 \widehat{D}_p \frac{\partial \widehat{C}_p}{\partial \widehat{r}}\right), \qquad&&\text{on}\qquad 0\le \widehat{r} \le \widehat{r}_p, \label{eq:dimensionlessn} \\
	\frac{\partial \widehat{J}_e}{\partial \widehat{x}} &=  a L \widehat{J}_{tot}, \qquad&&\text{on}\qquad 0\le \widehat{x} \le 1, \label{eq:dimensionlessd}\\
	\widehat{J}_e & = - \Xi_e \sigma_e B(x)\left[\frac{\partial \widehat{\phi}_e}{ \partial \widehat{x}} - 2(1-t^+) \frac{\partial \log \widehat{C}_e}{ \partial \widehat{x}}\right], \qquad&&\text{on}\qquad 0\le \widehat{x} \le 1, \label{eq:dimensionlesse}\\
	\frac{\mathcal{K}_e}{\gamma_e}\frac{\partial (\epsilon \widehat{C}_e)}{ \partial t}  &= - \frac{1}{\gamma_e}\frac{\partial \widehat{N}_e}{\partial \widehat{x}} + a L \mathcal{K}_e\widehat{J}_{tot}, \qquad&&\text{on}\qquad 0\le \widehat{x} \le 1, \label{eq:dimensionlessf}\\
	\widehat{N}_e &=  - \widehat{D}_e B(x) \frac{\partial \widehat{C}_e }{\partial \widehat{x}} + t^+ \mathcal{K}_e \gamma_e \widehat{J}_e, \qquad&&\text{on}\qquad 0\le \widehat{x} \le 1, \label{eq:dimensionlessg}\\
	\widehat{J}_{n,int} &= \widehat{j}_{int}\left[ \exp(\alpha_{a,int}\widehat{\eta}_{int}) - \exp(-\alpha_{c,int}\widehat{\eta}_{int}) \right], \label{eq:dimensionlessh} \\
	\widehat{J}_{n,sr} &= \widehat{j}_{sr}\left[ \exp(\alpha_{a,sr}\widehat{\eta}_{sr}) - \exp(-\alpha_{c,sr}\widehat{\eta}_{sr}) \right], \label{eq:dimensionlessi} \\
	\widehat{\eta}_{int} &= \lambda\left[ \widehat{\phi}_n - \widehat{U}_{n}\right] - \widehat{\phi}_e ,\qquad&&\text{at}\qquad \widehat{r}=\widehat{r}_n, \label{eq:dimensionlessj}\\
	\widehat{\eta}_{sr} &= \lambda \left[ \widehat{\phi}_n  - \widehat{U}_{sr}\right] - \widehat{\phi}_e ,\qquad&&\text{at}\qquad \widehat{r}=\widehat{r}_n,\label{eq:dimensionlessk}
	\end{alignat}
	\label{eq:dimensionless}
\end{subequations}
with following boundary and initial conditions 
\begin{subequations}
	\begin{alignat}{11}
	\frac{\partial \widehat{C}_n}{\partial \widehat{r}} &= 0, 
	&&&&&&&&\text{at}\qquad \widehat{r}&&=0,\label{eq:dimensionless_BCsa}\\
	- \widehat{D}_n \frac{\partial \widehat{C}_n}{\partial \widehat{r}} &= \mathcal{K}_n \gamma_n \widehat{J}_{n,tot}, 
	&&&&&&&&\text{at}\qquad \widehat{r}&&=\widehat{r}_n,\label{eq:dimensionless_BCsb}\\
	\frac{\partial \widehat{C}_p}{\partial \widehat{r}} &= 0, 
	&&&&&&&&\text{at}\qquad \widehat{r}&&=0,\label{eq:dimensionless_BCsg}\\
	- \widehat{D}_p \frac{\partial \widehat{C}_p}{\partial \widehat{r}} &= \mathcal{K}_p \gamma_p \widehat{J}_{p,tot}, 
	&&&&&&&&\text{at}\qquad \widehat{r}&&=\widehat{r}_p,\label{eq:dimensionless_BCsh}\\
	\widehat{J}_n &= \widehat{J}_{app}, &\qquad \widehat{J}_e &= 0, &\qquad \widehat{N}_e &= 0, &\qquad \widehat{\phi}_e &= 0, &\qquad
	&\text{at}\qquad \widehat{x}&&=0,\label{eq:dimensionless_BCsc}\\
	\widehat{J}_n &= 0,   &\qquad  \widehat{J}_e &= \widehat{J}_{app}, &\qquad
	&&&&&\text{at}\qquad \widehat{x}&&=l_n, \label{eq:dimensionless_BCsd}\\
	\widehat{J}_p &= 0  & \qquad \widehat{J}_e &= \widehat{J}_{app},  &\qquad
	&&&&&\text{at}\qquad \widehat{x}&&=l-l_p, \label{eq:dimensionless_BCse} \\
	\widehat{J}_p &= \widehat{J}_{app}, &\qquad \widehat{J}_e &= 0, &\qquad \widehat{N}_e &= 0, &\qquad \widehat{\phi}_e &= 0, &\qquad
	&\text{at}\qquad \widehat{x}&&=l,\label{eq:dimensionless_BCsf}\\
	\widehat{C}_e &= 1       &\qquad \widehat{C}_n &= \widehat{C}_{n_i},  &\qquad \widehat{C}_p &= \widehat{C}_{p_i},
	&&&&\text{at}\qquad \widehat{t}&&=0.
	\end{alignat}
	\label{eq:dimensionless_BCs}
\end{subequations}

To simplify the notation in the analysis below, we opt to drop the hat sign from the dimensionless 
variables from now on. In the next section, an asymptotic reduction and averaging techniques
will be used to reduce the full model \eqref{eq:dimensionless}-\eqref{eq:dimensionless_BCs} 
to a simpler time-dependent ODE system, 
where the evolution of some key averaged concentrations are tracked.

\subsection{Asymptotic Reduction and Averaging}\label{sec:asymptotic}
In the current study the objective is to develop a simplified dynamical model in the form of the 
system of ODEs, capable of tracking the evolution of key concentrations in the cell. Certain 
simplifying assumptions are to be made to develop a suitable mathematical model for our application.
There are two important techniques used in this analysis that help in simplifying 
system \eqref{eq:dimensionless}-\eqref{eq:dimensionless_BCs}. 
The first technique is asymptotic reduction which assumes that 
a certain parameter in the system takes a limiting value (either large or small), and the dynamics of 
the system are investigated in the vicinity of that limiting value by expanding each dependent variable in 
a Taylor series with respect to that parameter. 
Asymptotic reduction of the DFN model to the SP model has been considered by various authors 
\cite{marquis2019asymptotic, planella2023single,richardson2020generalised}, where different 
assumptions have been employed in each case to reduce the DFN model to a SP model. 
Marquis et al.~\cite{marquis2019asymptotic} derives an asymptotic reduction of the DFN model to find 
a simplified SPM with electrolyte. This new model is shown to diverge from the DFN 
model for charging rates greater than 1C (C referring to the capacity of the cell). 
Richardson et al.~\cite{richardson2020generalised} extends this work with a different 
assumption for performing the asymptotic reduction, to generate a simple SP model that 
can perform better under higher charging rates. Brosa Planella et al.~\cite{planella2023single} 
extend this work to account for side reactions in the cell, a study that inspired 
the current investigation. However, certain assumptions in our work are different from 
work of Brosa Planella et al. Note that only the relevant equations in the system of equations 
\eqref{eq:dimensionless} will be used in this analysis. In particular, electrolyte equations 
are not a matter of interest in this work and will not be used in this asymptotic analysis.

The second technique used in this analysis is the averaging 
of equations over their corresponding spatial domains in order to eliminate the spatial dependency.
Some quantities in the DFN model, e.g. concentrations, depend on both time and space (note that 
in the DFN model "space" means both the microscale variable $r$ and the macroscale variable $x$)
in contrast to our experimental data which is resolved only in time.
Hence, averaging space-dependent quantities over their domains will eliminate the spatial 
dependency, and we will be left with a time-dependent model describing the evolution of lumped quantities. 
The aforementioned two techniques are used in conjunction. First, we start with the following assumptions
needed for this analysis.

\paragraph*{Assumptions}
\begin{itemize}
	\item A1: The parameter $\lambda$ is large enough, so that the Taylor expansion of state variables
	in the vicinity of small $\lambda^{-1}$ remains a valid approximation. 
	Note that the parameter $\lambda$ is defined as the ratio of the typical potential in the electrodes 
	to the thermal voltage of the cell. At room temperature the thermal voltage	is approximately 
	$25 mV$, and represents the characteristic scale of the overpotential
	at the interface in BV relation. When the scale of the potential in the
	electrode is large (in the order of magnitude of $1$ Volt), the parameter $\lambda$ 
	remains large enough for the asymptotic analysis.
	This assumption refers to the physical case of small deviations from the equilibrium 
	potential (small overpotentials), when the typical voltage in the electrodes is much larger 
	than the thermal voltage. In this setting, the BV relations can be linearized.
	\item A2: The cathodic and anodic charge transfer coefficients for an interfacial reaction, 
	cf.~\eqref{eq:dimensionlessh}-\eqref{eq:dimensionlessi}, are 
	assumed to add up to one, namely,	$\alpha_a + \alpha_c = 1$.
	\item A3: Side reactions in the cell are weak and in the order of $\lambda^{-1}$ relative to the 
	main intercalation/deintercalation reactions. This assumption
	allows us to capture the side reaction dynamics as corrections to the main reactions, as explained below.
	\item A4: The equilibrium potentials of the intercalation and side reactions are dependent on the 
	concentration of the intercalated Lithium and of the Lithium participating in the side reactions, 
	respectively. The sensitivity of these equilibrium 
	potentials to changes in concentrations is small, thus, these nonlinear relationships can be linearized
	in the neighbourhood of certain reference values of concentrations. Note that concentrations might
	exhibit large variations while the cell is in operation, however, if the sensitivity of the equilibrium 
	potential to concentrations is small, this simplification remains valid.	
\end{itemize}
Note that the inherent assumption of the SP model states that the electrode particles behave in 
a similar manner, hence, one representative particle is sufficient to represent the microscale 
dynamics of the cell. This inherent assumption will be re-derived as part of the asymptotic analysis.

\paragraph*{Relaxation and Excitation Dynamics} Before delving into the asymptotic analysis of the DFN model,
we discuss the fundamental sources of dynamics within the cell. The primary driver of dynamics in the cell 
is the excitation induced by the current applied to it. When the cell is brought to rest or an
open-circuit state after an excitation period (charge/discharge), the system continues to evolve until 
it reaches an equilibrium state corresponding to the specific state-of-charge of the cell. 
The intensity of this phenomenon varies across different chemistries \cite{ovejas2019effects}. 
Consequently, we can distinguish two main regimes in the operation of the cell: excitation, 
representing the main process, and 
the relaxation dynamics of the cell in the absence of external influence.
In our modeling effort, both excitation and relaxation dynamics will be 
captured by the mathematical model.

\paragraph*{Expansion of variables} In order to perform the asymptotic reduction, 
according to assumption A1 we expand each of the dependent
variables in system \eqref{eq:dimensionless}-\eqref{eq:dimensionless_BCs} in the vicinity 
of $\lambda^{-1} \approx 0 $. The expansion of variables in powers of $\lambda^{-1} $ takes the form
\begin{equation*}
\phi_n = \phi_{n,0} + \lambda^{-1} \phi_{n,1} + \cdots,
\end{equation*}
where the subscripts $0$ and $1$ refer to the leading-order and first-order approximations, respectively. 
All other dependent variables in the system \eqref{eq:dimensionless}-\eqref{eq:dimensionless_BCs} 
are expanded in a similar manner. The expanded version of variables will be substituted 
into \eqref{eq:dimensionless}-\eqref{eq:dimensionless_BCs} to derive the leading-order and first-order 
approximation of equations.

\paragraph*{Electrode Potential} We start with the equations for the negative 
electrode potential. Averaging \eqref{eq:dimensionlessa} over the negative electrode domain, using the Gauss divergence 
theorem, and applying boundary conditions \eqref{eq:dimensionless_BCsc} and \eqref{eq:dimensionless_BCsd}
we get
\begin{equation}
\begin{alignedat}{3}
\frac{1}{l_n}\int_{0}^{l_n} \frac{\partial}{\partial x} J_n dx &= - \frac{1}{l_n}\int_{0}^{l_n} a_n L J_{n,tot} dx = 0 - J_{app},&&\\
\overline{J}_{n,tot}  &= \frac{J_{app}}{a_n L l_n}, \qquad \overline{J}_{n,tot} := \frac{1}{l_n}\int_{0}^{l_n} J_{n,tot} dx,
\label{eq:Jbartot}
\end{alignedat}
\end{equation}
where $\overline{J}_{n,tot}$ is the total current density averaged over the domain of the negative electrode. 
Hence, by averaging over the spatial domain 
the partial differential equation for the charge conservation in the solid phase reduces to 
an algebraic equation. This algebraic equation states that 
all the current applied to the anode during charge/discharge will be consumed at the solid-electrolyte 
interface for intercalation/deintercalation or side reactions, and acts as a constraint on the 
system of equations. Performing similar analysis for the positive electrode using 
\eqref{eq:dimensionlessl}, \eqref{eq:dimensionless_BCse} and \eqref{eq:dimensionless_BCsf} results in 
$\overline{J}_{p,tot} = -\frac{J_{app}}{a_p L l_p }$.
Asymptotic reduction of \eqref{eq:dimensionlessb} at the leading-order leads to
\begin{equation}
\begin{alignedat}{1}
\frac{\partial \phi_{n,0}}{\partial x} = 0, \qquad 0 < x < l_n.
\end{alignedat}
\label{eq:phi_n0}
\end{equation}
Thus, $\phi_{n,0} = \phi_{n,0}(t)$, and the leading-order potential is homogeneous in space.
Also, at the first-order, we have
\begin{equation}
\begin{alignedat}{1}
J_{n0} = - \Xi_n \frac{\partial \phi_{n,1}}{\partial x}.
\end{alignedat}
\label{eq:phi_n1}
\end{equation}
\paragraph*{Interfacial Kinetics} In the next step, we simplify the BV relations introduced in 
\eqref{eq:dimensionlessh} and \eqref{eq:dimensionlessi}. For this purpose, 
we first linearize the BV relation, and second, we also linearize the relations between the equilibrium potentials and
concentrations. Using assumption A1, the BV relation \eqref{eq:dimensionlessh} can be linearized as
\begin{equation*}
\begin{alignedat}{1}
J_{int} &\cong j_{int} (\alpha_{a,int}+\alpha_{c,int}) \eta_{int},
\end{alignedat}
\end{equation*}
and similarly for \eqref{eq:dimensionlessi}. Invoking assumption A2, they can be further 
simplified to
\begin{subequations}
	\begin{alignat}{1}
	J_{int} &= j_{int} \eta_{int},\label{eq:BV_linearizeda}\\
	J_{sr} &= j_{sr} \eta_{sr}.\label{eq:BV_linearizedb}
	\end{alignat}
	\label{eq:BV_linearized}
\end{subequations}
Also, the overpotentials in the description of 
BV relations involve terms related to equilibrium potentials of intercalation and side reaction. As
stipulated by assumption A4, the equilibrium potentials are expanded as 
\begin{subequations}
	\begin{alignat}{1}
	U_{n} (C_{n}) = U_{n}\big\rvert_{C_{n,0}} + \lambda^{-1}\frac{d U_{n}}{d C_n}\big\rvert_{C_{n,0}} C_{n,1} + \cdots,\label{eq:Un}\\
	U_{sr} (C_{sr}) = U_{sr}\big\rvert_{C_{sr,0}} + \lambda^{-1}\frac{d U_{sr}}{d C_{sr}}\big\rvert_{C_{sr,0}} C_{sr,1} + \cdots,\label{eq:Usr}
	\end{alignat}
\end{subequations}
where $C_{n,0}$ and $C_{sr,0}$ are the leading-order concentrations used as the reference states for 
linearization, and $C_{n,1}$ and $C_{sr,1}$ are the first-order approximations of concentrations
i.e. $C_{n} \cong C_{n,0} + \lambda^{-1}C_{n,1}$ and $C_{sr} \cong C_{sr,0} + \lambda^{-1}C_{sr,1}$. Hence, 
performing asymptotic reduction on BV relation \eqref{eq:BV_linearizeda}, 
and using \eqref{eq:dimensionlessj} and \eqref{eq:Un}, we get
\begin{equation*}
\begin{alignedat}{1}
J_{int,0} &= \left(j_{int,0} +\lambda^{-1} j_{int,1} \right)\left(\lambda\left[ \phi_{n,0} + \lambda^{-1}\phi_{n,1} - U_{n}\big\rvert_{C_{n,0}} - \lambda^{-1}\frac{d U_{n}}{d C_n}\big\rvert_{C_{n,0}} C_{n,1} \right] -\phi_{e,0} - \lambda^{-1}\phi_{e,1} + \cdots\right).\\
\end{alignedat}
\end{equation*}
Thus, at the leading-order we have
\begin{equation}
\begin{alignedat}{1}
J_{int,0} &= 
\underbrace{j_{int,0} \left( \phi_{n,1}-\phi_{e,0} - \frac{d U_{n}}{d C_n}\big\rvert_{C_{n,0}} C_{n,1} \right)}_{J_n^\dagger}
+ 
\underbrace{j_{int,1} \left(\phi_{n,0} - U_{n}\big\rvert_{C_{n,0}}\right)}_{J_n^\ddagger},\\
\end{alignedat}
\label{eq:J_int}
\end{equation}
where $J_n^\dagger$ and $J_n^\ddagger$ represent (at the leading-order) the excitation and relaxation dynamics 
of the cell, respectively. This choice of excitation and relaxation dynamics in \eqref{eq:J_int} is justified in
two ways. First, as explained in Section \ref{sec:comparison}, the second term is assumed to be zero in the study by
Brosa Planella et al.~\cite{planella2023single}. However, their assumption leads to the lack of relaxation dynamics in the 
positive electrode. For this reason, the second term in \eqref{eq:J_int} is assumed to take into account the
relaxation dynamics of the cell. Second, the second term represents the deviation of the leading-order potential of the 
negative particle from its equilibrium potential. Setting this term to zero eliminates the relaxation dynamics of 
the negative particle. Equation \eqref{eq:J_int} will be 
used in subsequent analysis for describing $(\phi_{n,1}-\phi_{e,0})$, as
\begin{equation}
\begin{alignedat}{1}
\phi_{n,1}-\phi_{e,0} &= \frac{J_n^\dagger}{j_{int,0}} + \frac{d U_{n}}{d C_n}\big\rvert_{C_{n,0}} C_{n,1}.\\
\end{alignedat}
\label{eq:phi}
\end{equation}

Next, we focus our attention on the side reaction current density. Assuming that the side reactions in the 
cell are weak as stipulated by assumption A3, we postulate that 
$j_{sr} = \lambda^{-1}\widetilde{j}_{sr}$, where $\widetilde{j}_{sr}$ is of a different order of 
magnitude than $j_{sr}$. This choice allows us to capture the side 
reaction effect at the order of $\lambda^{-1}$ (smaller order of magnitude than the 
intercalation). At the leading-order, the side reaction is 
not observed due to this choice reflecting the assumption of weak side reactions. 
Therefore, performing asymptotic reduction on \eqref{eq:BV_linearizedb}, and using 
\eqref{eq:dimensionlessk} and \eqref{eq:Usr}, we get
\begin{equation}
\begin{alignedat}{11}
J_{sr,0} &=\lambda^{-1} \widetilde{j}_{sr,0} \left(\lambda \left[ \phi_{n,0}  - U_{sr} \big\rvert_{C_{sr,0}}\right] + \phi_{n,1} - \phi_{e,0} - \frac{d U_{sr}}{d C_{sr}}\big\rvert_{C_{sr,0}} C_{sr,1}\right),
\end{alignedat}
\label{eq:Jhat_LeadingOrderb}
\end{equation}
and by rearranging this equation we get
\begin{equation}
\begin{alignedat}{11}
J_{sr,0} &= \widetilde{j}_{sr,0}\left( \phi_{n,0}  - U_{sr} \big\rvert_{C_{sr,0}}\right) 
+   
\lambda^{-1} \widetilde{j}_{sr,0} \left( \phi_{n,1} - \phi_{e,0} - \frac{d U_{sr}}{d C_{sr}}\big\rvert_{C_{sr,0}} C_{sr,1}\right).
\end{alignedat}
\label{eq:Jhat_LeadingOrderc}
\end{equation}
As already mentioned, the side reaction is only considered at the first-order approximation and vanishes at the 
leading-order. Hence, to eliminate the term of order $\mathcal{O}(1)$ in \eqref{eq:Jhat_LeadingOrderc}, we set 
$\phi_{n,0} = U_{sr}\big\rvert_{C_{sr,0}}$. Therefore, $U_{sr}\big\rvert_{C_{sr,0}}$ is also uniform in 
space (as is $\phi_{n,0}$), which refines the underlying assumption of the SP model in which 
the behaviour of all solid particles is assumed uniform in space at the macroscale. Note that this assumption will impose
the uniformity of $U_{n}\big\rvert_{C_{n,0}}$ in space as well. Hence, starting with particles with the same initial concentrations, they will evolve in exactly same manner. Consequently, solving for one 
representative particle suffices to capture the dynamics of all solid particles. Note that quantities 
that are concentration-dependent will then be uniform in space, and can be easily averaged. Thus, the expression for 
the relaxation dynamic in \eqref{eq:J_int} becomes (after averaging quantities)
\begin{equation}
\begin{alignedat}{1}
\overline{J}_n^\ddagger &= j_{int,1} \left(U_{sr}\big\rvert_{C_{sr,0}} - U_{n}\big\rvert_{C_{n,0}}\right).
\end{alignedat}
\label{eq:intrinsic}
\end{equation}
By substituting \eqref{eq:phi} into \eqref{eq:Jhat_LeadingOrderc} for the side reaction current density, 
we get
\begin{equation}
\begin{alignedat}{1}
J_{sr,0} &=\lambda^{-1} \widetilde{j}_{sr,0} \left( \frac{J_n^\dagger}{j_{int,0}} + \frac{d U_{n}}{d C_n}\big\rvert_{C_{n,0}} C_{n,1}  - \frac{d U_{sr}}{d C_{sr}}\big\rvert_{C_{sr,0}} C_{sr,1}\right).
\label{eq:J_sr_raw}
\end{alignedat}
\end{equation}
The averaged current density for the side reaction $\overline{J}_{sr,0}$ can be computed by averaging 
\eqref{eq:J_sr_raw} over the negative electrode domain. Note that the exchange current density is 
a function of the concentration at the interface, however, due to the inherent assumption in the SP model,
where particles are uniform in space, the exchange current density will also be uniform in 
electrode domain. Also, the excitation current density
averaged over the negative electrode domain becomes $\overline{J}_n^\dagger  = \frac{J_{app}}{a_n L l_n}$ according
to \eqref{eq:Jbartot}. We can then average the expression \eqref{eq:J_sr_raw} as 
\begin{equation}
\begin{alignedat}{1}
\overline{J}_{sr,0} &= \frac{j_{sr,0}}{a L l_n j_{int,0}} J_{app} + j_{sr,0} \frac{d U_{n}}{d C_n}\big\rvert_{C_{n,0}} \overline{C}_{n,1} - j_{sr,0} \frac{d U_{sr}}{d C_{sr}}\big\rvert_{C_{sr,0}} \overline{C}_{sr,1}.
\end{alignedat}
\label{eq:J_sr_bar}
\end{equation}
Equations \eqref{eq:intrinsic} and \eqref{eq:J_sr_bar} will be used in subsequent analysis.

\paragraph*{Conservation of Charge}
The conservation of charge within the cell implies that the quantity of charge entering 
the cell is equivalent to the amount exiting the cell at each instance of time. This fundamental 
principle reflects the balance of electrical charge within the cell, ensuring that 
the net charge in the cell remains constant throughout the cell processes. The total current density 
on the negative electrode can be split into two components 
$J_{n,tot} = J_{n,tot,0} + \lambda^{-1}J_{n,tot,1}$, and $J_{n,tot,0} = J_{n,int,0} + J_{sr,0}$.
At the leading-order $J_{sr,0}$ vanishes (due to assumption A3), and the leading-order 
interfacial current density is given entirely by the intercalation current density,
$J_{n,tot,0} = J_{n,int,0} $. This implies that 
at the leading-order, the dynamics are driven merely by intercalation (and there are no side reactions).
The side reaction will enter as a correction term in the first-order approximation.
Also, the intercalation current density at the leading-order can be split into two components, namely, 
the excitation ($J_n^\dagger$) and relaxation dynamics ($J_n^\ddagger$). Hence, the total current 
density on the negative electrode becomes 
$J_{n,tot} = J_n^\dagger + J_n^\ddagger + \lambda^{-1} (J_{int,1} + J_{sr,1})$.
A similar analysis for the positive electrode can be performed, yielding
$J_{p,tot} = J_p^\dagger + J_p^\ddagger$. Note that in the positive electrode there is no side
reaction, and hence the correction to intercalation process is absent for this electrode.

The Li concentration on the interface of the electrode particle ($r=1$)
is homogeneous over the electrode spatial domain due to the macroscale uniformity of electrode particles assumed in the 
SP model. 
Hence the total current density is uniform over space, and is equal to its average value.
Averaging each of these relations over the corresponding electrode domains, we get
$\overline{J}_{n,tot} = \overline{J}_n^\dagger + \overline{J}_n^\ddagger + \lambda^{-1} (\overline{J}_{int,1} + \overline{J}_{sr,1})$, and $\overline{J}_{p,tot} = \overline{J}_p^\dagger + \overline{J}_p^\ddagger$. 
It is also known that the total current density driven by the excitation dynamics in each electrode is 
proportional to the applied current density as $\overline{J}_n^\dagger  = \frac{J_{app}}{a_n L l_n}$,
and $\overline{J}_p^\dagger  = - \frac{J_{app}}{a_p L l_p}$. For the conservation of charge to hold 
in the cell, the total charge flux in the cell must be zero, namely,
$\overline{J}_{n,tot} l_n + \overline{J}_{p,tot} l_p = 0$. This implies that at the leading
order, the current densities driven by the relaxation dynamics for the positive and negative electrodes
should interact as  $\overline{J}_p^\ddagger = -\frac{l_n}{l_p}\overline{J}_n^\ddagger$, and
at the first order approximation as $\overline{J}_{sr,1} = -\overline{J}_{int,1}$.

With this definition of relaxation dynamics for negative and positive electrodes interfacial current density, 
the total charge in the cell is conserved.
We note that the first-order approximation terms in the negative electrode serve as a correction factor
to the intercalation process occurring at the leading-order.

\paragraph*{Transport of Lithium in Particles}
We perform asymptotic analysis and averaging on microscale equations of electrode particles to 
describe the evolution of concentration of intercalated Li. 
Introducing the asymptotic expansion in \eqref{eq:dimensionlessc} and using the boundary 
conditions in \eqref{eq:dimensionless_BCsa} and \eqref{eq:dimensionless_BCsb}, followed by 
averaging over the spherical domain, gives at the leading-order (note the boundary condition at $r_n=1$) 
\begin{equation}
\begin{alignedat}{3}
\int_{0}^{r_n}\mathcal{K}_n \frac{\partial C_{n,0}}{\partial t} r^2 dr &=  \int_{0}^{r_n}\frac{1}{r^2} \frac{\partial}{\partial r} \left(r^2 D_{n0} \frac{\partial C_{n,0}}{\partial r}\right) r^2 dr, \qquad&&\qquad t\ge 0,
\end{alignedat}
\end{equation}
with the boundary conditions
\begin{equation}
\begin{alignedat}{3}
\frac{\partial C_{n,0}}{\partial r} &= 0, \qquad&&\text{at}\qquad r=0,\\
- D_{n0} \frac{\partial C_{n,0}}{\partial r} &= \mathcal{K}_n \gamma_n J_{n,int,0}, \qquad&&\text{at}\qquad r=r_n.
\end{alignedat}
\end{equation}
Applying the Gauss divergence theorem, the average rate of growth of concentration at the leading-order is 
obtained as the net flux out of the boundary, hence
\begin{equation}
\begin{alignedat}{1}
\frac{d \overline{C}_{n,0}}{d t} &= \frac{\gamma_n}{r_n} \overline{J}_{n,int,0} = \frac{\gamma_n}{r_n} (\overline{J}_n^\dagger + \overline{J}_n^\ddagger) = \frac{\gamma_n}{r_n} (\frac{J_{app}}{a_n L l_n} + \overline{J}_n^\ddagger), \qquad \text{where} \quad \overline{C}_{n,0} =  \int_{0}^{r_n} C_{n,0} r^2 dr.
\label{eq:Cn0}
\end{alignedat}
\end{equation}
On the other hand, at the first-order approximation we get
\begin{equation}
\begin{alignedat}{3}
\int_{0}^{r_n}\mathcal{K}_n \frac{\partial C_{n,1}}{\partial t} r^2 dr &=  \int_{0}^{r_n}\frac{1}{r^2} \frac{\partial}{\partial r} \left(r^2 D_{n}\big\rvert_{C_{n,0}} \frac{\partial C_{n,1}}{\partial r} + r^2 D_{n}^\prime\big\rvert_{C_{n,0}} C_{n,1}\frac{\partial C_{n,0}}{\partial r}  \right) r^2 dr, \qquad&&\qquad  t\ge 0,
\end{alignedat}
\end{equation}
with the boundary conditions
\begin{equation}
\begin{alignedat}{3}
\frac{\partial C_{n,1}}{\partial r} &= 0, \qquad&&\text{at}\qquad r=0,\\
\mathcal{K}_n \gamma_n J_{int,1} &= - \left(r^2 D_{n}\big\rvert_{C_{n,0}} \frac{\partial C_{n,1}}{\partial r} + r^2 D_{n}^\prime\big\rvert_{C_{n,0}} C_{n,1}\frac{\partial C_{n,0}}{\partial r}  \right) , \qquad&&\text{at}\qquad r=r_n.
\end{alignedat}
\end{equation}
Note that the boundary condition on the interface of the electrode particle is computed using 
$J_{int,1} $. Applying the 
boundary conditions and the Gauss divergence theorem, and knowing that $J_{int,1} = - J_{sr,0}$, we obtain
\begin{equation}
\begin{alignedat}{1}
\frac{d \overline{C}_{n,1}}{d t} &=  \frac{\gamma_n}{r_n} \overline{J}_{int,1} = - \frac{\gamma_n}{r_n} \overline{J}_{sr,0}, \qquad \text{where} \quad \overline{C}_{n,1} =  \int_{0}^{r_n} C_{n,1} r^2 dr.
\end{alignedat}
\end{equation}
Noting that $\overline{C}_{n} \approx \overline{C}_{n,0} + \lambda^{-1} \overline{C}_{n,1}$, the growth rate of Li concentration in the negative electrode is governed by 
\begin{equation}
\begin{alignedat}{1}
\frac{d \overline{C}_{n}}{d t} &=  \frac{\gamma_n}{r_n} \left(\frac{J_{app}}{a_n L l_n} + \overline{J}_n^\ddagger - \lambda^{-1} \overline{J}_{sr,0}\right),
\end{alignedat}
\label{eq:Cn}
\end{equation}
where the expression for $\overline{J}_{sr,0}$ is computed in \eqref{eq:J_sr_bar}. 

A similar analysis can be performed for the positive particle. Introducing the asymptotic expansion in 
\eqref{eq:dimensionlessn} and using the boundary 
conditions \eqref{eq:dimensionless_BCsg} and \eqref{eq:dimensionless_BCsh}, followed by 
averaging over the spherical domain gives at the leading-order we get
\begin{equation}
\begin{alignedat}{1}
\frac{d \overline{C}_{p0}}{d t} &= \frac{\gamma_p}{r_p} \overline{J}_{p,int,0} = \frac{\gamma_p}{r_p} (\overline{J}_p^\dagger + \overline{J}_p^\ddagger), \qquad \text{where} \quad \overline{C}_{p0} =  \int_{0}^{r_p} C_{p0} r^2 dr.
\label{eq:Cp0}
\end{alignedat}
\end{equation}

\paragraph*{Conservation of Lithium}
As the total inventory of Lithium in the cell is conserved, Li assumed to occur in four different 
phases (anode intercalation, anode side reaction, electrolyte, and cathode solid phase) such that the
corresponding rates of change should add up to zero, namely,
\begin{equation}
\begin{alignedat}{1}
l_n\frac{d}{dt}\overline{C}_n(t) + l_n\frac{d}{dt}\overline{C}_{sr}(t) + l\frac{d}{dt}\overline{C}_e(t) + l_p\frac{d}{dt}\overline{C}_p(t)  &= 0,
\end{alignedat}
\label{eq:normalization}
\end{equation}
where $A$ denotes the cross-sectional area of the electrode.
This normalization condition should be satisfied by the derived system of equations. However, the 
computation of $\overline{C}_e(t)$ necessitates information about the concentration gradient at the 
boundary (after asymptotic reduction and averaging of \eqref{eq:dimensionlessd}), which is absent in the 
time-dependent model. Also, the amount of Lithium in the electrolyte is always conserved 
as noted in \cite{planella2023single}, meaning that $\frac{d}{dt}\overline{C}_e(t) = 0$. This implies
that the Li ions will enter the electrolyte at the same rate that they exit the electrolyte phase in 
different domains of the cell. Referring to \eqref{eq:Cn} and \eqref{eq:Cp0},
we conclude that the side reaction dynamic becomes
\begin{equation}
\begin{alignedat}{1}
\frac{d \overline{C}_{sr,1}}{d t} &=  \frac{\gamma_n}{r_n} \overline{J}_{sr,0},
\end{alignedat}
\label{eq:Csr}
\end{equation}
in order to retain the Li conservation in the cell.

\paragraph*{Dynamical Model}
The concentration 
evolution in time of the two key averaged concentrations in the cell can be computed as
\begin{equation}
\begin{alignedat}{1}
\frac{d \overline{C}_{n}}{d t} &= \frac{\gamma_n}{r_n} \left(\frac{J_{app}}{a_n L l_n} + \overline{J}_n^\ddagger - \lambda^{-1}\overline{J}_{sr}\right),\\
\frac{d \overline{C}_{sr,1}}{dt} &=  \frac{\gamma_n}{r_n} \overline{J}_{sr},
\end{alignedat}
\label{eq:ode1}
\end{equation}
where $\overline{J}_n^\ddagger$ and $\overline{J}_{sr}$ are obtained from \eqref{eq:intrinsic} 
and \eqref{eq:J_sr_bar}, respectively. Before moving on to the formulation of the inverse problem, we need to prepare
the ground by making the following comments about \eqref{eq:ode1}.
\begin{itemize}
	\item As discussed in Section \ref{sec:interfacial}, upon 
	consideration of the relations governing the 
	intercalation and plating current densities in the BV equation \eqref{eq:BVs}, it becomes apparent that these 
	equations are both dependent on the concentrations of the intercalated Li and  Li in side reactions , 
	namely, $j_{sr,0} = j_{sr,0} (\overline{C}_{n}, 
	\overline{C}_{sr})$ and $j_{int,0} = j_{int,0} (\overline{C}_{n}, 
	\overline{C}_{sr})$, and overpotential $\eta$. In our SPM modeling framework, the need for solving for the potential
	profile and the overpotential is eliminated using equation \eqref{eq:phi}. The dependency of the exchange current
	densities on concentrations is unknown, and needs to be determined using data-driven calibration strategies,
	cf.~Section \ref{sec:inverse}. As both exchange current densities are concentration dependent, we close 
	the model by introducing a variable $\omega = \omega(\overline{C}_{n}, \overline{C}_{sr}) = \frac{j_{sr,0}}{j_{int,0}}$
	representing a constitutive relation describing the competition 
	between the side reaction and intercalation exchange current densities. This relation
	controls how the total current density is split between side reaction and intercalation at each particular 
	state of the cell.
	\item The concentrations $C_n$ and $C_{sr}$ introduced in the asymptotic analysis are expanded 
	up to the first-order in $\lambda^{-1}$. Knowing that side reactions are not observed at the leading-order 
	$\overline{C}_{sr,0} = 0$, we conclude that $\overline{C}_{sr} = \overline{C}_{sr,1}$. The concentration 
	of the intercalated Lithium can then be expressed as 
	$\overline{C}_{n} = \overline{C}_{n,0} + \lambda^{-1}\overline{C}_{n,1} $.
	Note that when we expand $\overline{C}_{sr}$ in \eqref{eq:ode1}, only the first-order 
	approximation of concentration	$\overline{C}_{n,1}$ appears in the expressions (with the 
	leading term $\overline{C}_{n,0}$ absent). In this case we make the assumption that 
	$\overline{C}_{n,1} = \zeta \overline{C}_{n}$ in order to close the 
	mathematical model, where $\zeta$ is a scalar parameter, $0 < \zeta \ll 1$. Note that
	this assumption is not true, as the parameter $\zeta$ could be concentration-dependent. However, 
	in order to close the mathematical model we opt to simplify the expression to reduce the computational 
	complexity of the inverse modeling.
	\item The exchange current density in the cell is defined as the interfacial current density
	while the cell is in an equilibrium state, for both the forward and the backward
	interfacial reactions. $j_{int,0}$ refers to the interfacial current density 
	for Li intercalation or deintercalation on the negative particle surface at the leading-order 
	(assuming no side reactions) while at equilibrium. While the cell is in an equilibrium state, 
	the dynamics are driven by two physical mechanisms active at the electrode-electrolyte interface. 
	The leading one is the Li intercalation/deintercalation at a specific rate ($j_{int,0}$). 
	The second mechanism are the side reactions occurring at the interface and represented by the 
	first-order correction terms.
	This mechanism can be regarded as the interaction
	between the intercalated Li and plated Li. In mathematical terms, $j_{int,1}$ represents the 
	rate at which the intercalated Li is contributing to the growth of the plated
	Li phase, and vice versa, $j_{sr,0}$ represents the rate at which the plated Li is contributing
	to the growth of intercalated Li. This interaction can be viewed as the forward/backward reactions
	between the two phases. As the local concentrations in each phase must 
	remain stationary at equilibrium, we conclude that $j_{int,1} = j_{sr,0}$. We denote
	this exchange current density by $j_{sr}$, and note that the exchange current density 
	is a function of concentration, $j_{sr} = j_{sr}(\overline{C}_{n}, \overline{C}_{sr})$.
\end{itemize}
Taking into account these considerations, and substituting \eqref{eq:intrinsic} 
and \eqref{eq:J_sr_bar} into the system of equations \eqref{eq:ode1}, we finally get
\begin{equation}
\begin{alignedat}{2}
\frac{d \overline{C}_{n}}{d t} &= 
\frac{\gamma_n}{r_n a_n L l_n} \left[ 1 - \lambda^{-1}\omega \right] J_{app}
&&+ \frac{\gamma_n}{r_n} j_{sr} \left[U_{sr,0} - U_{n,0}\right]
- \frac{\gamma_n}{r_n} j_{sr} U_n^\prime  \overline{C}_{n} 
+ \frac{\gamma_n}{r_n}\lambda^{-1} j_{sr} U_{sr}^\prime \overline{C}_{sr},\\
\frac{d \overline{C}_{sr}}{dt} &=  
\underbrace{
	\frac{\gamma_n}{r_n a_n L l_n} \omega J_{app} ~~~~~~~~~~~~~~~
}_{\text{Excitation Dynamics}}
&&+ 
\underbrace{
	\frac{\gamma_n}{r_n} j_{sr} U_n^\prime \zeta \overline{C}_{n} 
	- \frac{\gamma_n}{r_n} j_{sr} U_{sr}^\prime \overline{C}_{sr},~~~~~~~~~~~~~~~~~~~~~~~~~~~~~~~~~~~~~
}_{\text{Relaxation Dynamics}}
\end{alignedat}
\label{eq:ode_full}
\end{equation}
where $U_{sr,0}$ and $U_{n,0}$ are scalar reference potentials. Note that the right-hand-side
of this system consists of two parts corresponding to the relaxation dynamics and the 
excitation dynamics. The first term on the right-hand-side of each equation represents the 
excitation dynamics of the cell. The remaining terms are linear in concentrations and 
represent the relaxation dynamics of the cell. Thus, this 
simplified model is capable of both taking into account the relaxation when excitation is absent 
($J_{app} = 0$), and also to track the dynamics of the cell when the excitation is present. 
This concludes the derivation of the ODE model.

\subsection{Comparison to the SPMe+SR Model}\label{sec:comparison}
This modeling framework is inspired by the SPMe+SR model of Brosa Planella et al.~\cite{planella2023single}. 
However, certain assumptions in our modeling approach differ from their work to better suit our specific 
configuration, particularly in tracking time-dependent concentrations without spatial resolution. After 
careful consideration of the SPMe+SR model, it is evident that this model has the following drawbacks.
\begin{enumerate}
	\item A one-sided BV relation is used for modeling plating in the cell, 
	with one exponential term in the corresponding expression. As the output of the exponential 
	term is always positive, the current density of Li plating at the solid-electrolyte interface
	is always negative. This implies that the model is only capable of predicting Li plating (and not 
	stripping). As noted by Sahu et al.~\cite{sahu2023continuum}, a two-sided BV relation must 
	be used to account for both plating and stripping in the cell. In our framework, we have used a 
	two-sided BV relation in \eqref{eq:dimensionless} to prevent this issue.
	\item Once averaged over the spatial domain, the model fails to take into account the relaxation 
	dynamics for the positive electrode, as evidenced by Equations (23)-(25) in 
	\cite{planella2023single}. While the space-averaged model adequately accounts for Lithium 
	conservation within the cell, it fails to capture the relaxation dynamics 
	on the positive electrode, and its dynamics are solely driven by excitation.
	\item On the negative electrode the terms corresponding to relaxation dynamics of plated Li and intercalated Li 
	possess opposite signs (once the cell is set to rest), meaning that intercalated Li phase and 
	plated Li phase will converge to equilibrium state in different directions. If Li in the intercalated
	phase becomes intercalated (deintercalated) in relaxation regime, the Li in plated phase gets stripped 
	(plated). This contradicts the evidence from experimental data, 
	cf.~Section \ref{sec:experiment}, in which the deintercalation process is accompanied by the 
	stripping process in relaxation the regime.
\end{enumerate}
These inconsistencies in the SPMe+SR model stem from two key factors: 
\begin{enumerate}
	\item The one-sided BV relation prevents the model from predicting Li stripping, as discussed earlier. The
	solution to this issue is to use a two-sided BV relation as in \eqref{eq:dimensionless}.
	\item The relaxation dynamics of the SPMe+SR model are not consistent with the dynamics of the cell due to
	the underlying assumptions of the asymptotic reduction framework of Brosa Planella et al.~\cite{planella2023single}.
\end{enumerate}
In order to address the second issue (inconsistency in relaxation dynamics), we need to understand the source of this
inconsistency in the SPMe+SR model. Referring to this model, if we assume there is no side 
reaction in the cell, the interfacial current density for the side reaction becomes zero, and hence, the intercalated
Li dynamics will only be driven by excitation. In other words, the relaxation dynamics of the Li in negative electrode
particles is only accounted for when a side reaction is present, and it is indeed in the reverse direction to the 
side reaction. In simpler terms, the relaxation dynamics of the Li in the negative electrode is dependant on the 
side reaction dynamics and this dependence is also observed for the Li dynamics in the positive electrode (no side reaction
on the positive electrode results in no relaxation dynamics for Li in positive electrode).
This dependency of the relaxation dynamics on the presence of a side reaction is the source of the inconsistency. 

This inconsistency arises due to the fact that the relaxation dynamics is not accounted 
for in the leading-order terms of the interfacial current density (recall that leading-order terms refer to vanishing 
side reaction, cf.~Assumption A3). 
Specifically, Brosa Planella et al.~\cite{planella2023single} assume that
$\phi_{n,0} = U_{n}\big\rvert_{C_{n,0}}$. Consequently, the second term on the right-hand side of
Equation \eqref{eq:J_int} vanishes, refining the underlying assumption of the SP model
that all particles exhibit uniform behaviour in space (as $\phi_{n,0}$ is spatially uniform). 
We elucidate how this assumption leads to loss of relaxation dynamics at the leading order.

In our modeling framework, we decompose the intercalation process at the leading-order 
into two components, capturing both the excitation and relaxation dynamics of the cell.
This formulation ensures that both dynamics are present in the space-averaged model, and that the
relaxation dynamics is independent of the presence of side reactions in the cell.
We highlight that failure to include relaxation dynamics terms will result in the space-averaged 
model to exhibit non-trivial behavior only when a current is applied to the cell.
To achieve that, we relax the assumption that $\phi_{n,0} = U_{n}\big\rvert_{C_{n,0}}$, and introduce the 
relaxation dynamics as the second term on the right-hand side in \eqref{eq:J_int}.
In order to refine the inherent assumption of the SPM regarding the uniformity of particles in space, 
we introduce the assumption $\phi_{n,0} = U_{sr}\big\rvert_{C_{sr,0}}$, as detailed in 
\eqref{eq:Jhat_LeadingOrderc}.
While one might argue that this assumption neglects the relaxation dynamics of the side reaction,
it is important to consider the relative magnitudes of the plated Li concentrations compared to the
intercalated Li concentrations. The relaxation dynamics is primarily driven by the intercalation 
process rather than the side reaction. Consequently, this assumption remains valid and
allows us to close the mathematical model. 
We also note that the aforementioned assumptions imply that the relaxation dynamics of the Li in the negative electrode
is driven by the difference between the equilibrium potentials of the intercalated Li and plated Li.

By comparing our model to the one introduced by Sahu et al.~\cite{sahu2023continuum}, we 
remark that they introduced a Heaviside step function in the definition of the Li plating dynamics. The intention
of this step function is to ensure that Li stripping process is stopped once the concentration 
of the plated Li becomes zero. Our mathematical model does not take this into account, however, it could be 
easily added to the model to ensure proper operation of the model at all concentrations. As we will see
in Section \ref{sec:results_fullmodel}, the results of model fitting to experimental data demonstrate the 
desired behaviour, meaning no stripping occurs below zero concentration. Thus, in the interest of simplicity, 
we have decided not to include the Heaviside function in our model.

\subsection{Linearizing Relaxation Dynamics}\label{sec:forward}
For simplicity, from now on, we will be using $C_1$ and $C_2$ to denote $\overline{C}_{n}$ and 
$\overline{C}_{sr}$, respectively. Also, the hat sign on top of variables will be dropped for simplicity 
of notation. Aggregating all constants, the ODE system \eqref{eq:ode_full} becomes
\begin{equation}
\begin{alignedat}{1}
\frac{d C_1}{d t} &= 
\alpha \left[ 1 - \lambda^{-1}\omega(C_1,C_2) \right] J_{app}
+ \beta_1 j_{sr}(C_1,C_2) 
+ \beta_2 j_{sr}(C_1,C_2) C_1
+ \beta_3\lambda^{-1} j_{sr}(C_1,C_2) C_2,\\
\frac{d C_2}{dt} &=  
\alpha \omega(C_1,C_2) J_{app}
+ \beta_4 j_{sr}(C_1,C_2) C_1
- \beta_3 j_{sr}(C_1,C_2) C_2,
\end{alignedat}
\label{eq:ode2}
\end{equation}
where $j_{sr} = j_{sr} (C_1,C_2)$, $\omega = \omega(C_1,C_2)$,
$\alpha \in \RR$, and $\beta_i\in\RR, ~ i=1,\cdots,4$. 
As can be observed, the system of equations \eqref{eq:ode2} has many unknown parameters (five scalar 
parameters and two functions), which makes the inverse modeling formulation 
complicated to solve, as discussed in Section \ref{sec:inverse}. Also, both the relaxation and 
the excitation dynamics components of the mathematical model are nonlinear, adding to the complexity 
of the inverse problem. As both the relaxation part and the excitation part of the model are 
functions of the exchange current densities, one needs to solve the inverse problem by matching the model 
output against the experimental data for the entire charge/discharge cycle, fitting all unknown 
parameters and constitutive relation simultaneously. 
The resulting infinite-dimensional optimization problem is therefore very difficult to solve.
One simplifying assumption can break this problem 
down into two simpler sub-problems. When the cell is excited, specifically through high charge/discharge 
rates, the contribution of excitation dynamics is much larger than the relaxation dynamics portion. 
Thus, the problem can be segmented into two parts as follows: (i) solve the inverse problem for the relaxation 
dynamics when the excitation is zero (eliminating the excitation part from the equations), 
and (ii) solve the inverse problem for the full model when the cell is 
excited (with the relaxation part calibrated in step (i)). 
This formulation will break the inverse problem into two separate sub-problems, each 
involving a smaller number of unknown parameters, hence decreasing the overall computational complexity. 
It has however one caveat, namely, the relaxation dynamics part is also a function 
of the exchange current density (which in turn is a function of concentration).
Calibrating the relaxation dynamics first requires finding
an optimal form of the constitutive relation $j_{sr}(C_1,C_2)$, which will then be used for the excitation dynamics. 
However, we know that the excitation dynamics plays a stronger role in determining the behavior of 
the system, and hence, the constitutive relation needs to be determined from the excitation dynamics. 
This brings us to another simplifying assumption. When the cell is not excited, changes in 
concentrations are small in comparison to when the cell is excited. This means 
that the concentrations will exhibit small changes with respect to some reference state, and accordingly, 
the change in the exchange current density $j_{sr}$ is also negligible. 
Therefore, the second assumption is to linearize the relaxation 
dynamics part around a reference state of the cell. We thus define 
$C_1 = \widehat{C}_1 + C_1^\prime$, $C_2 = \widehat{C}_2 + C_2^\prime$, 
and linearize the constitutive relation as 
$j_{sr} \approx \widehat{j}_{sr}(\widehat{C}_1,\widehat{C}_2) + \frac{\partial j_{sr}}{\partial C_1} 
\big\rvert_{\widehat{C}_1} (C_1-\widehat{C}_1) + \frac{\partial j_{sr}}{\partial C_2}\big\rvert_{\widehat{C}_2} (C_2-\widehat{C}_2)$. 
Note that $\widehat{C}_1$ and $\widehat{C}_2$ denote a reference state of concentrations and 
deviations from the reference state $\widehat{C}_1$ and $\widehat{C}_2$ are small. 
We substitute these equations into the relaxation portion of equation 
\eqref{eq:ode2}, aggregate all constants and after eliminating high-order terms, 
we get (the constants are again named 
as $\beta$, however, these are different constants than before)
\begin{equation}
\begin{alignedat}{1}
\frac{d C_1}{d t} &= 
\beta_1
+ \beta_2 C_1
+ \lambda^{-1} \beta_3 C_2,\\
\frac{d C_2}{dt} &=  
\beta_4  C_1
- \beta_3 C_2,
\end{alignedat}
\label{eq:ode_relaxation}
\end{equation}
for the relaxation dynamics of the cell.
Therefore, with this linearization, the full model takes the form (in vector notation)
\begin{equation}
\begin{alignedat}{1}
\frac{d}{dt} \bC(t) &= \bA\bC(t) + \bB + \bF( J_{app} (t)), \\
\bC(0) &= \bC_0, \\
\bC(t) &= \begin{bmatrix}
C_1 (t)\\
C_2 (t)
\end{bmatrix},\\
\bB &= \begin{bmatrix}
\beta_1 \\
0
\end{bmatrix},\\
\bA &= \begin{bmatrix}
\beta_2 & \lambda^{-1}\beta_3\\
\beta_4& - \beta_3
\end{bmatrix},\\
\bF &= \begin{bmatrix}
\alpha \left[ 1 - \lambda^{-1}\omega \right]J_{app}  \\
\alpha \omega J_{app}  
\end{bmatrix},\\
\end{alignedat}
\label{eq:ODE}
\end{equation}
where $\bbeta= \left[ \beta_1,\beta_2,\beta_3, \beta_4\right] \in \RR^4 $ are the parameters
of the relaxation dynamics, and $\omega = \omega(C_1,C_2)$ and $\alpha\in \RR$ are the 
unknown parameters and functions for the excitation dynamics of the cell. 
There are five scalar parameters and one constitutive relation given in terms of a function of two variables 
to be determined using inverse 
modeling. It is notable that the concentrations of different Li phases obtained from NMR spectroscopy 
experiments do not have a physical unit due to the nature of this methodology and the complexities of 
the computational post-processing of its data. Hence, it is impossible to match the concentrations of 
the physical model i.e., $C_1(t)$ and $C_2(t)$, to the Li content obtained from NMR spectroscopy. The 
inverse modeling approach will need to be designed to account for the conversion between physical variables
in the model and the experimental quantities. The parameters of the model will be tuned from the experimental data, 
which automatically takes care of this conversion between variables and experimental quantities.

\section{Inverse modeling}\label{sec:inverse}

The system of equations in \eqref{eq:ODE} is not closed due to the
dependence of $\omega$ on the state variables, which is unknown. To
address this challenge, one can explore the relationship between
$\omega$ and the two state variables $C_1$ and $C_2$ through 
data-driven calibration techniques.  In this methodology, the function
$\omega(C_1,C_2)$ could be determined through either a parametric or a
non-parametric approach. In the parametric approach, the functional
form describing the dependence of $\omega$ on the state variables is
assumed and its parameters are calibrated via data-driven calibration
techniques.  Conversely, in the non-parametric approach, this
relationship can be inferred without explicitly assuming any functional
form describing how the constitutive relation $\omega$ depends on the
state variables.  The only assumptions imposed on the constitutive
relation are the regularity of the function $\omega(C_1,C_2)$ and its
behaviour at the boundaries of the domain.  The latter technique is
superior, as it removes the assumptions about the underlying
functional form of the constitutive relation. In the current
investigation, our focus will be on the latter approach.

The inverse problem will be defined as follows: given a set of
time-dependent measurements of state variables, $\widetilde{C}_1(t)$
and $\widetilde{C}_2(t)$, within the time window $t \in [0,T]$,
cf.~Figure \ref{fig:experimental}, we seek to reconstruct the
constitutive relation $\omega = \omega(C_1,C_2)$ such that the
solution to the ODE system \eqref{eq:ODE} will best fit the
experimental measurements.  Note that in this formulation, no
\textit{a priori} assumption regarding the functional form of the
constitutive relation is made other than its regularity and behavior
for limiting values of the state variables.  The dynamics of the
system is split into two parts: (i) the relaxation dynamics and (ii)
the excitation dynamics. In Section \ref{sec:relaxation} we present
the formulation of the inverse problem for the relaxation dynamics
with details deferred to Appendix \ref{sec:relaxation_appendix} due to
their similarity to the formulation of the computational framework for
the excitation dynamics which is presented in full in Section
\ref{sec:excitation}.  Finally, in Section \ref{sec:robust} a more
robust framework is introduced for the inverse problem.

\subsection{Relaxation Dynamics}\label{sec:relaxation}

In this section, we aim to calibrate model \eqref{eq:ODE} for the
relaxation dynamics only.  When the cell is set to rest, the applied
current is zero, and the excitation term $\bF( J_{app} (t))$ on the
right-hand-side of the model vanishes. Hence, the problem reduces to
finding $\bbeta$ and one can formulate a suitable inverse problem to
calibrate each of the parameters in $\bbeta$ using cell data by
minimizing a cost functional $\JJ_1:\RR^4\rightarrow\RR$ defined as 
\begin{equation}
\begin{alignedat}{1}
\JJ_1(\bbeta) &= \frac{1}{2} \int_{0}^{T} \Big\lvert\Big\lvert \bW \, \br(t;\bbeta) \Big\rvert\Big\rvert_2^2 dt, \qquad \text{where}\\
\br (t;\bbeta) &= \bC (t;\bbeta) - \widetilde{\bC} (t),\\
\bC (t;\bbeta) &= \begin{bmatrix}
C_1 (t;\bbeta)\\
C_2 (t;\bbeta)
\end{bmatrix},\\
\widetilde{\bC}(t) &= \begin{bmatrix}
\widetilde{C}_1 (t)\\
\widetilde{C}_2 (t)
\end{bmatrix},\\
\bW &= \begin{bmatrix} 1&0 \\ 0&\sqrt{w}\end{bmatrix}
\end{alignedat}
\label{eq:cost_OCV}
\end{equation}
in which $\bW$ is a weight matrix, $T$ is the final time of the cycle,
$\lvert\lvert\cdot\rvert\rvert_2$ represents the Euclidean norm, and
the dependence of the state variables ($C_1$ and $C_2$) on the
parameters $\bbeta$ is governed by Eq.~\eqref{eq:ODE}.  As the typical
magnitudes of the state variables differ by one order of magnitude,
the weight matrix $\bW$ is designed to introduce a suitable
normalization.  Optimal parameter values can be found by solving the
minimization problem
\begin{equation}
\begin{alignedat}{1}
\overline{\bbeta}  &= \underset{\bbeta  \in \RR^4 }{\arg\min \JJ_1(\bbeta)}.
\end{alignedat}
\label{eq:argmin_OCV}
\end{equation}
For the purpose of solving this problem, a gradient-based optimization approach can be used,
defined by the iterative procedure as 
\begin{equation}
\begin{alignedat}{1}
\bbeta^{(n+1)} &= \bbeta^{(n)} - \tau^{(n)} \bnab_{\bbeta} \JJ_1(\bbeta^{(n)}), \qquad n=1,2,\cdots,\\
\bnab_{\bbeta} \JJ_1(\bbeta^{(n)}) &= \begin{bmatrix}
\frac{\partial}{\partial \beta_1}\JJ_1(\bbeta^{(n)}) & \quad \frac{\partial}{\partial \beta_2}\JJ_1(\bbeta^{(n)}) &\quad \frac{\partial}{\partial \beta_3}\JJ_1(\bbeta^{(n)}) &\quad \frac{\partial}{\partial \beta_4}\JJ_1(\bbeta^{(n)})\\
\end{bmatrix},
\end{alignedat}
\label{eq:iterative_OCV}
\end{equation}
where $n$ refers to the iteration number, $\tau^{(n)}$ refers to the
step length along the descent direction at each iteration, and
$\bnab_{\bbeta} \JJ_1(\bbeta)$ represents the gradient of cost
functional with respect to the each of the unknown parameters.  Note
that this optimization problem can be solved in two ways. First, the
step length $\tau^{(n)}$ could be computed once in each iteration for
the gradient of the cost functional, which gives rise to the standard
gradient descent technique \cite{nw00}. The second approach is to
update each of the parameters one after another in each iteration of
the algorithm and the step length is to be computed for each of them
independently, which is referred to as the coordinate descent
technique.  In the present study we use the standard gradient descent
technique.  Following the steps presented in Appendix
\ref{sec:relaxation_appendix}, the gradient of the cost functional is
obtained as
\begin{equation}
\begin{alignedat}{1}
\bnab_\bbeta \JJ_1
&= \begin{bmatrix}
- \int_{0}^{T} {\bC^\ast}^{\top} \bI_0 ~dt 
&\quad
- \int_{0}^{T} {\bC^\ast}^{\top} \bI_2 \widehat{\bC} ~ dt 
&\quad
- \int_{0}^{T} {\bC^\ast}^{\top} \bI_3 \widehat{\bC} ~ dt 
&\quad
- \int_{0}^{T} {\bC^\ast}^{\top} \bI_4 \widehat{\bC} ~ dt 
\end{bmatrix}.
\end{alignedat}
\label{eq:gradient_OCV}
\end{equation}
Now that the gradient is computed, we can use the iterative scheme
\eqref{eq:iterative_OCV} to minimize the cost functional to find the
optimal parameters values $\overline{\bbeta} $. The computational
framework is summarized as Stage I in Algorithm \ref{alg:optimal}.
When solving problem \eqref{eq:argmin_OCV}, the piece of each cycle
that corresponds to the relaxation dynamics is used as data
$\widetilde{\bC}(t)$, cf.~\eqref{eq:cost_OCV}.

\subsection{Excitation Dynamics}\label{sec:excitation}
In this section, we assume that the optimal parameter values of the linear dynamics corresponding 
relaxation dynamics are determined. Consequently, we would like to train a model that can 
predict the excitation dynamics of the cell using a nonlinear constitutive relation, via minimizing 
the mismatch between model predictions and experimental data. The 
nonlinear dynamics of the excitation consists of a constitutive relation 
$\omega(C_1,C_2)$ (representing the competition between intercalation and plating), 
and a scalar parameter $\alpha$. Before introducing the optimization framework, 
we need to define two intervals on which the state variables are defined:
\begin{itemize}
	\item $\I := \biggl[ C_1,C_2 \in \RR \Big\vert C_1\in[C_1^{\alpha}, C_1^{\beta}], C_2\in [C_2^{\alpha}, C_2^{\beta}] \biggr]$ is referred to as the identifiability interval, which is the region of 
	state variables spanned by the solution of Eq.~\eqref{eq:ODE}, note that this interval is a function
	of iterations of the iterative algorithm \ref{alg:optimal}, 
	\item $\L := \biggl[   C_1,C_2 \in \RR \Big\vert C_1\in [C_1^a, C_1^b], C_2\in [C_2^a, C_2^b]  \biggr]$, 
	where $C_1^a \leq C_1^{\alpha}$, $C_1^b \geq C_1^{\beta}$, $C_2^a \leq C_2^{\alpha}$ 
	and $C_2^b \geq C_2^{\beta}$; this will be the interval we seek to reconstruct the constitutive 
	relation on, which is generally larger than the identifiability region, i.e.,  $\I \subseteq \L$; 
	the aim is to reconstruct the constitutive relation on this larger interval than spanned by the 
	solution of the ODE system in order to make it possible to reconstruct the constitutive relation on a fixed domain.
\end{itemize}
The constitutive relation defined over $\L$ is considered to be an element of a Hilbert space $\mathcal{X}$. 
Note that the function $\omega$ depends on two state variables which is 
an extension to the problems considered in 
\cite{bukshtynov20111228, bukshtynov2013889, pnm14, sethurajan2015accurate}, in which
the constitutive relation is a function of one state variable only. 
This will add another layer of complexity to the problem of identifying constitutive 
relation. The complexity arises in converting the directional derivative of the objective function 
\eqref{eq:directional} to its Riesz 
form by a change of variables in two dimensions, as will be explained below. Hence, 
to simplify the problem, we will assume that the constitutive relation depending on two 
state variables has a separable form, i.e.,
\begin{equation}
\begin{alignedat}{1}
\omega(C_1,C_2) = \omega_1(C_1)\cdot \omega_2(C_2).
\end{alignedat}
\label{eq:separable}
\end{equation}
Consequently, one can reconstruct each of these factors separately, and then merge the 
results. The functions $\omega_1(C_1)$ and $\omega_2(C_2)$, and the parameter $\alpha$
need to be identified 
from data by solving a suitable inverse problem to minimize the mismatch
between the experimental and true measurements of the system by defining the cost 
functional $\JJ_2:\RR\times\mathcal{X}\times\mathcal{X}\rightarrow\RR$ as
\begin{equation}
\begin{alignedat}{1}
\JJ_2(\alpha, \omega_1,\omega_2) &= \frac{1}{2} \int_{0}^{T} \Big\lvert\Big\lvert \bW \, \br(t;\alpha,\omega_1,\omega_2) \Big\rvert\Big\rvert_2^2 dt,\\
\br (t;\alpha,\omega_1,\omega_2) &= \bC (t;\alpha,\omega_1,\omega_2) - \widetilde{\bC} (t),\\
\bC (t;\alpha, \omega_1,\omega_2) &= \begin{bmatrix}
C_1 (t;\alpha,\omega_1,\omega_2)\\
C_2 (t;\alpha, \omega_1,\omega_2)
\end{bmatrix},\\
\end{alignedat}
\label{eq:cost}
\end{equation}
where the dependence of the state variables ($C_1$ and $C_2$) on the constitutive 
relation $\omega$ is governed by Eq.~\eqref{eq:ODE}. The optimal 
reconstructions of the constitutive relations are obtained by solving the minimization 
problem
\begin{equation}
\begin{alignedat}{1}
\left[\overline{\omega}_1, \overline{\omega}_2, \overline{\alpha}\right] &= \underset{\omega_1 \in \mathcal{X},~ \omega_2 \in \mathcal{X},~ \alpha  \in \RR}{\arg\min \JJ_2(\alpha,\omega_1,\omega_2)},\\
\end{alignedat}
\label{eq:inverse}
\end{equation}
where $\mathcal{X}$ is a suitable Hilbert function space where $\omega_1$ and 
$\omega_2$ belong to. Note that the cost functional $\JJ_2(\alpha,\omega_1,\omega_2)$ 
is a function of two constitutive relations and a parameter. Hence, 
when solving the inverse problem, three parallel problems need to be solved 
simultaneously. For simplicity, these three sub-problems are decoupled and solved. 
In each problem, two of the unknowns are kept constant and 
the third one is optimized.
For the purpose of solving this problem, a 
gradient-based optimization approach can be used with an iterative procedure as 
\begin{equation}
\begin{alignedat}{1}
\omega_1^{(n+1)} &= \omega_1^{(n)} - \tau_1^{(n)} \bnab_{\omega_1}^{\mathcal{X}}  \JJ_2(\alpha^{(n)}, \omega_1^{(n)},\omega_2^{(n)}) \qquad n=1,2,\cdots, \\
\omega_2^{(n+1)} &= \omega_2^{(n)} - \tau_2^{(n)} \bnab_{\omega_2}^{\mathcal{X}}  \JJ_2(\alpha^{(n)}, \omega_1^{(n+1)},\omega_2^{(n)}) \qquad n=1,2,\cdots,\\
\alpha^{(n+1)} &= \alpha^{(n)} - \tau_3^{(n)} \frac{\partial}{\partial \alpha} \JJ_2(\alpha^{(n)}, \omega_1^{(n+1)},\omega_2^{(n+1)}) \qquad n=1,2,\cdots,\\
\end{alignedat}
\label{eq:iterative}
\end{equation}
where $n$ refers to the iteration number, 
$\tau_i^{(n)}, i\in\{1,2,3\}$ refers to the step length along 
the descent direction at each iteration, and 
$\bnab_{\omega_1}^{\mathcal{X}} \JJ_2(\alpha,\omega_1,\omega_2)$ and 
$\bnab_{\omega_2}^{\mathcal{X}} \JJ_2(\alpha,\omega_1,\omega_2)$ 
represent the gradients of cost functional with respect to the each 
of the constitutive relations, and 
$\frac{\partial}{\partial \alpha}\JJ_2(\alpha,\omega_1,\omega_2)$ is 
the partial derivative of the cost functional with respect to the unknown parameter. 
Note that relation \eqref{eq:iterative} represents the 
steepest-descent optimization algorithm, however, in practice, one can use 
more sophisticated techniques such as the conjugate-gradients method. The 
Polak-Ribiere conjugate-gradient formulation has been used for this study.
In all cases, the key 
ingredient of the optimization algorithm is the information about the gradient 
of the cost functional with respect to the constitutive relation. Note that the 
constitutive relation $\omega(C_1,C_2)$ is a continuous function of state 
variables over $\L$, hence the gradients 
$\bnab_{\omega_1}^{\mathcal{X}} \JJ_2(\alpha,\omega_1,\omega_2)$ and 
$\bnab_{\omega_2}^{\mathcal{X}} \JJ_2(\alpha,\omega_1,\omega_2)$ are 
infinite-dimensional sensitivities of the cost functional to the perturbations 
of these constitutive relations. In order to compute these gradients, adjoint 
sensitivity analysis is leveraged \cite{bukshtynov20111228, bukshtynov2013889, pnm14}. 
An application of this adjoint sensitivity analysis to reconstruction of 
constitutive relations in electrochemistry field can also be found in 
\cite{sethurajan2015accurate}. One needs to reconstruct each of the factors
$\omega_1(C_1)$ and $\omega_2(C_2)$ in \eqref{eq:separable}
as the elements of the Sobolev space $H^1(\L)$ 
to ensure the continuity of the reconstructed constitutive relation, thus, the gradient needs 
to be obtained with respect to the corresponding inner product. However, to 
simplify the derivation, we will first obtain the gradient in the space 
$\mathcal{X}(\I) = L^2(\I)$, and we will use the results of this 
derivation to find the Sobolev gradient. Note that the following mathematical derivation 
focuses solely on the gradient of the cost functional with respect to $\omega_1$. 
The derivation of the gradient with respect to $\omega_2$ and the partial derivative of 
the cost functional with respect to $\alpha$ follow a similar process.
In order to obtain convenient expression for 
the gradient, we begin by computing the Gateaux (directional) derivative with respect to 
perturbation of $\omega_1$ as 
\begin{equation}
\begin{alignedat}{3}
\JJ_2^\prime(\alpha,\omega_1,\omega_2;\omega_1^\prime) &= \lim_{\epsilon\to 0}\epsilon^{-1} \left[\JJ_2(\alpha, \omega_1+\epsilon\omega_1^\prime,\omega_2)-\JJ_2(\alpha, \omega_1,\omega_2)\right] \\&= \int_{0}^{T} (\bw \,  \br(t;\alpha,\omega_1,\omega_2))^\top \bC^\prime(\alpha,\omega_1,\omega_2;\omega_1^\prime) dt,\\
\bC^\prime(\alpha,\omega_1,\omega_2;\omega_1^\prime) &= \begin{bmatrix}
C_{1}^\prime(\alpha,\omega_1,\omega_2;\omega_1^\prime)\\
C_{2}^\prime(\alpha,\omega_1,\omega_2;\omega_1^\prime)
\end{bmatrix},
\end{alignedat}
\label{eq:directional}
\end{equation}
where $\bC^\prime (\alpha,\omega_1,\omega_2;\omega_1^\prime)$ is the solution to 
the system of perturbation equations. In order to derive this system, the state 
variables are perturbed with respect to $\omega_1$ as 
\begin{equation}
\begin{alignedat}{1}
\bC(\alpha,\omega_1,\omega_2) &= \widehat{\bC}(\widehat{\alpha},\widehat{\omega}_1,\widehat{\omega}_2) + \epsilon \left[ \bC^\prime(\alpha,\omega_1,\omega_2;\omega_1^\prime) \right] + \mathcal{O}(\epsilon^2).
\end{alignedat}
\label{eq:Cprime}
\end{equation}
The constitutive relations are perturbed with respect to $\omega_1$ as (the arguments are dropped for brevity)
\begin{equation}
\begin{alignedat}{1}
\omega_1  &= \widehat{\omega}_1 
+ \epsilon\left[ \omega_1^\prime 
+ \frac{d\omega_1}{d C_1}C_{1}^\prime \right]  + \mathcal{O}(\epsilon^2) ,\\
\omega_2 &= \widehat{\omega}_2 
+ \epsilon\left[ \frac{d\omega_2}{d C_2}C_{2}^\prime \right]  + \mathcal{O}(\epsilon^2).
\end{alignedat}
\end{equation}
Note that perturbation of one constitutive relation will affect both concentrations 
(as $C_1$ and $C_2$ are not decoupled).
The perturbation of the constitutive relation takes the form 
\begin{equation}
\begin{alignedat}{1}
\omega &= \widehat{\omega}_1 \, \widehat{\omega}_2
+ \epsilon \left[ \widehat{\omega}_2 \omega_1^\prime 
+ \widehat{\omega}_2\frac{d \omega_1}{dC_1} C_{1}^\prime
+ \widehat{\omega}_1 \frac{d\omega_2}{dC_2}C_{2}^\prime
\right] + \mathcal{O}(\epsilon^2).
\end{alignedat}
\label{eq:omegaprime}
\end{equation}
Substituting \eqref{eq:Cprime} and \eqref{eq:omegaprime} into \eqref{eq:ODE}, 
and collecting terms proportional to $\epsilon$,
we get the perturbation system of equations corresponding to $\omega_1^\prime$ as 
\begin{equation}
\begin{alignedat}{1}
\frac{dC_{1}^\prime }{dt}  &=  
\beta_2 C_{1}^\prime + \lambda^{-1}\beta_3 C_{2}^\prime - \widehat{\alpha} \lambda^{-1} J_{app}\left[ 
\widehat{\omega}_2 \omega_1^\prime 
+ \widehat{\omega}_2\frac{d \omega_1}{dC_1}C_{1}^\prime
+ \widehat{\omega}_1 \frac{d\omega_2}{dC_2}C_{2}^\prime \right]\\
\frac{d C_{2}^\prime }{dt}  &= 
\beta_4 C_{1}^\prime - \beta_3 C_{2}^\prime  
+ \widehat{\alpha} J_{app} \left[ 
\widehat{\omega}_2 \omega_1^\prime 
+ \widehat{\omega}_2\frac{d \omega_1}{dC_1}C_{1}^\prime
+ \widehat{\omega}_1 \frac{d\omega_2}{dC_2}C_{2}^\prime  \right]\\
C_1^\prime(\omega_1^\prime) (0) &= C_2^\prime(\omega_1^\prime)(0) = 0.
\end{alignedat}
\label{eq:perturb_1}
\end{equation}
Following similar procedure the perturbation system of equations corresponding to $\omega_2^\prime$ 
and $\alpha^\prime$ will be obtained.
In matrix form, we get the perturbed system of equations as  
\begin{subequations}
	\begin{alignat}{1} 
	\frac{d}{dt} \bC^\prime (t) &= \bA \bC^\prime(t) + \bD \bC^\prime(t) + \widehat{\omega}_2 \widehat{\alpha} \brho \omega_1^\prime, \label{eq:perturb1a}\\
	\bC^\prime  (0) &= \boldsymbol{0},\\
	\bD &= \begin{bmatrix}
	- \widehat{\alpha} \lambda^{-1}J_{app} \widehat{\omega}_2\frac{d \omega_1}{dC_1}  & - \widehat{\alpha} \lambda^{-1}J_{app} \widehat{\omega}_1 \frac{d\omega_2}{dC_2}\\
	\widehat{\alpha} J_{app} \widehat{\omega}_2\frac{d \omega_1}{dC_1} & \widehat{\alpha} J_{app}\widehat{\omega}_1 \frac{d\omega_2}{dC_2}
	\end{bmatrix},\\
	\brho &= \begin{bmatrix}
	- \lambda^{-1} J_{app}\\
	J_{app}
	\end{bmatrix}.
	\end{alignat}
	\label{eq:perturb1}
\end{subequations}
Note that the first term in the right-hand-side of the ODE \eqref{eq:perturb1a} is the linear sub-problem 
corresponding to the relaxation dynamics (cf.~Eq.~\eqref{eq:ode_relaxation}), 
and the second and third terms correspond to the excitation dynamics.
We will obtain one ODE system for the perturbation of each unknown.
Also, in all scenarios of perturbation of $\alpha$, $\omega_1$ and $\omega_2$
the matrix $\bD$ appears to be identical, with differences occurring in the definition of 
the third term in the right-hand-side of \eqref{eq:perturb1a}.
The directional derivative of the cost functional can be computed in a different manner than 
\eqref{eq:directional} by invoking the Riesz representation 
theorem to the directional derivatives in the functional space as 
\begin{equation}
\begin{alignedat}{1}
\JJ_2^\prime(\alpha,\omega_1,\omega_2;\omega_1^\prime) &= \langle \bnab_{\omega_1}^{\mathcal{X}} \JJ_2, \omega_1^\prime \rangle_{\mathcal{X}(\L)},
\end{alignedat}
\end{equation}
and similarly for $\JJ_2^\prime(\alpha,\omega_1,\omega_2;\omega_2^\prime)$,
where $\langle\cdot,\cdot\rangle_{\mathcal{X(\L)}}$ represents the inner product 
in the Hilbert space $\mathcal{X}$ over $\L$ interval. Note that the Riesz representer
in a functional space will reduce to the partial derivative in a finite-dimensional Euclidean space, namely, 
$\JJ_2^\prime(\alpha,\omega_1,\omega_2;\alpha^\prime) =  \frac{\partial \JJ_2}{\partial \alpha}\cdot \alpha^\prime$. 
Assuming $\mathcal{X}(\L) = L^2(\L)$, the directional derivative will be expressed in terms of 
the $L^2$ inner product as
\begin{equation}
\begin{alignedat}{1}
\JJ_2^\prime(\alpha,\omega_1,\omega_2;\omega_1^\prime) &= \int_{C_1^{a}}^{C_1^{b}} \bnab_{\omega_1}^{L^2} \JJ_2 \cdot \omega_1^\prime ds.
\end{alignedat}
\label{eq:riesz}
\end{equation}
Note that the Gateaux derivative \eqref{eq:directional} is not consistent 
with the Riesz form \eqref{eq:riesz}, as the expression for the perturbation of the 
constitutive relations is hidden in the perturbations of the state variables 
$C_1^\prime(\alpha,\omega_1,\omega_2;\omega_1^\prime)$ and
$C_2^\prime(\alpha,\omega_1,\omega_2;\omega_1^\prime)$ in Eq.~\eqref{eq:directional}. 
Also, the integration variable in Gateaux form is time, whereas the Riesz form
uses the state variable as the integration variable. In order 
to tackle the first issue (introducing an explicit dependence on the perturbation of 
the constitutive relation into the Gateaux differential, as in \eqref{eq:riesz}), 
we will leverage adjoint analysis, in which an adjoint 
problem is defined in a judicious manner so that expression for directional 
derivative becomes consistent with its Riesz form \eqref{eq:riesz}. 
Whereas, to overcome the latter issue (inconsistency in integration variable) 
a change of variables is used. 

We begin with adjoint analysis. We multiply 
\eqref{eq:perturb1} by the vector of adjoint variables $\bC^\ast(t) = \left[C_{1}^\ast(t), C_{2}^\ast(t) \right]^\top$, and 
integrating in time, we obtain
\begin{equation}
\begin{alignedat}{1}
\int_{0}^{T} \bC^{\ast\top} \frac{d}{dt} \bC^\prime dt 
- \int_{0}^{T} \bC^{\ast\top} \bA \bC^\prime dt 
- \int_{0}^{T} \bC^{\ast\top} \bD \bC^\prime dt 
- \int_{0}^{T} \bC^{\ast\top}  \widehat{\omega}_2 \widehat{\alpha} \brho \omega_1^\prime dt &= 0.
\end{alignedat}
\end{equation}
Performing integration by parts for the first term and 
applying the initial conditions of the perturbation system \eqref{eq:perturb_1}, we get
\begin{equation}
\begin{alignedat}{1}
- \bC^{\ast\top}(T)  \bC^\prime(T)
+ \int_{0}^{T} \frac{d}{dt}\bC^{\ast\top}  \bC^\prime dt 
+ \int_{0}^{T} \bC^{\ast\top} \bA \bC^\prime dt \\
+ \int_{0}^{T} \bC^{\ast\top} \bD \bC^\prime dt 
+ \int_{0}^{T} \bC^{\ast\top}  \widehat{\omega}_2 \widehat{\alpha} \brho \omega_1^\prime dt &= 0.
\end{alignedat}
\end{equation}
Factoring out $\bC^\prime$, we get
\begin{equation}
\begin{alignedat}{1}
- \bC^{\ast\top}(T)  \bC^\prime(T)
+ \int_{0}^{T} \left[\frac{d}{dt}\bC^{\ast\top} + \bC^{\ast\top} \bA  + \bC^{\ast\top} \bD \right]  \bC^\prime dt 
+ \int_{0}^{T} \bC^{\ast\top}  \widehat{\omega}_2 \widehat{\alpha} \brho \omega_1^\prime dt &= 0.
\end{alignedat}
\label{eq:derivation1}
\end{equation}
We define the adjoint system of equations in a judicious manner as
\begin{equation}
\begin{alignedat}{1}
\frac{d}{dt}\bC^{\ast} (t) + \bA^\top \bC^{\ast}(t) + \bD^\top \bC^{\ast}(t)  &= \bw \, \br(t;\omega_1,\omega_2), \\
\bC^\ast(T) &= \boldsymbol{0}.
\end{alignedat}
\label{eq:adjoint}
\end{equation}
Note that when performing adjoint analysis for system of equations with respect to perturbation 
of $\omega_2$ and $\alpha$, the evolution of adjoint variables $\bC^{\ast} (t)$ is governed by 
exactly the same system of equations and terminal conditions and the difference is in how this 
information is used to determine the corresponding gradient. With this definition of the 
adjoint system, Eq.~\eqref{eq:derivation1} becomes
\begin{equation}
\begin{alignedat}{1}
\int_{0}^{T} \left[(\bw \, \br)^\top\right]  \bC^\prime dt  &= - \int_{0}^{T} \bC^{\ast\top}  \widehat{\omega}_2 \widehat{\alpha} \brho \omega_1^\prime dt.
\end{alignedat}
\end{equation}
Thus, the directional derivative with respect to $\omega_1^\prime$ becomes
\begin{equation}
\begin{alignedat}{1}
\JJ_2^\prime(\alpha,\omega_1,\omega_2;\omega_1^\prime) &= -\int_{0}^{T} \widehat{\omega}_2 \widehat{\alpha} {\bC^\ast}^\top \brho \omega_1^\prime dt,
\end{alignedat}
\label{eq:derivation2}
\end{equation}
due to the choice of the source term in the adjoint system \eqref{eq:adjoint} so that the 
expression of Gateaux differential appears in the equation. Likewise, following similar procedure 
of adjoint analysis for $\omega_2^\prime$ and $\alpha^\prime$, the directional derivatives with 
respect to each of these unknowns become
\begin{subequations}
	\begin{alignat}{1}
	\JJ_2^\prime(\alpha,\omega_1,\omega_2;\omega_2^\prime) &= -\int_{0}^{T} \widehat{\omega}_1 \widehat{\alpha} {\bC^\ast}^\top \brho \omega_2^\prime dt,\label{eq:derivation3a}\\
	\JJ_2^\prime(\alpha,\omega_1,\omega_2;\alpha^\prime) &= -\int_{0}^{T} \widehat{\omega}_1 \widehat{\omega}_2 {\bC^\ast}^\top \boldsymbol{\varrho} \alpha^\prime dt,\label{eq:derivation3b}
	\end{alignat}
	\label{eq:derivation3}
\end{subequations}
where 
\begin{equation*}
\begin{alignedat}{1}
\boldsymbol{\varrho} &= \begin{bmatrix}
( \widehat{\omega}^{-1} - \lambda^{-1}) J_{app}\\
J_{app}
\end{bmatrix}.\\
\end{alignedat}
\end{equation*}

As can be observed, the Gateaux differential \eqref{eq:derivation3a}-\eqref{eq:derivation3b} is expressed in terms of 
perturbation of the constitutive relation, which is consistent with 
Riesz form \eqref{eq:riesz}. However, the integration variable in 
relations \eqref{eq:derivation3a}-\eqref{eq:derivation3b} (time) is different than the integration variable in 
Riesz form (state variable). To make them consistent, a change of 
variables must be used, namely,
\begin{equation}
\begin{alignedat}{1}
dt &= \frac{dC_1}{\beta_1 + \beta_2 C_1 + \lambda^{-1} \beta_3 C_2 + \widehat{\alpha} (1-\lambda^{-1} \widehat{\omega} )J_{app}} = \frac{dC_2}{\beta_4 C_1 - \beta_3 C_2 + \widehat{\alpha}\widehat{\omega} J_{app}},
\end{alignedat}
\end{equation}
which is obtained by rearrangement of the forward model \eqref{eq:ODE}.
This makes it possible to change the integration variable in \eqref{eq:derivation2} 
from time ($dt$) to the state ($dC_1$ and $dC_2$), as required by the Riesz 
representation \eqref{eq:riesz}, as the
mapping from time to state variable is unique, 
$\mathcal{K}:=\{\cup_{t\in\left[0,T\right]} [C_1(t),C_2(t)]\}$.
As the mapping from time to state variables is unique, the 
integral over the $\L$ interval can be expressed as an integral over the 
contour $\mathcal{K}$. Hence, applying this change of variables to \eqref{eq:derivation2}
and \eqref{eq:derivation3b}, we obtain
\begin{equation}
\begin{alignedat}{1}
\JJ_2^\prime(\alpha,\omega_1,\omega_2;\omega_1^\prime) &= -\int_{C_1^\alpha}^{C_1^\beta} \frac{\widehat{\omega}_2 \widehat{\alpha} {\bC^\ast}^\top \brho }{\beta_1 + \beta_2 C_1 + \lambda^{-1} \beta_3 C_2 + \widehat{\alpha} (1-\lambda^{-1} \widehat{\omega} )J_{app}} \omega_1^\prime ds,\\
\JJ_2^\prime(\alpha,\omega_1,\omega_2;\omega_2^\prime) &= -\int_{C_2^\alpha}^{C_2^\beta} \frac{\widehat{\omega}_1 \widehat{\alpha} {\bC^\ast}^\top \brho }{\beta_4 C_1 - \beta_3 C_2 + \widehat{\alpha}\widehat{\omega} J_{app}} \omega_2^\prime ds,\\
\JJ_2^\prime(\alpha,\omega_1,\omega_2;\alpha^\prime) &= \left[-\int_{0}^{T} \widehat{\omega}_1 \widehat{\omega}_2 {\bC^\ast}^\top \boldsymbol{\varrho} dt\right] \cdot \alpha^\prime.
\end{alignedat}
\end{equation}
Note that $\alpha^\prime$ is independent of time and is taken out of integral. 
Hence, the $L^2$ gradients and the partial derivative are computed as
\begin{equation}
\begin{alignedat}{1}
\bnab_{\omega_1}^{L^2} \JJ_2 &= -\frac{\widehat{\omega}_2 \widehat{\alpha} {\bC^\ast}^\top \brho}{\beta_1 + \beta_2 C_1 + \lambda^{-1} \beta_3 C_2 + \widehat{\alpha} (1-\lambda^{-1} \widehat{\omega} )J_{app}},\\
\bnab_{\omega_2}^{L^2} \JJ_2 &= -\frac{\widehat{\omega}_1 \widehat{\alpha} {\bC^\ast}^\top \brho}{\beta_4 C_1 - \beta_3 C_2 + \widehat{\alpha}\widehat{\omega} J_{app}},\\
\frac{\partial \JJ_2}{\partial \alpha} &= -\int_{0}^{T} \widehat{\omega}_1 \widehat{\omega}_2 {\bC^\ast}^\top \boldsymbol{\varrho} dt.
\label{eq:L2}
\end{alignedat}
\end{equation}
Above, we derived gradient expressions with respect to constitutive relations
in the $L^2$ functional space. However, as noted in earlier studies 
\cite{bukshtynov20111228, bukshtynov2013889, pnm14}, these gradients are 
not a suitable choice for reconstruction of constitutive relations as 
they are generally discontinuous and are undefined outside the 
identifiability region $\I$. Thus, to ensure the regularity and the 
smoothness of the reconstructed relations over the domain of definition $\L$, 
we will redefine them in the $H^1$ Sobolev space of functions of the 
concentrations $C_1$ and $C_2$ with square-integrable derivatives. 
A natural choice is to construct the Sobolev gradients for both constitutive 
relations by assuming $\mathcal{X} = H^1(\L)$. 
Since the constitutive relation in the governing system \eqref{eq:ODE} depends on the 
product $\omega_1\cdot\omega_2$, an optimization formulation in which these two factors 
are determined independently as in \eqref{eq:inverse} is underdetermined, because the 
mean of the product $\omega_1\cdot\omega_2$ can be changed by each of the factors, 
which can lead to numerical complications. We will therefore amend the formulation such 
that the mean value of one of the factors will be fixed (for example, at zero).
One can achieve this by imposing hard constraints on the mean 
of the functions so that their mean stays stationary in the optimization 
framework. In this work, the mean of one of the factors is set to remain constant during 
the optimization process 
by ensuring the Sobolev gradients are defined such that they do not modify the mean. 
This will leave the first factor to capture the mean value of the entire constitutive relation.
Also, the physical constraints of the problem imply that the 
constitutive relations should be bounded between zero and one. For such 
reasons, one is required to constrain the functions in order to ensure 
physically plausible solutions. In this framework, we do not impose any restrictions
on the mean of the constitutive relation $\omega$, thus, the physical constraint is 
not guaranteed to be satisfied.
Thus, two functional spaces will be used in this framework for extending the $L^2$ 
gradients to $\L$ interval, namely, $\mathcal{X} = H^1(\L)$ and 
$\mathcal{X} = H^1_0(\L)$ (where the subscript $0$ denotes a space of functions of zero mean). 
The $H^1$ Sobolev space is endowed with the inner product as
\begin{equation}
\begin{alignedat}{1}
\langle \bnab_{\omega_1}^{H^1} \JJ_2, \omega_1^\prime \rangle_{H^1} &= \int_{C_1^{a}}^{C_1^{b}} \left(\bnab_{\omega_1}^{H^1} \JJ_2 \cdot \omega_1^\prime + l^2 \frac{d \bnab_{\omega_1}^{H^1} \JJ_2}{ds}\frac{d \omega_1^\prime}{ds} \right) ds,
\label{eq:Sobolev}
\end{alignedat}
\end{equation}
for computing the directional derivative $\JJ_2^\prime(\alpha,\omega_1,\omega_2;\omega_1^\prime)$,
where $0<l<\infty$ is the length-scale parameter, controlling the intensity of 
smoothness of gradients. Setting this parameter to zero recovers the $L^2$ 
inner product, cf.~Eq.~\eqref{eq:riesz}. 
A similar $H^1$ inner product is also used for the computation of 
$\JJ_2^\prime(\alpha,\omega_1,\omega_2;\omega_2^\prime)$ with the difference that 
$\bnab_{\omega_2}^{H^1} \JJ_2$ is replaced with $\PP_0\bnab_{\omega_2}^{H^1_0} \JJ_2$ to ensure
the zero mean of the reconstructed function.
The operator $\PP_0: H^1 \rightarrow H^1_0$ represents the orthogonal projection
on the subspace of functions with zero mean and is defined as 
$\PP_0 u = u-\overline{u}$, where $\overline{u}$ in the mean of the function over the domain. 
Here we assume that $\omega_1^\prime \in H^1(\L)$ and  $\omega_2^\prime \in H^1_0(\L)$. 
So by invoking the Riesz theorem, we obtain
\begin{equation}
\begin{alignedat}{2}
\JJ_2^\prime(\alpha,\omega_1,\omega_2;\omega_1^\prime) &= \langle \bnab_{\omega_1}^{L^2} \JJ_2, \omega_1^\prime \rangle_{L^2(\L)} = \langle \bnab_{\omega_1}^{H^1} \JJ_2, \omega_1^\prime \rangle_{H^1(\L)},
\label{eq:riesz1}
\end{alignedat}
\end{equation}
and similarly for $\JJ_2^\prime(\alpha,\omega_1,\omega_2;\omega_2^\prime)$.
Considering \eqref{eq:Sobolev} and \eqref{eq:riesz1}, and performing 
integration by parts with respect to $s$, we obtain
\begin{equation}
\begin{alignedat}{1}
\int_{C_1^{a}}^{C_1^{b}} \bnab_{\omega_1}^{L^2} \JJ_2 \cdot \omega_1^\prime  ds  
&= \int_{C_1^{a}}^{C_1^{b}} \left(\bnab_{\omega_1}^{H^1} \JJ_2 \cdot \omega_1^\prime 
- l^2 \frac{d^2 \bnab_{\omega_1}^{H^1} \JJ_2}{ds^2}\omega_1^\prime \right) ds +
\frac{d \bnab_{\omega_1}^{H^1} \JJ_2}{ds} \omega_1^\prime \Big\lvert_{C_1^{a}}^{C_1^{b}},
\end{alignedat}
\end{equation}
noting that the perturbations $\omega_1^\prime$ and $\omega_2^\prime$ 
are arbitrary. A similar analysis can be performed for perturbation with respect to 
$\omega_2^\prime$. By imposing the Neumann boundary conditions on the Sobolev 
gradients, we obtain the following inhomogeneous elliptic boundary-value 
problems defining the smoothed gradients in the $H^1$ and $H^1_0$ space 
based on the $L^2$ gradients as
\begin{equation}
\begin{alignedat}{3}
\bnab_{\omega_1}^{H^1} \JJ_2 - l^2 \frac{d^2 \bnab_{\omega_1}^{H^1} \JJ_2}{ds^2} &= \bnab_{\omega_1}^{L^2} \JJ_2, \qquad&&\text{on}\qquad \L,\\
\frac{d \bnab_{\omega_1}^{H^1} \JJ_2}{ds} &= 0,\qquad&&\text{at}\qquad s= C_1^{a},C_1^{b},
\end{alignedat}
\label{eq:BVP_1}
\end{equation}
and 
\begin{equation}
\begin{alignedat}{3}
\bnab_{\omega_2}^{H^1_0} \JJ_2 - \frac{1}{C_2^{b}-C_2^{a}}\int_{C_2^{a}}^{C_2^{b}} \bnab_{\omega_2}^{H^1_0} \JJ_2 \, ds  - l^2 \frac{d^2 \bnab_{\omega_2}^{H^1_0} \JJ_2}{ds^2} &= \bnab_{\omega_2}^{L^2} \JJ_2,\qquad&&\text{on}\qquad \L,\\
\frac{d \bnab_{\omega_2}^{H^1_0} \JJ_2}{ds} &= 0,\qquad&&\text{at}\qquad s= C_2^{a},C_2^{b}.
\end{alignedat}
\label{eq:BVP_2}
\end{equation}
This framework ensures that the gradient of the cost 
functional with respect to $\omega_2$ has a zero mean at each step of 
the algorithm, hence, the mean of the function $\omega_2$ remains 
unchanged during the iteration process. Note that the behaviour of the Sobolev 
gradients on the boundaries needs to be specified via suitable boundary 
conditions. The choice of the boundary 
condition is nontrivial. In this case, the homogeneous Neumann boundary condition is 
adopted which preserves the values of the derivatives of the functions $\omega_1$ and 
$\omega_2$ at the boundaries, but allows the gradient to modify their values at the boundary. 
Some other choices of boundary condition are possible based on the physics 
of the problem. For example, imposing homogeneous Dirichlet boundary conditions would preserve 
the values of $\omega_1$ and $\omega_2$ at the boundaries, but would make it possible to 
modify their derivatives. Also, extending the gradients to a Sobolev space can be 
seen as an extrapolation of gradients to the regions of the state space where the sensitivity 
information is not available, i.e., the L2 gradient vanishes identically 
\cite{bukshtynov20111228}. The computational framework 
for the solution of optimization problems \eqref{eq:iterative_OCV} and \eqref{eq:iterative}
is summarized in Algorithm \ref{alg:optimal}. It is also notable that for solving the forward 
system \eqref{eq:ODE} as part of the computational framework in Algorithm \ref{alg:optimal} throughout 
this study the MATLAB routine {\tt ODE45} is used with a loose tolerance. As can be observed in 
Figure \ref{fig:raw_data}, the current applied to the cell is discontinuous, hence making the 
forward problem stiff. However, some analysis revealed that when the tolerance of the ODE 
solver is loose, the accuracy of the results is satisfactory as the step size of the ODE solver 
will be large and the effect of sharp changes in current profile will not be pronounced by the solver. 
On the other hand, using stiff ODE solvers requires very tight tolerances to be able to achieve the 
required accuracy from the algorithm. Hence, for the sake of saving computational time, 
the non-stiff solver ({\tt ODE45}) with loose tolerance is used in this work. 
\begin{algorithm}\footnotesize
	\singlespacing
	\SetAlgoLined
	\vspace{-1em}
	\KwIn{~~$\bbeta^{(0)}, \alpha^{(0)}, \omega_1^{(0)}, \omega_2^{(0)}$ \textbf{---} Initial guesses for parameters and constitutive relations\\
		~~~~~~~~~~~~~$N$ \textbf{---} Maximum iteration number\\
		~~~~~~~~~~~~~$TOL$ \textbf{---} Tolerance}
	\KwOut{$\overline{\bbeta},\overline{\alpha},\overline{\omega}(C_1,C_2)$ \textbf{---} Optimally constructed parameters and constitutive relations\\
	}
	\SetAlgoVlined
	\vspace{0.2cm}\hrule\vspace{0.05cm}\hrule\vspace{0.2cm}
	{\bf Stage I:} Optimal reconstruction of $\bbeta$:\\
	\Indp \Indp 
	Initialization:\\
	\qquad set $n=0$,\\
	\qquad set $\bbeta^{(0)}$ as initial guess,\\
	\Repeat(){$ \frac{\JJ_1(\bbeta^{(n)})}{\JJ_1(\bbeta^{(n-1)})} < \text{TOL}$ or $n>N$  }{
		\vspace{0.1cm}
		$\bullet$ set $n = n+1$, \\
		$\bullet$ solve forward problem \eqref{eq:ODE} based on prior estimation of $\widehat{\bbeta}$ to obtain $C_1(t;\bbeta^{(n-1)})$ and $C_2(t;\bbeta^{(n-1)})$, assuming $J_{app} = 0$, \\
		$\bullet$ solve adjoint problem \eqref{eq:adjoint_OCV} to obtain $C_1^\ast(t;\bbeta^{(n-1)})$ and $C_2^\ast(t;\bbeta^{(n-1)})$, \\
		$\bullet$ compute gradient of cost functional with respect to parameters, $\bnab_{\bbeta} \JJ_1$ via \eqref{eq:gradient_OCV},\\
		$\bullet$ determine step length $\tau^{(n)}$ of optimization iterative scheme \eqref{eq:iterative_OCV} via Brent's line search scheme as outlined in \cite{pftv86},\\
		$\bullet$ compute the updated parameters for $\bbeta^{(n)}$ via \eqref{eq:iterative_OCV} as the posterior estimation of $\widehat{\bbeta}$,
	}()
	\vspace{0.2cm}\hrule\vspace{0.05cm}\hrule\vspace{0.2cm}
	\Indm\Indm
	{\bf Stage II:} Optimal reconstruction of $\alpha$, $\omega_1$ and $\omega_2$:\\
	\Indp \Indp 
	Initialization:\\
	\qquad set $n=0$,\\
	\qquad set $\bbeta = \overline{\bbeta}$,\\
	\qquad set $\alpha^{(0)}$, $\omega_1^{(0)}(C_1)$ and $\omega_2^{(0)}(C_2)$ as initial guesses,\\
	\Repeat(){$ \frac{\JJ_2(\alpha^{(n)},\omega_1^{(n)},\omega_2^{(n)})}{\JJ_2(\alpha^{(n-1)},\omega_1^{(n-1)},\omega_2^{(n-1)})} < \text{TOL}$ or $n>N$ }{
		\vspace{0.1cm}
		$\bullet$ set $n = n+1$, \\
		$\bullet$ solve forward problem \eqref{eq:ODE} to obtain $C_1(t;\alpha^{(n-1)},\omega_1^{(n-1)},\omega_2^{(n-1)})$ and $C_2(t;\alpha^{(n-1)},\omega_1^{(n-1)},\omega_2^{(n-1)})$, \\
		$\bullet$ solve adjoint problem \eqref{eq:adjoint} to obtain $C_1^\ast(t;\alpha^{(n-1)},\omega_1^{(n-1)},\omega_2^{(n-1)})$ and $C_2^\ast(t;\alpha^{(n-1)},\omega_1^{(n-1)},\omega_2^{(n-1)})$, \\
		$\bullet$ compute $L^2$ gradients of cost functional with respect to constitutive relations,  $\bnab_{\omega_1}^{L^2} \JJ$ and $\bnab_{\omega_2}^{L^2} \JJ$, and $\frac{\partial\JJ}{\partial\alpha}$ via \eqref{eq:L2},\\
		$\bullet$ solve the boundary-value problems \eqref{eq:BVP_1} and \eqref{eq:BVP_2} to obtain Sobolev gradients of cost functionals $\bnab_{\omega_1}^{H^1} \JJ$ and $\bnab_{\omega_2}^{H^1_0} \JJ$,\\
		$\bullet$ determine step length $\tau^{(n)}$ of optimization iterative scheme \eqref{eq:iterative} via Brent's line search scheme as outlined in \cite{pftv86},\\
		$\bullet$ compute the updated relations for $\alpha^{(n)}$, $\omega_1^{(n)}$ and $\omega_2^{(n)}$ via \eqref{eq:iterative},
	}()
	Compute $\overline{\omega}(C_1,C_2) = \overline{\omega}_1^{(n)} \cdot \overline{\omega}_2^{(n)}$
	\caption{ \textsc{Computational Framework for Optimal Reconstruction of Constitutive Relations}\label{alg:cap}}
	\label{alg:optimal}
\end{algorithm}

\subsection{Formulation with Aggregated Data}\label{sec:robust}

The computational framework outlined in Algorithm \ref{alg:optimal} could be utilized to train models 
for both relaxation and excitation dynamics based on a single cycle of the cell. In other words, 
each sequence of data 
$\mathcal{D}_i, i\in \mathcal{C}, \mathcal{C}= \{\text{C3},\text{C2},\text{1C},\text{2C},\text{3C}\}$,
could be used as the training data for optimal reconstruction of parameters and 
constitutive relations. In this scenario, the parameters and the constitutive relations would be adjusted 
to minimize the mismatch between predictions of the model and the experimental concentrations for a specific cycle.
However, it is known that such models suffer from robustness issues, as the trained model tends to 
exhibit acceptable performance only over a limited range of cycles (C-rates) close to the 
cycle used for training, cf.~Section \ref{sec:results_fullmodel}. To enhance the
robustness of the optimal reconstruction framework, one can train the models on a wider range 
of C-rates by concatenating different sequences of data, each corresponding to a particular C-rate,
$\mathcal{D}_t = \bigoplus_i \mathcal{D}_i, i\in \mathcal{C}$. In this scenario, the cost functional
would be defined as the sum of cost functionals for each sequence of data for relaxation dynamics as
$\JJ_1(\bbeta;\mathcal{D}_t^{ocv}) = \sum_{i\in \mathcal{C}}^{} \JJ_1(\bbeta;\mathcal{D}_i^{ocv})$, 
where $\JJ_1(\bbeta;\mathcal{D}_i^{ocv})$ denotes the cost functional computed by using $\mathcal{D}_i^{ocv}$
as the experimental data. The cost functional for the excitation dynamics would be defined as
$\JJ_2(\alpha,\omega_1,\omega_2;\mathcal{D}_t^j) = \sum_{i\in \mathcal{C}}^{} 
\JJ_2(\alpha,\omega_1,\omega_2;\mathcal{D}_i^j)$, where $j\in\{ch,dch\}$.
With this revised definition of cost functionals for optimization, the gradients need to 
be computed accordingly. Since the gradient is a linear operator, the gradient of the combined
cost functional reduces to the sum of the gradients of cost functionals for 
each sequence of data $\mathcal{D}_i$. More precisely, for the relaxation dynamics we get
$\bnab_{\bbeta} \JJ_1(\bbeta;\mathcal{D}_t^{ocv}) = \sum_{i\in \mathcal{C}}^{} \bnab_{\bbeta} \JJ_1(\bbeta;\mathcal{D}_i^{ocv})$,
and for the excitation dynamics
$\bnab_{\omega_1} \JJ_2(\alpha,\omega_1,\omega_2;\mathcal{D}_t^j) = \sum_{i\in \mathcal{C}}^{} 
\bnab_{\omega_1} \JJ_2(\alpha,\omega_1,\omega_2;\mathcal{D}_i^j)$, where $j\in\{ch,dch\}$, and similarly
for $\bnab_{\omega_2} \JJ_2(\alpha,\omega_1,\omega_2;\mathcal{D}_t^j)$ and 
$\frac{\partial}{\partial \alpha} \JJ_2(\alpha,\omega_1,\omega_2;\mathcal{D}_t^j)$.
After the computation of the cost functionals and gradients for optimization, the remainder of 
the computational framework remains unchanged.

\section{Results} \label{sec:results}
In this section, we first present the results for the relaxation dynamics part of the model, 
as outlined in Section \ref{sec:relaxation}. Once the parameters $\bbeta$ of the relaxation dynamics
are determined, we solve the inverse problem to compute the optimal forms of the constitutive relations 
and parameters for the excitation dynamics, namely, $\omega_1$, $\omega_2$ and $\alpha$, using the parameters 
describing the relaxation dynamics obtained earlier, according to Section \ref{sec:excitation}. 
Before doing so, one needs to validate the methodology proposed in Sections \ref{sec:relaxation} 
and \ref{sec:excitation} for the gradients computed using the
adjoint analysis. One can design a computational test that verifies the 
validity of all the steps involved, and hence the validity of the gradients 
computed using the proposed methodology. Also, the computational framework presented in 
Algorithm \ref{alg:optimal} is validated using synthetic data that is manufactured, in order to reconstruct 
some known constitutive relations from manufactured data. The results of computational 
tests are presented in Appendix \ref{sec:validate}.

\subsection{Relaxation Dynamics}\label{sec:results_OCV}
\begin{figure}[!ht]
	\centering
	\begin{subfigure}[b]{0.45\textwidth}
		\centering
		\includegraphics[width=1\textwidth]{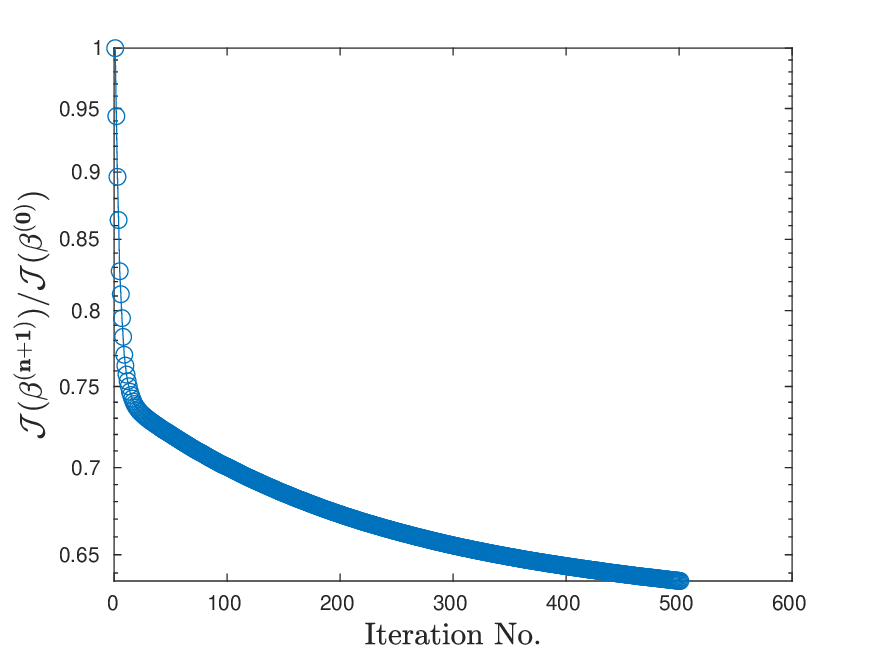}
	\end{subfigure}
	\caption{Cost functional history $\JJ_1({\bbeta})$ relative to its initial value as a function of iteration number.}
	\label{fig:J_history_OCV}
\end{figure}

\begin{figure}[!ht]
	\centering
	\begin{subfigure}[b]{0.31\textwidth}
		\centering
		\includegraphics[width=1\textwidth]{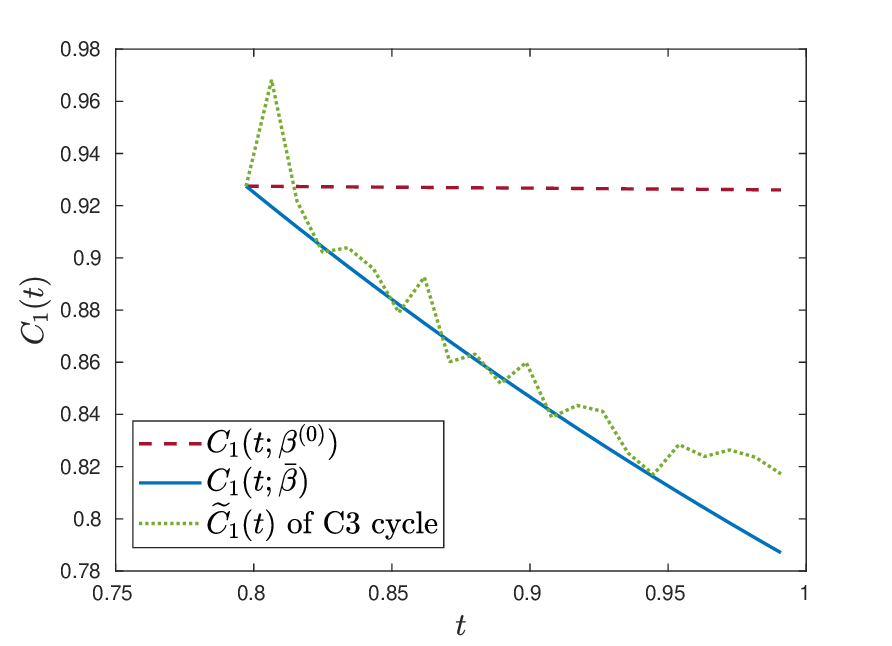}
		\subcaption{C3 Cycle}
	\end{subfigure}
	\begin{subfigure}[b]{0.31\textwidth}
		\centering
		\includegraphics[width=1\textwidth]{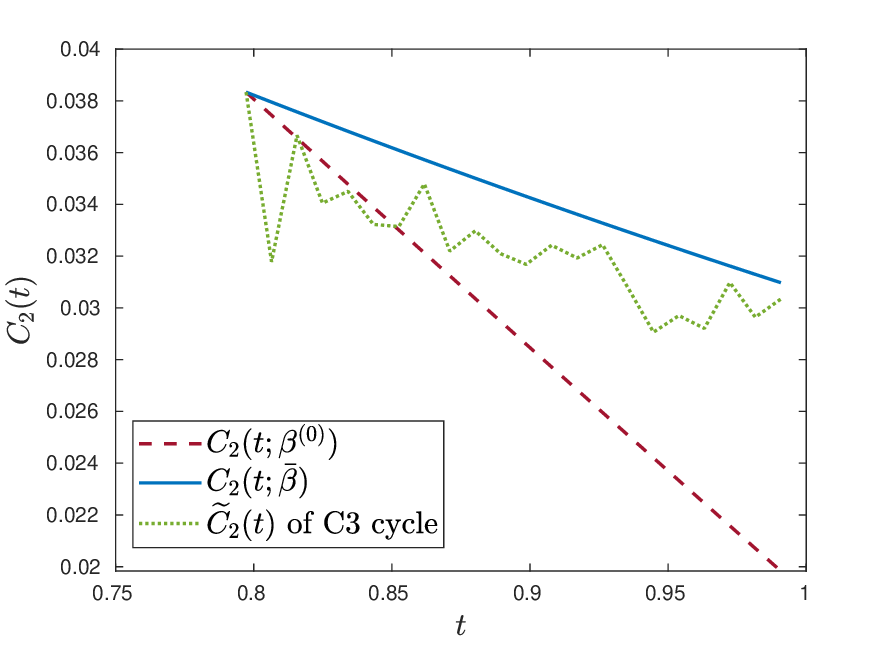}
		\subcaption{C3 Cycle}
	\end{subfigure}
	\begin{subfigure}[b]{0.31\textwidth}
		\centering
		\includegraphics[width=1\textwidth]{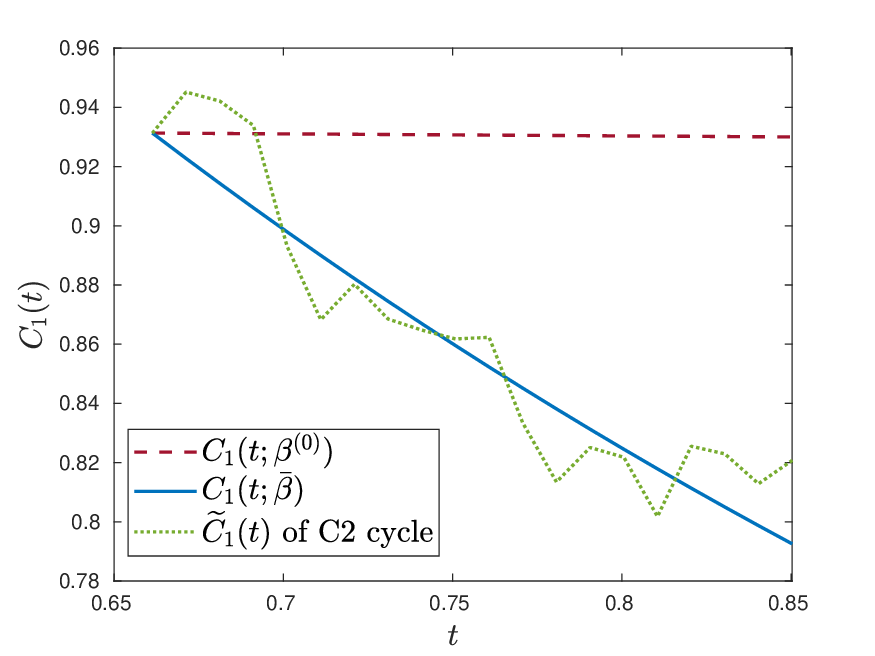}
		\subcaption{C2 Cycle}
	\end{subfigure}
	\begin{subfigure}[b]{0.31\textwidth}
		\centering
		\includegraphics[width=1\textwidth]{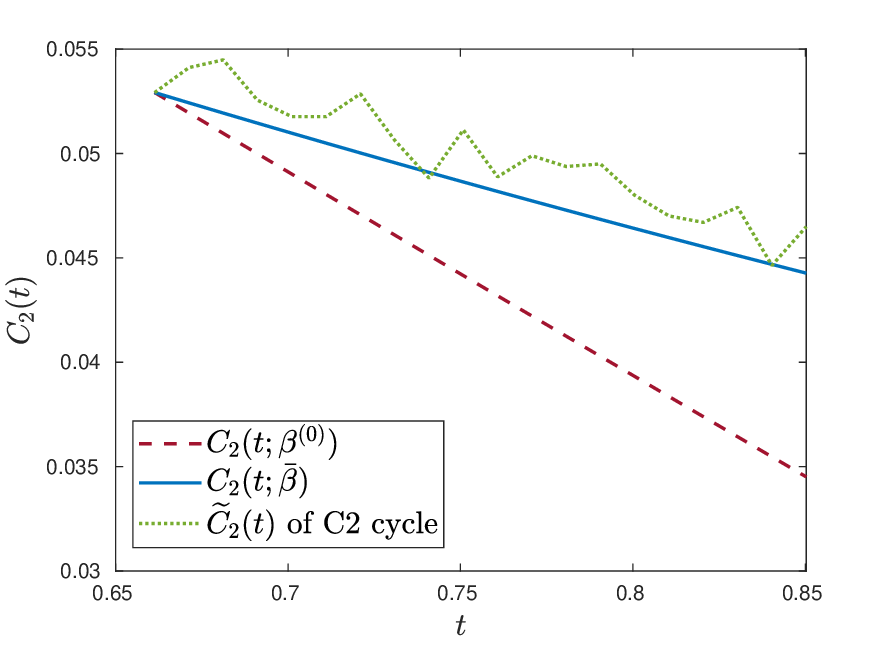}
		\subcaption{C2 Cycle}
	\end{subfigure}
	\begin{subfigure}[b]{0.31\textwidth}
		\centering
		\includegraphics[width=1\textwidth]{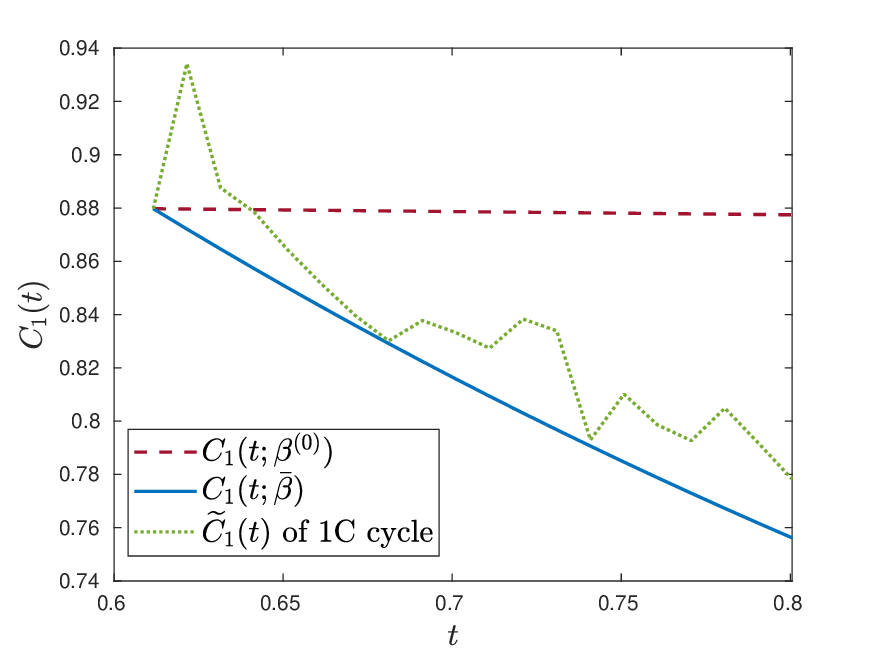}
		\subcaption{1C Cycle}
	\end{subfigure}
	\begin{subfigure}[b]{0.31\textwidth}
		\centering
		\includegraphics[width=1\textwidth]{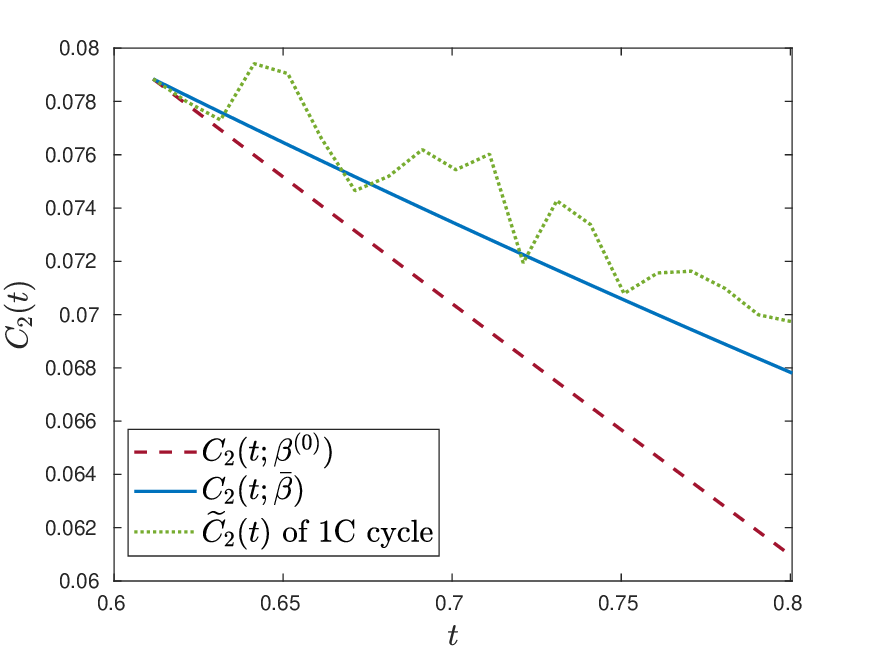}
		\subcaption{1C Cycle}
	\end{subfigure}	
	\begin{subfigure}[b]{0.31\textwidth}
		\centering
		\includegraphics[width=1\textwidth]{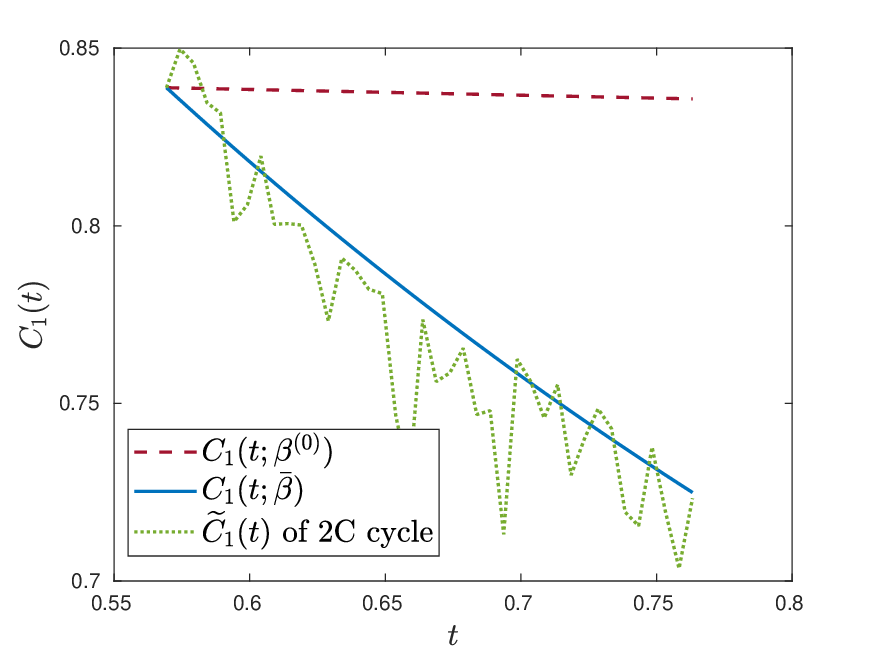}
		\subcaption{2C Cycle}
	\end{subfigure}
	\begin{subfigure}[b]{0.31\textwidth}
		\centering
		\includegraphics[width=1\textwidth]{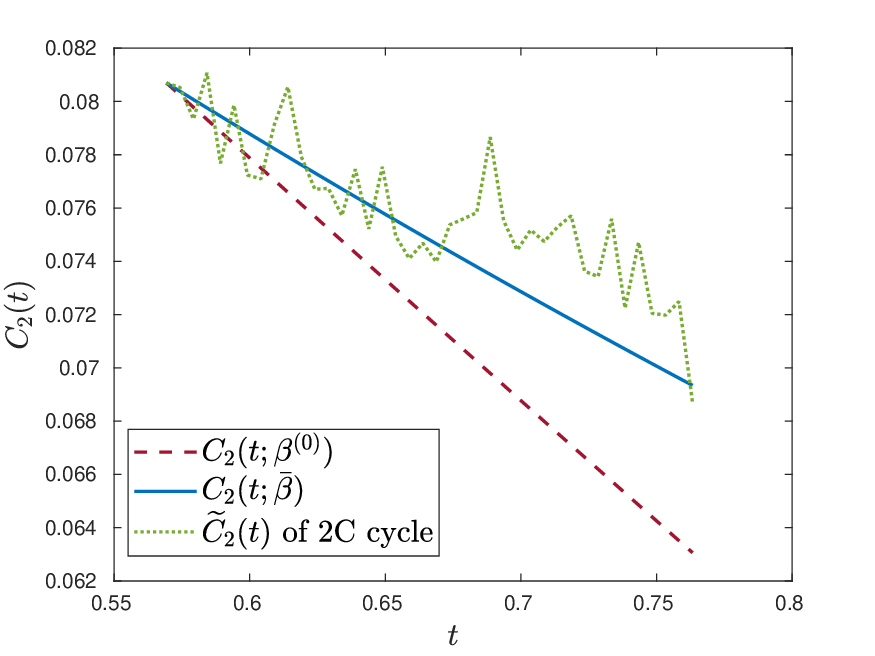}
		\subcaption{2C Cycle}
	\end{subfigure}
	\begin{subfigure}[b]{0.31\textwidth}
		\centering
		\includegraphics[width=1\textwidth]{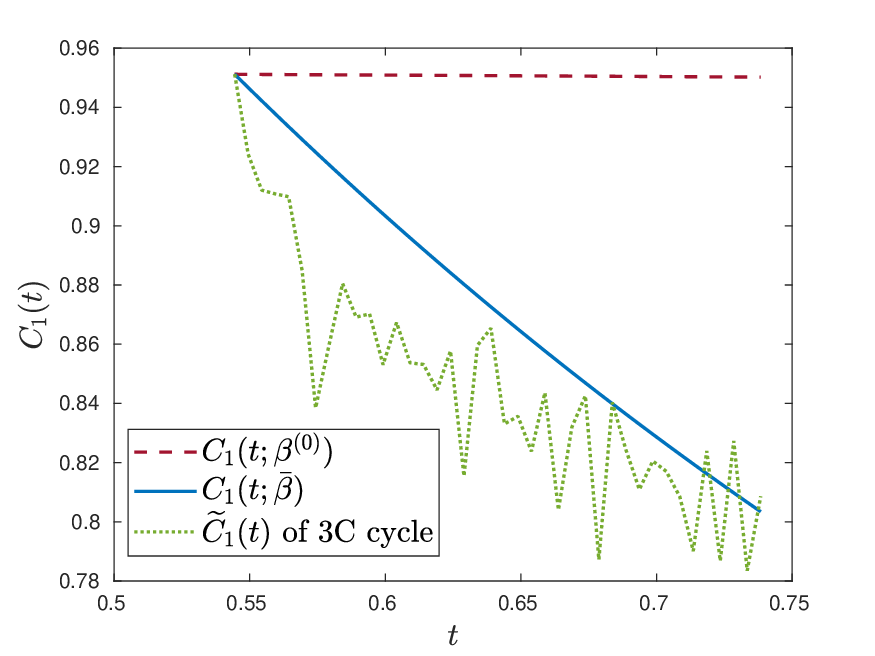}
		\subcaption{3C Cycle}
	\end{subfigure}
	\begin{subfigure}[b]{0.31\textwidth}
		\centering
		\includegraphics[width=1\textwidth]{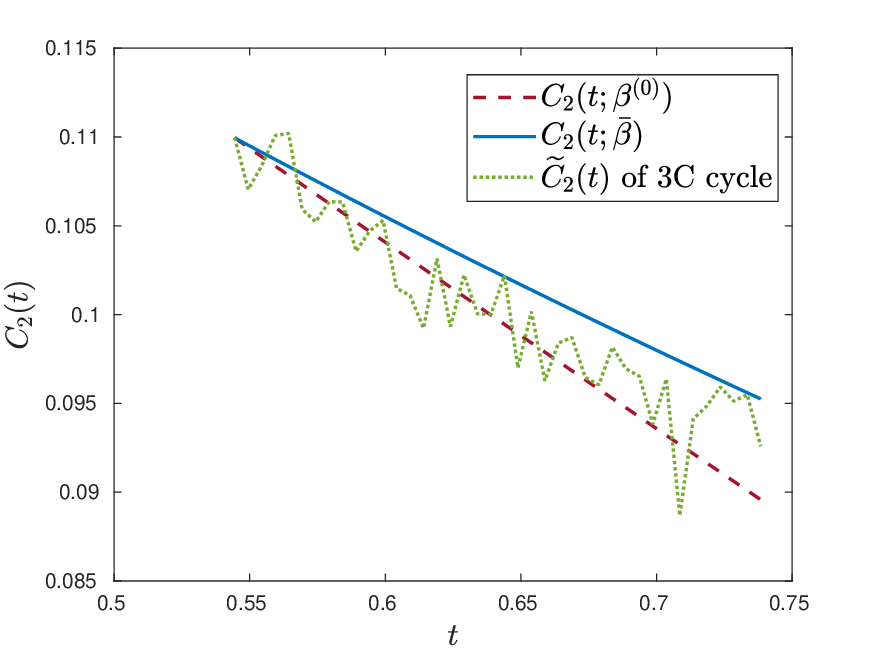}
		\subcaption{3C Cycle}
	\end{subfigure}
	\caption{Dependence of concentrations $C_1(t)$ and $C_2(t)$ on time for different cycles 
		of the cell, using the initial guess for parameters $\bbeta^{(0)}$ (dashed red line), 
		and the optimal values of parameters $\overline{\bbeta}$ (solid blue line). 
		The experimental concentrations $\widetilde{C}_1(t)$ and $\widetilde{C}_2(t)$ for each 
		cycle are shown as dotted green lines.}
	\label{fig:trajectory_OCV}
\end{figure}

The computational framework outlined in Stage I of Algorithm \ref{alg:optimal} is used to 
find optimal parameter values of the relaxation dynamics of the cell. 
The parameters to initialize the optimization algorithm are chosen as 
$\bbeta^{(0)} = \left[-0.1,-0.1,-0.1,-0.1\right]$, $N=500$, and $TOL = 10^{-6}$.
The interval $\L$ for the optimization framework is 
$(C_1, C_2) \in\left[-0.5, 1.5\right] \times \left[-0.2, 0.5\right]$. This choice has 
been made based on the magnitude of the state variables in different cycles.
Also, the optimization framework with aggregated data, cf.~Section \ref{sec:robust}, 
has been used here. In other words, $\mathcal{D}_t = \bigoplus_i \mathcal{D}_i, i\in \mathcal{C}$ 
has been used as the training data. The relative decay of the cost functional for the iterative 
scheme, cf.~\eqref{eq:cost_OCV}, is presented in Figure \ref{fig:J_history_OCV}. As can be observed, 
the cost functional value is decaying significantly relative to its initial value. The rate of decay 
is becoming slow at later iterations.
The optimal solution found as the result of the iterative scheme is 
$\bar{\bbeta} \approx \left[0.85,-1.85, 0.55, -0.02\right]$. According to 
\eqref{eq:ODE}, the matrices in system \eqref{eq:ode_relaxation} become 
\begin{equation}
\begin{alignedat}{1}
\bB &= \begin{bmatrix}
0.85\\
0
\end{bmatrix},\\
\bA &= \begin{bmatrix}
-1.85&0.003\\
-0.02 & -0.55
\end{bmatrix}.
\end{alignedat}
\end{equation}
Note that matrix $\bA$ has two
real negative eigenvalues as $\sigma(\bA) = -1.85,-0.55$. This implies that 
the linear part of the system corresponding to the relaxation dynamics of the cell has the form 
of a delay towards an an equilibrium point, which is consistent with the behaviour of the cell at 
relaxation mode, cf.~Figure \ref{fig:experimental}. The results of predicting
the evolution of concentrations for different cycles of the cell using the optimal parameter
values are shown in Figure \ref{fig:trajectory_OCV}. As can be observed, the model performs relatively
well on a wide range of C-rates. The optimal parameter values for the relaxation dynamics will be 
used when solving the optimization problem for excitation dynamics.

\subsection{Excitation Dynamics}\label{sec:results_fullmodel}
\begin{figure}[!ht]
	\centering
	\begin{subfigure}[b]{0.45\textwidth}
		\centering
		\includegraphics[width=1\textwidth]{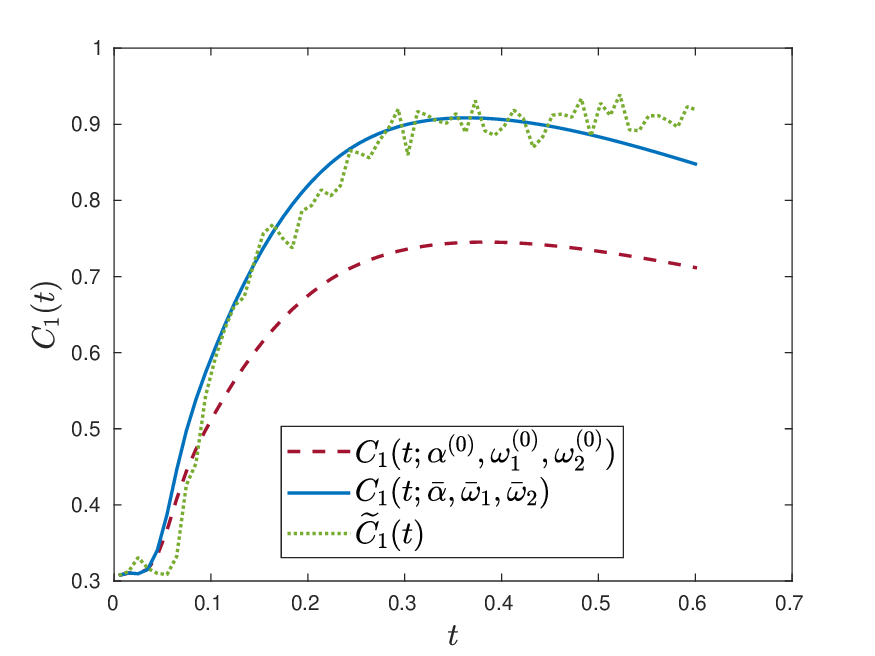}
		\subcaption{Charge regime - 1C cycle}
	\end{subfigure}
	\begin{subfigure}[b]{0.45\textwidth}
		\centering
		\includegraphics[width=1\textwidth]{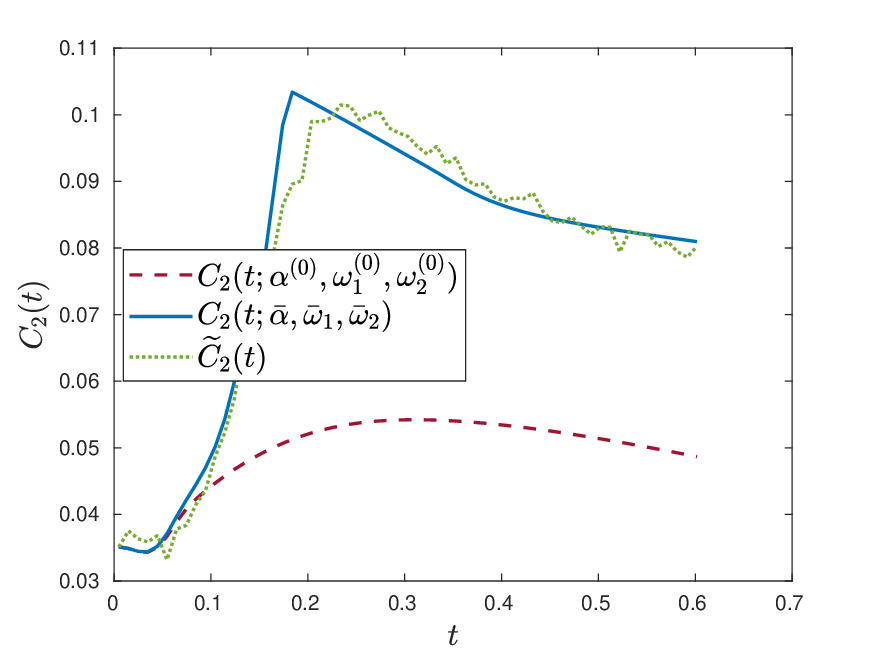}
		\subcaption{Charge regime - 1C cycle}
	\end{subfigure}
	\begin{subfigure}[b]{0.45\textwidth}
		\centering
		\includegraphics[width=1\textwidth]{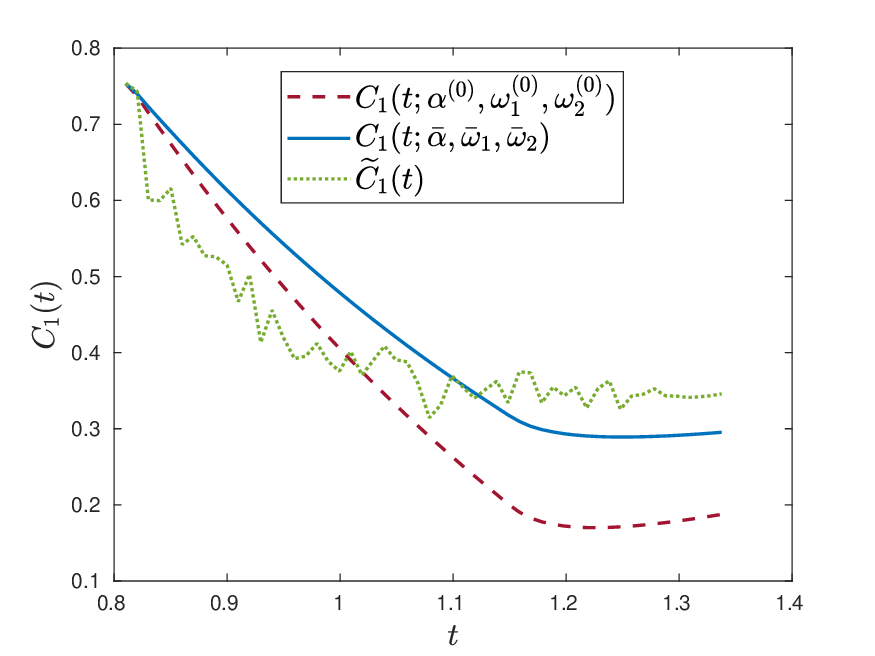}
		\subcaption{Discharge regime - 1C cycle}
	\end{subfigure}
	\begin{subfigure}[b]{0.45\textwidth}
		\centering
		\includegraphics[width=1\textwidth]{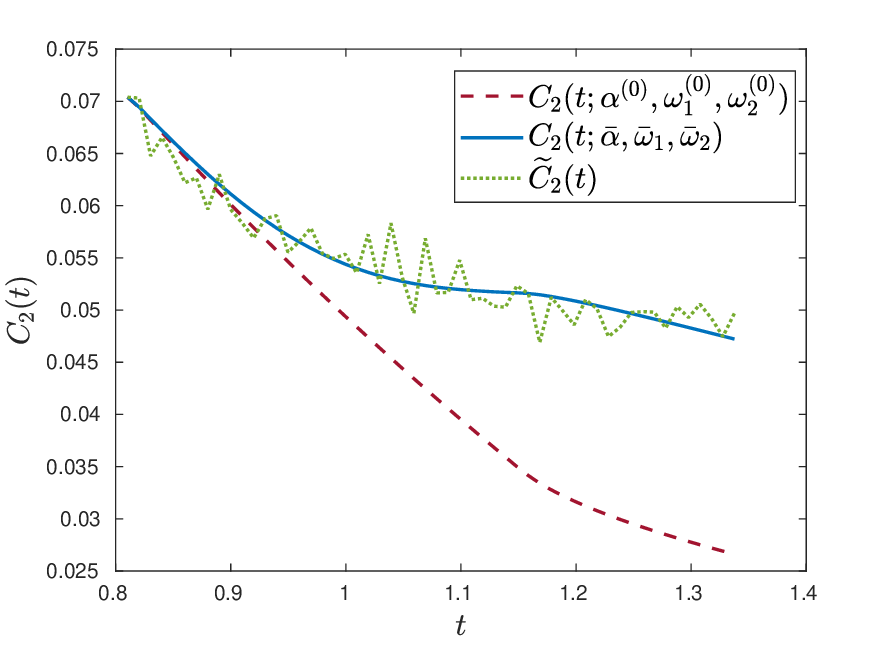}
		\subcaption{Discharge regime - 1C cycle}
	\end{subfigure}
	\caption{The dependence of concentrations $C_1(t)$ and $C_2(t)$ on time for the charge regime of the 1C 
		cycle (a,b), and the discharge regime of the 1C cycle (c,d), using the initial guess
		for the parameter and constitutive relations $(\alpha^{(0)},\omega_1^{(0)},\omega_2^{(0)})$ (dashed red line), 
		and the optimal parameter and constitutive relations 
		$(\overline{\alpha},\overline{\omega}_1,\overline{\omega}_2)$ (solid blue line) obtained by solving the 
		inverse problem \eqref{eq:inverse} using the data for the charge and discharge regimes 
		of 1C cycle, namely, $\mathcal{D}_{1C}^{ch}$ and $\mathcal{D}_{1C}^{dch}$, respectively.
		The experimental concentrations $\widetilde{C}_1$ and $\widetilde{C}_2(t)$ are shown using green dotted line.}
	\label{fig:1C_charge}
\end{figure}

In this section, we present the results of the inverse modeling approach presented in 
Algorithm \ref{alg:optimal} using the machinery developed in Section \ref{sec:excitation}.
Note that in this section, the parameters of the relaxation dynamics are assumed known, and are given 
by the results of Section \ref{sec:results_OCV}. First, we begin by fitting the unknown constitutive relations 
and parameter in \eqref{eq:ODE} describing the excitation dynamics to the data corresponding to individual 
cycles, namely, $\mathcal{D}_i^j, i\in \mathcal{C}, j\in\{ch,dch\}$. 
Also, two different regimes are used for solving the inverse problem \eqref{eq:inverse},
namely, charge and discharge regimes. Thus, a separate inverse problem is solved pertaining to each regime 
and the results are compared. In order to initialize the stage II of the Algorithm 
\ref{alg:optimal}, the initial guesses for constitutive relations are set to be 
$\omega_1^{(0)}(C_1) = 0.25$, $\omega_2^{(0)}(C_2) = 0.25$, $\alpha^{(0)} = 5$. The 
choice of this initial guess for constitutive relations is dictated by our knowledge of the 
physics of the cell, in which the constitutive relation $\omega(C_1,C_2)$ (defining the 
competition between intercalation vs. plating) is dominated by the intercalation process, 
hence attains a value between zero to one, closer to zero.
As mentioned in Section \ref{sec:excitation}, the function $\omega_1(C_1)$ 
is reconstructed in space $H^1$, however, the function $\omega_2(C_2)$ is reconstructed in 
space $H^1_0$, where the mean of the function remains stationary. Algorithm \ref{alg:optimal} 
is allowed to run for a maximum of $N=30$ iterations. The smoothing parameter in the $H^1$ inner product is 
set $l = 1$. The interval $\L$ is set as $(C_1,C_2) \in \left[-0.5,1.5\right] \times\left[-0.2,0.5\right]$. 
The wide choice of interval $\L$ for each state variable ensures that the choice of the somewhat arbitrary 
boundary conditions satisfied by the Sobolev gradient, cf.~\eqref{eq:Sobolev}, has little effect on the 
behavior of the gradient for concentrations of interest. 
In other words, if the interval $\L$ is chosen to be too close to the identifiability region bounds, 
the behaviour of the function at the end points of the identifiability region will be affected by 
the choice of boundary conditions in $H^1$ reconstruction. The results obtained by solving 
optimization problem \eqref{eq:inverse} for the charge and discharge regimes of the 1C cycle are 
presented in Figure \ref{fig:1C_charge}.
As can be observed, the large-scale details of the measurement data $\widetilde{C}_1(t)$ and $\widetilde{C}_2(t)$ are well 
captured by the model equipped with the optimally reconstructed constitutive relations and parameter. 
Note that the fine details of the measurements result from 
the noise in the NMR measurements, and hence, it is preferable for the model not to resolve such details. 
The optimal constitutive relations and parameter as the result of fitting \eqref{eq:ODE} to individual cycles
are not presented here for brevity, as such results are similar with minor differences.

\begin{figure}[t]
	\centering
	\begin{subfigure}[b]{0.45\textwidth}
		\centering
		\includegraphics[width=1\textwidth]{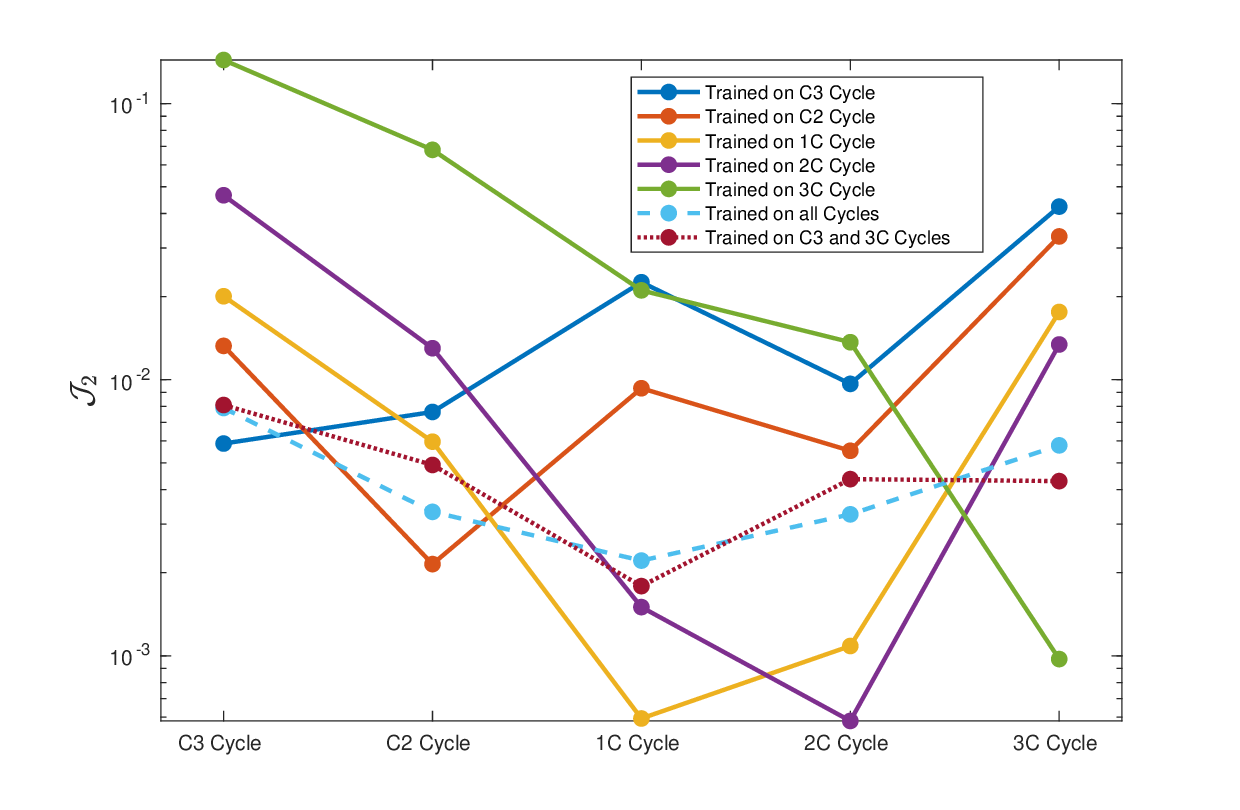}
		\subcaption{}
	\end{subfigure}
	\begin{subfigure}[b]{0.45\textwidth}
		\centering
		\includegraphics[width=1\textwidth]{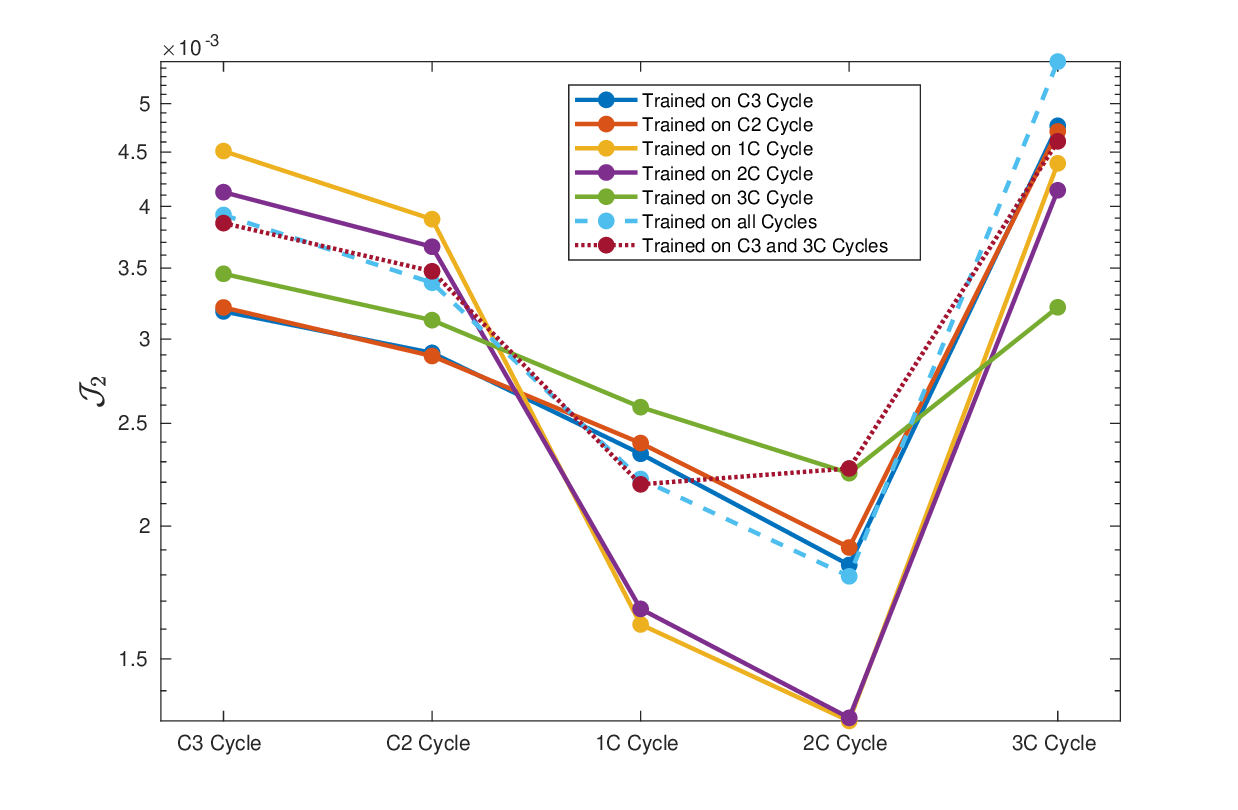}
		\subcaption{}
	\end{subfigure}
	\caption{Dependence of the least-squares error
		$\JJ_2(\overline{\alpha},\overline{\omega}_1,\overline{\omega}_2;\mathcal{D}_i^{ch})$
		between the experimental data from different cycles
		$i\in\mathcal{C}$, and the predictions of model
		\eqref{eq:ODE} using the optimally reconstructed parameter
		and constitutive relations
		$(\overline{\alpha},\overline{\omega}_1,\overline{\omega}_2)$
		obtained by solving inverse problem \eqref{eq:inverse} using
		the data corresponding to a given cycle, for the charge
		regime (a) and the discharge regime (b).  For each line in
		the plots, model \eqref{eq:ODE} is calibrated with Algorithm
		\ref{alg:optimal} using the data from the cycle indicated in
		the legend and then tested against data from all other
		cycles (indicated on the horizontal axis).  Also, the
		performance of the model calibrated using aggregated data
		$\mathcal{D}_t^{ch}$ for (a) and $\mathcal{D}_t^{dch}$ for
		(b), is demonstrated by dashed line. Additionally,
		performance of model \eqref{eq:ODE} calibrated based on the
		data from the C3 and 3C cycles only, i.e.,
		$\bigoplus_i\mathcal{D}_i^{ch}, i\in \{\text{C3},
		\text{3C}\}$ for (a) and
		$\bigoplus_i\mathcal{D}_i^{dch}, i\in \{\text{C3},
		\text{3C}\}$ for (b), is represented by dotted lines.}
	\label{fig:all_cycles}
\end{figure}

\begin{figure}[!ht]
	\centering
	\begin{subfigure}[b]{0.45\textwidth}
		\centering
		\includegraphics[width=1\textwidth]{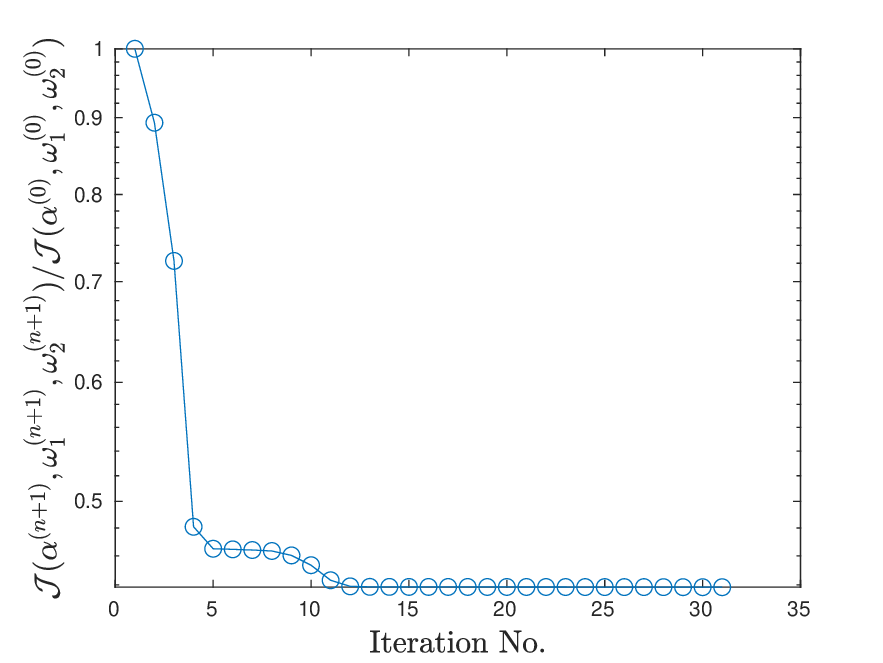}
		\subcaption{}
	\end{subfigure}
	\begin{subfigure}[b]{0.45\textwidth}
		\centering
		\includegraphics[width=1\textwidth]{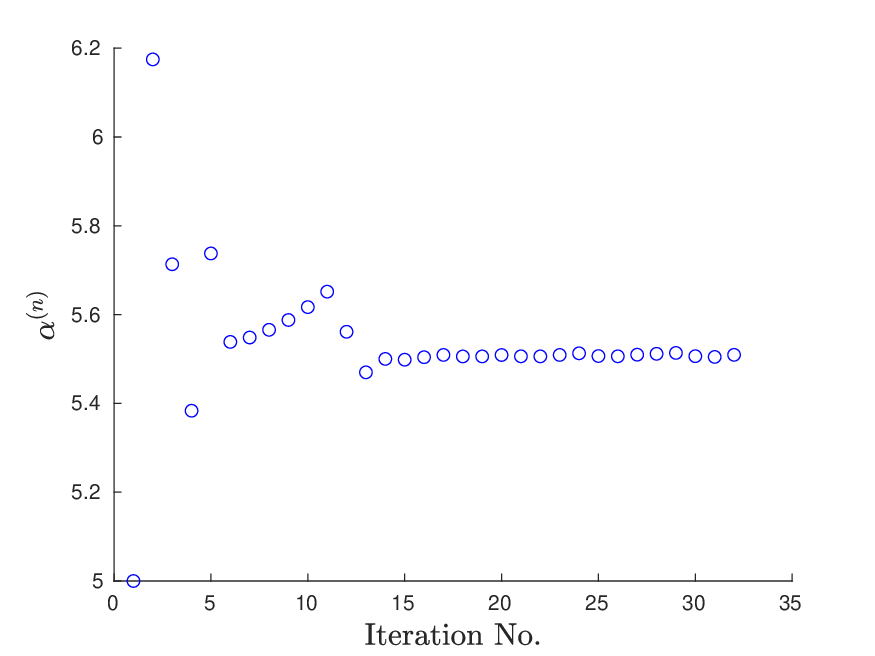}
		\subcaption{}
	\end{subfigure}
	\begin{subfigure}[b]{0.45\textwidth}
		\centering
		\includegraphics[width=1\textwidth]{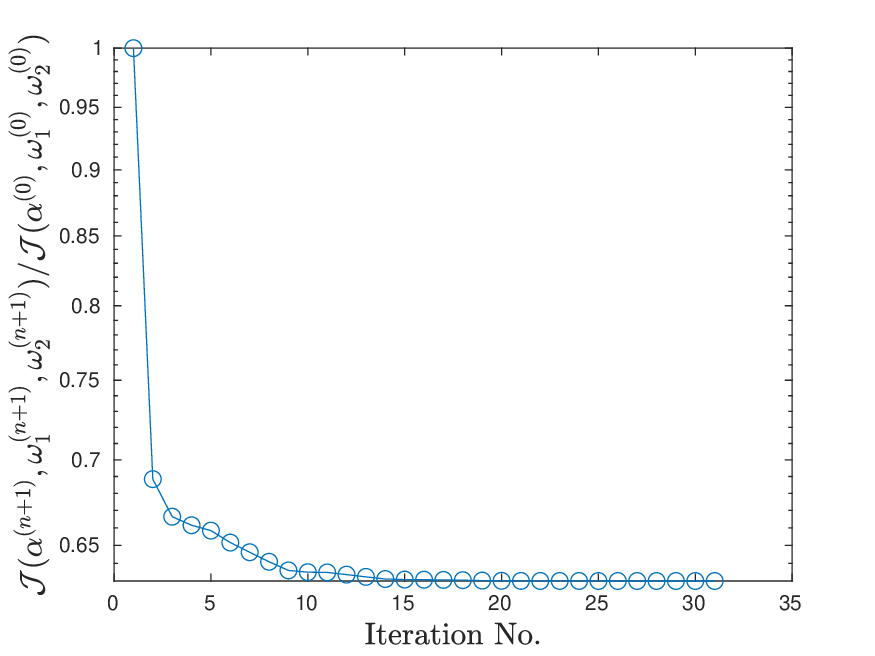}
		\subcaption{}
	\end{subfigure}
	\begin{subfigure}[b]{0.45\textwidth}
		\centering
		\includegraphics[width=1\textwidth]{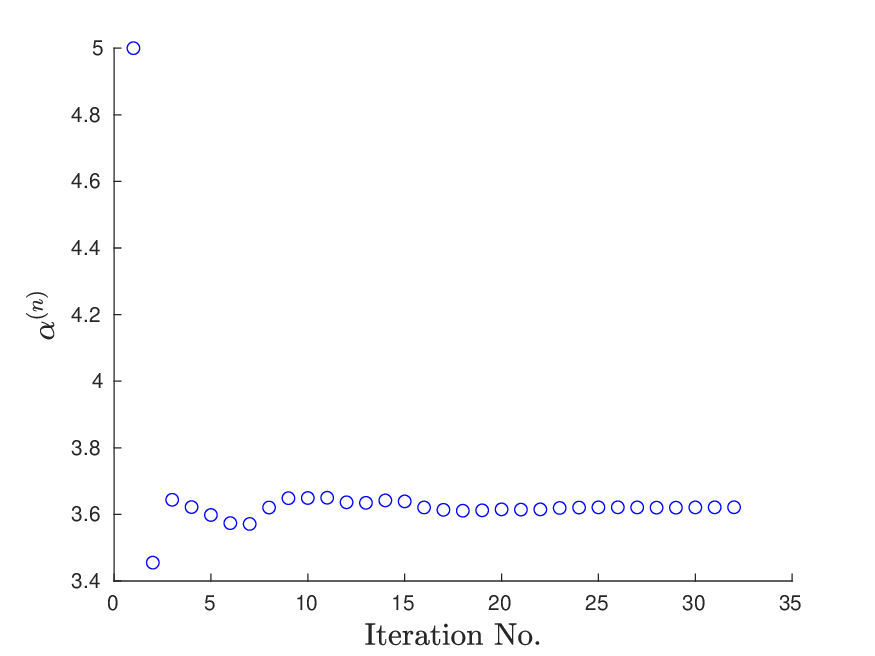}
		\subcaption{}
	\end{subfigure}
	\caption{The decay of cost functional $\JJ_2(\alpha,\omega_1,\omega_2)$ normalized with respect to its 
		initial value with iterations (a,c),
		and the evolution of parameter $\alpha$ with iterations (b,d) in the solution of the 
		inverse problem \eqref{eq:inverse} 
		where model \eqref{eq:ODE} is calibrated with stage II of Algorithm \ref{alg:optimal} 
		using aggregated data for the charge regime $\mathcal{D}_t^{ch}$ 
		(a,b), and discharge regime $\mathcal{D}_t^{dch}$ (c,d).}
	\label{fig:all_history}
\end{figure}

\begin{figure}[!ht]
	\centering
	\begin{subfigure}[b]{0.45\textwidth}
		\centering
		\includegraphics[width=1\textwidth]{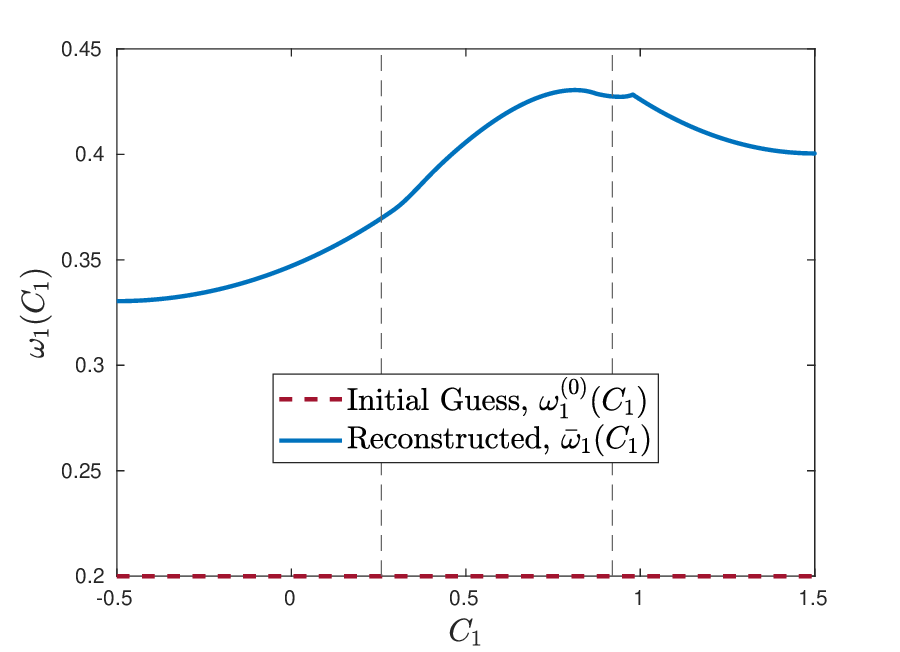}
		\subcaption{}
	\end{subfigure}
	\begin{subfigure}[b]{0.45\textwidth}
		\centering
		\includegraphics[width=1\textwidth]{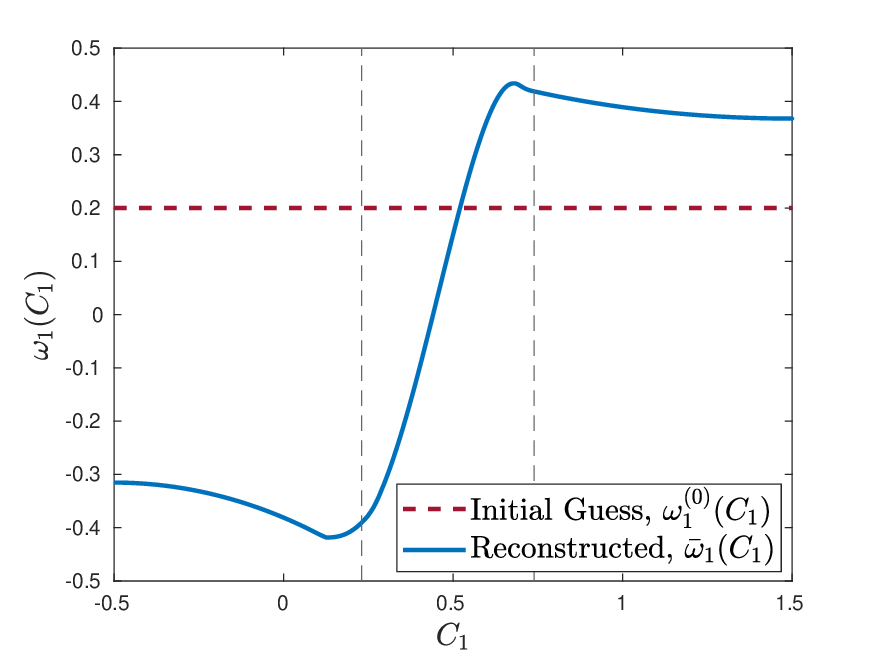}
		\subcaption{}
	\end{subfigure}
	\begin{subfigure}[b]{0.45\textwidth}
		\centering
		\includegraphics[width=1\textwidth]{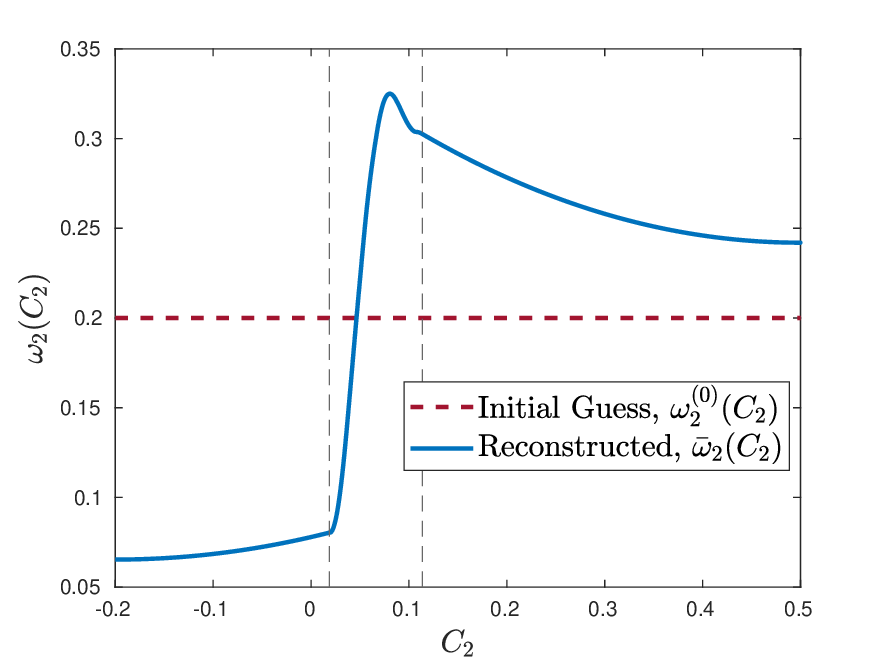}
		\subcaption{}
	\end{subfigure}
	\begin{subfigure}[b]{0.45\textwidth}
		\centering
		\includegraphics[width=1\textwidth]{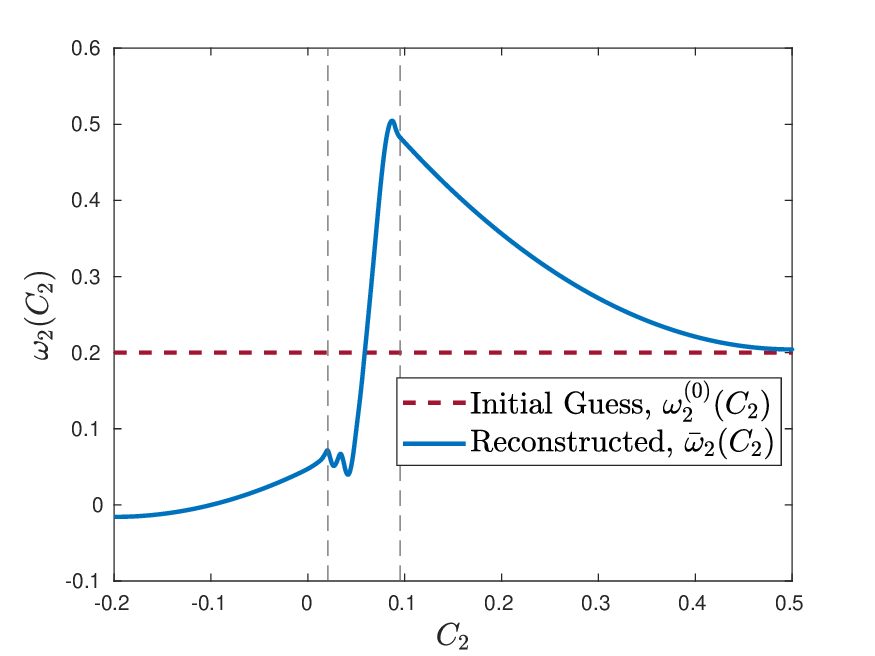}
		\subcaption{}
	\end{subfigure}
	\begin{subfigure}[b]{0.45\textwidth}
		\centering
		\includegraphics[width=1\textwidth]{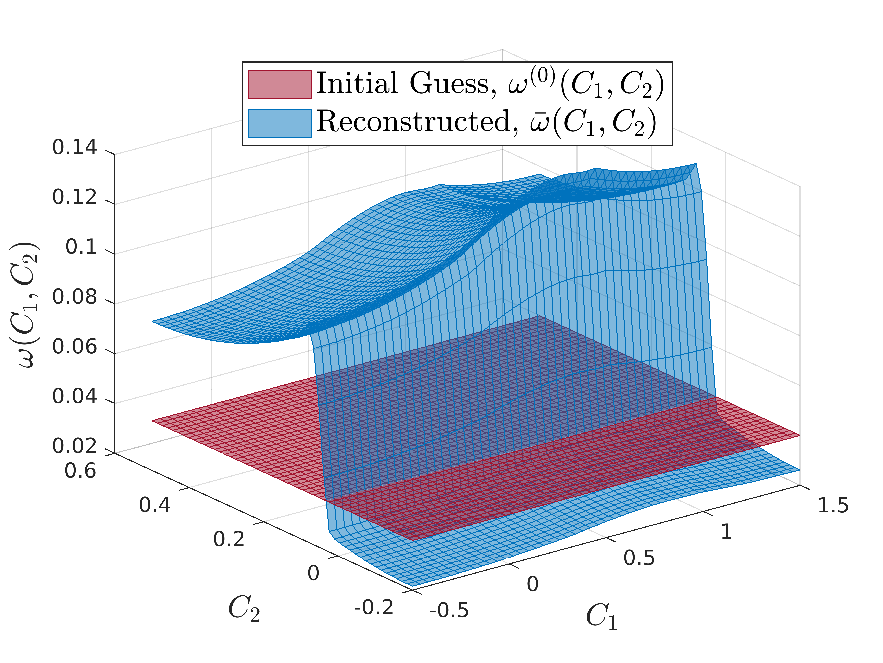}
		\subcaption{}
	\end{subfigure}
	\begin{subfigure}[b]{0.45\textwidth}
		\centering
		\includegraphics[width=1\textwidth]{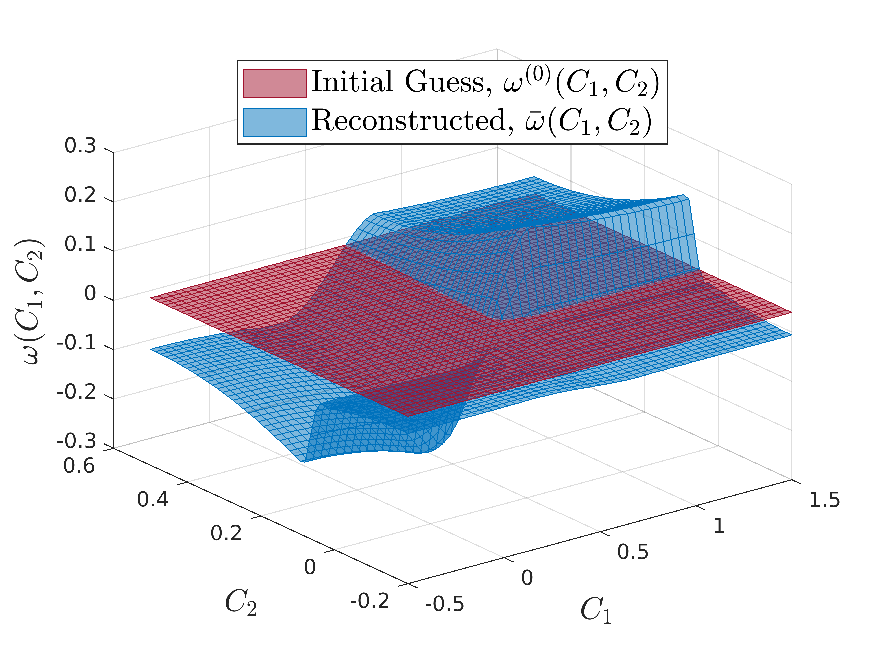}
		\subcaption{}
	\end{subfigure}
	\caption{The initial guess for the constitutive relations $\omega_1^{(0)}(C_1)$, $\omega_2^{(0)}(C_2)$, 
		and $\omega^{(0)}(C_1,C_2)$ (red), and the optimal form of the reconstructed constitutive relations 
		$\overline{\omega}_1(C_1)$, $\overline{\omega}_2(C_2)$, and $\overline{\omega}(C_1,C_2)$ (blue) 
		where model \eqref{eq:ODE} is calibrated with stage II of Algorithm \ref{alg:optimal} using 
		aggregated data for the charge regime $\mathcal{D}_t^{ch}$ (a,c,e), 
		and discharge regime $\mathcal{D}_t^{dch}$ (b,d,f).}
	\label{fig:all_omega}
\end{figure}

\begin{figure}[!ht]
	\centering
	\begin{subfigure}[b]{0.32\textwidth}
		\centering
		\includegraphics[width=1\textwidth]{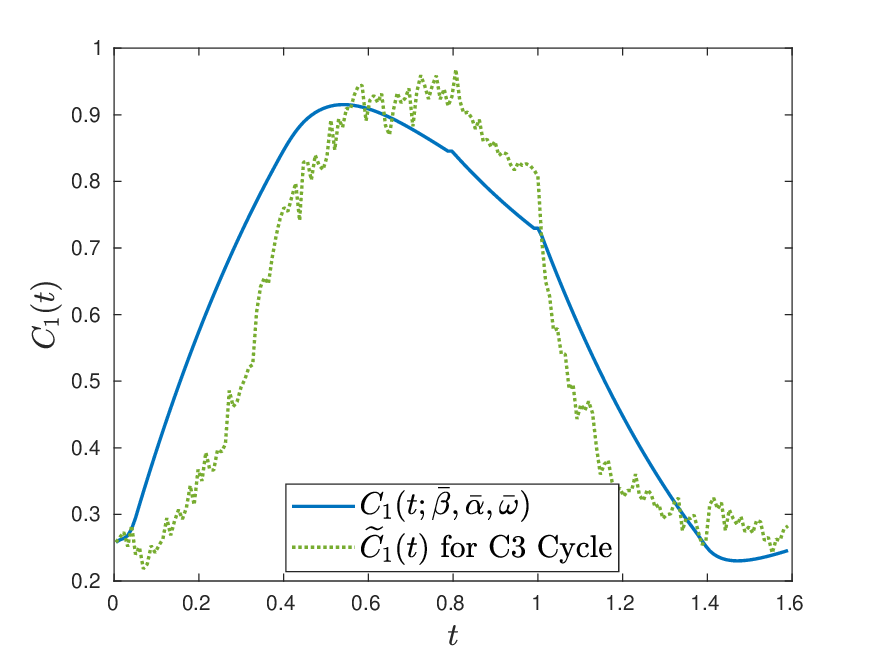}
		\subcaption{C3 cycle}
	\end{subfigure}
	\hfill
	\begin{subfigure}[b]{0.32\textwidth}
		\centering
		\includegraphics[width=1\textwidth]{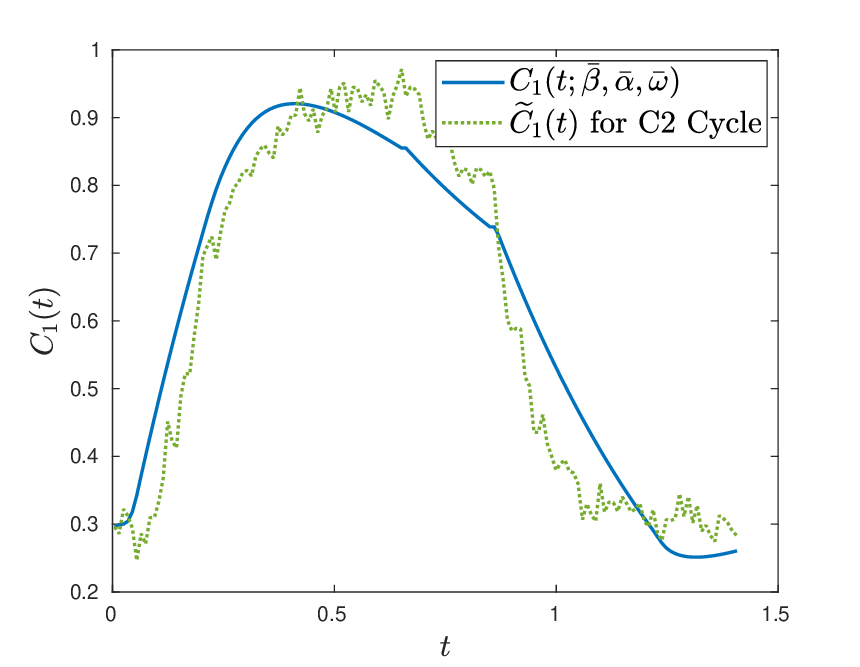}
		\subcaption{C2 cycle}
	\end{subfigure}
	\hfill
	\begin{subfigure}[b]{0.32\textwidth}
		\centering
		\includegraphics[width=1\textwidth]{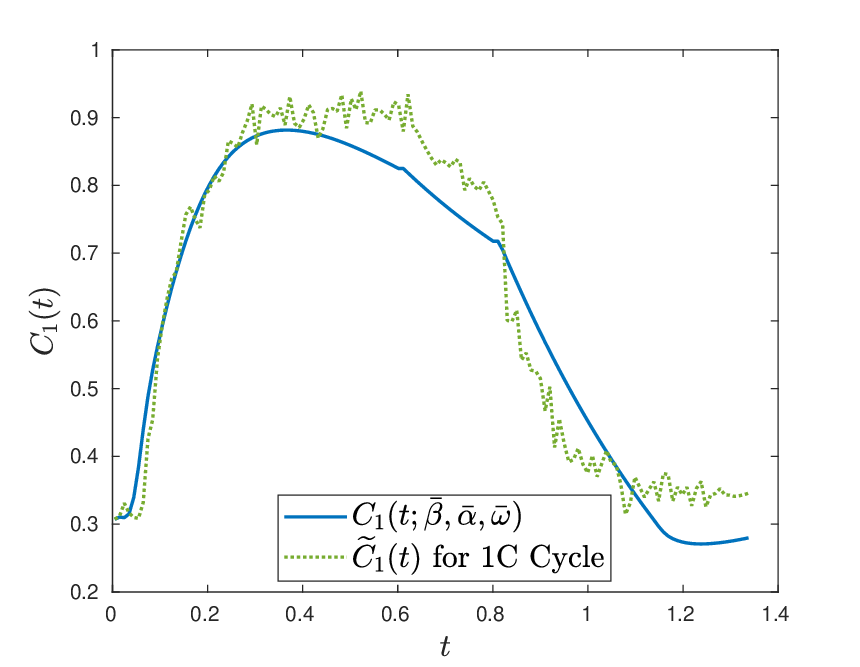}
		\subcaption{1C cycle}
	\end{subfigure}
	\hfill
	\begin{subfigure}[b]{0.32\textwidth}
		\centering
		\includegraphics[width=1\textwidth]{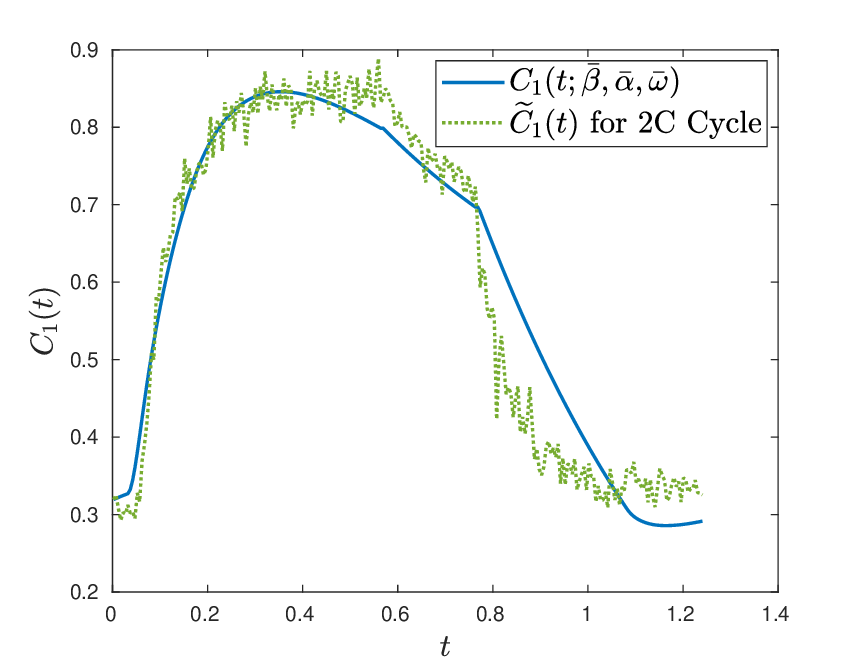}
		\subcaption{2C cycle}
	\end{subfigure}
	\begin{subfigure}[b]{0.32\textwidth}
		\centering
		\includegraphics[width=1\textwidth]{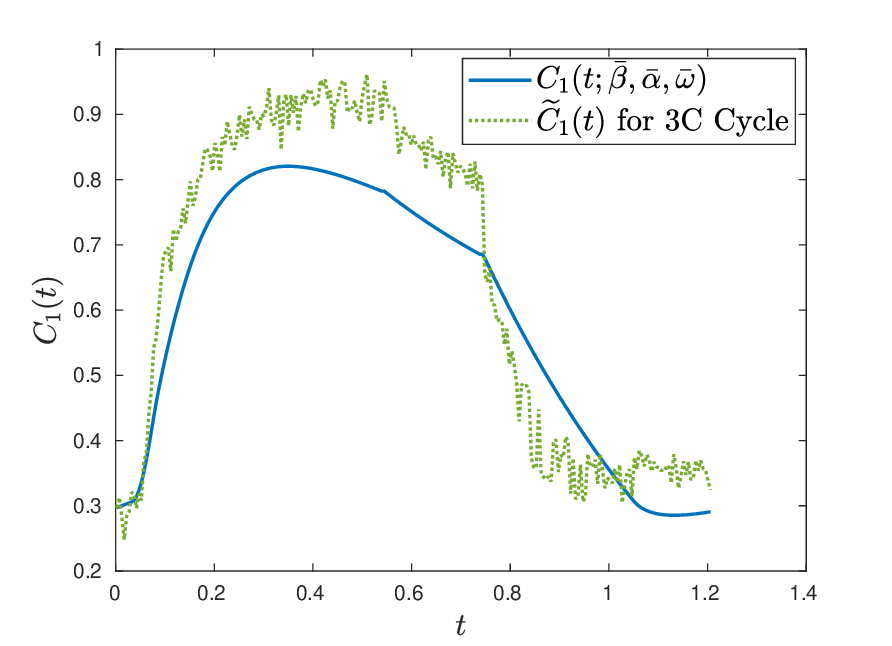}
		\subcaption{3C cycle}
	\end{subfigure}
	\caption{The dependence of the state variable $C_1(t)$ on time in the solution of the forward problem \eqref{eq:ODE}
		using the optimal parameters values and optimal forms of the constitutive 
		relation ($\overline{\bbeta},\overline{\alpha},\overline{\omega}$)
		reconstructed by calibrating system \eqref{eq:ODE} using aggregated data $\mathcal{D}_t$ for charge, 
		discharge and OCV regimes, cf.~Section \ref{sec:robust}. The dashed green and the solid blue lines 
		represent the experimental concentrations and the solution of the forward problem \eqref{eq:ODE} 
		using optimal parameters and constitutive relations, respectively.}
	\label{fig:all_c1_trajectory}
\end{figure}

\begin{figure}[!ht]
	\centering
	\begin{subfigure}[b]{0.32\textwidth}
		\centering
		\includegraphics[width=1\textwidth]{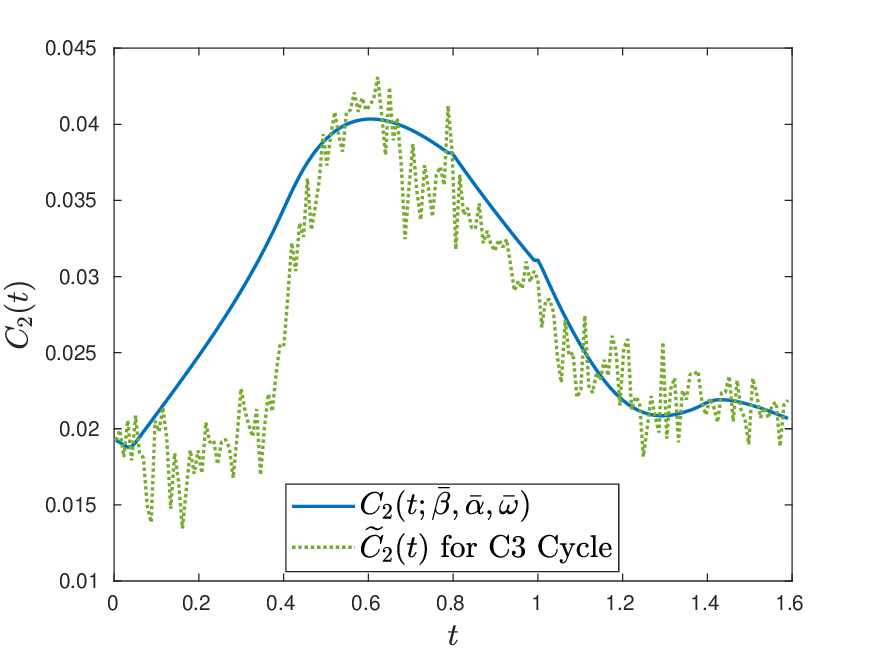}
		\subcaption{C3 cycle}
	\end{subfigure}
	\hfill
	\begin{subfigure}[b]{0.32\textwidth}
		\centering
		\includegraphics[width=1\textwidth]{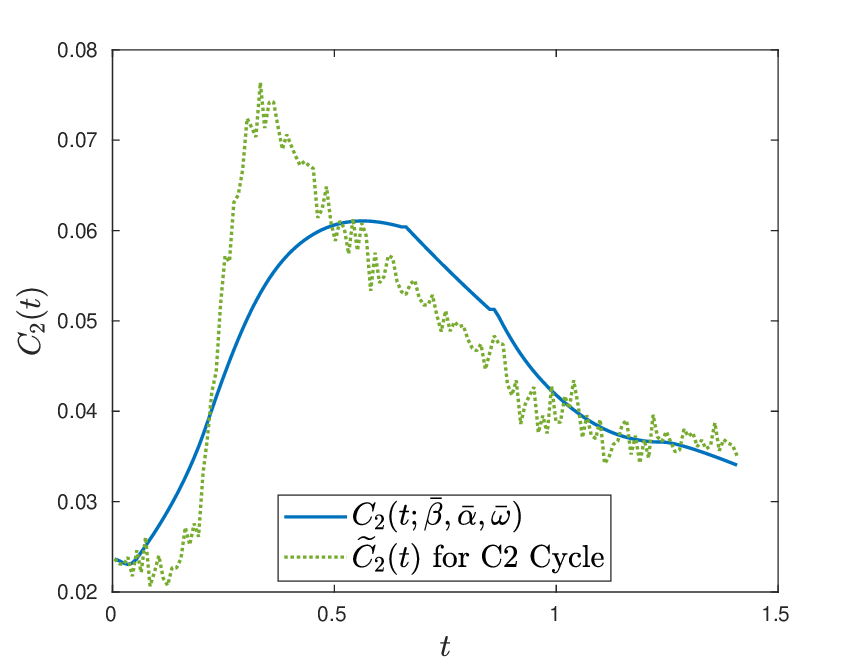}
		\subcaption{C2 cycle}
	\end{subfigure}
	\hfill
	\begin{subfigure}[b]{0.32\textwidth}
		\centering
		\includegraphics[width=1\textwidth]{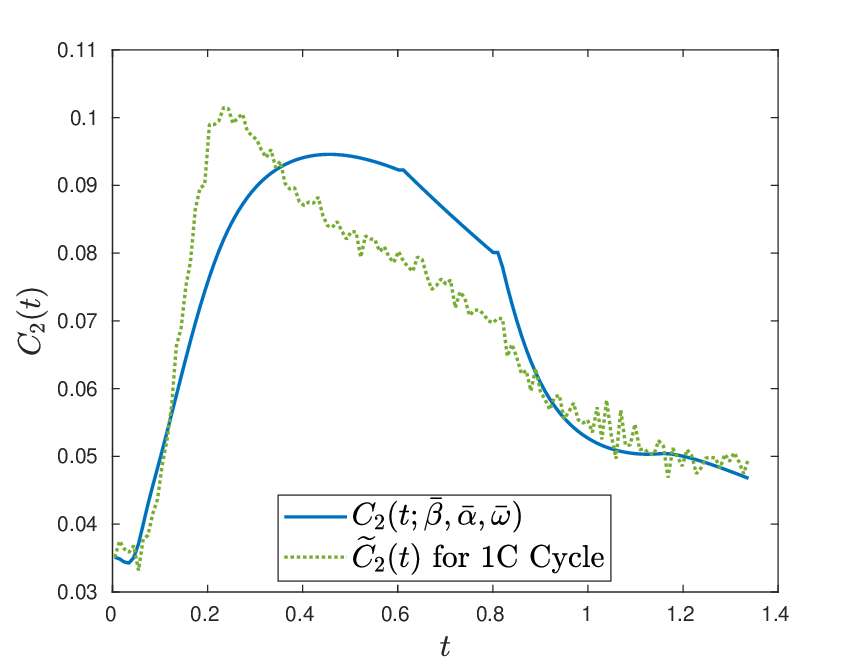}
		\subcaption{1C cycle}
	\end{subfigure}
	\hfill
	\begin{subfigure}[b]{0.32\textwidth}
		\centering
		\includegraphics[width=1\textwidth]{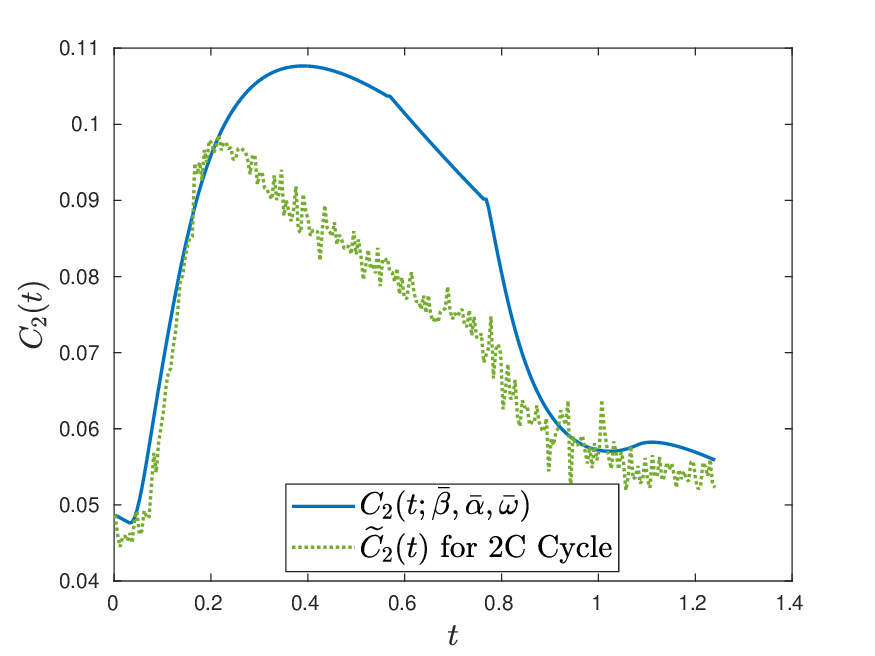}
		\subcaption{2C cycle}
	\end{subfigure}
	\begin{subfigure}[b]{0.32\textwidth}
		\centering
		\includegraphics[width=1\textwidth]{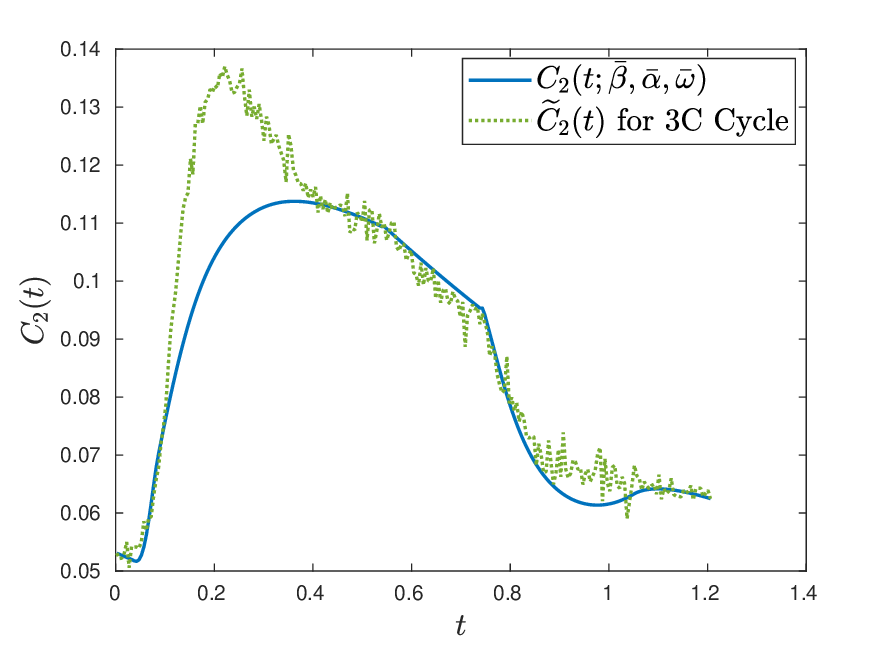}
		\subcaption{3C cycle}
	\end{subfigure}
	\caption{The dependence of the state variable $C_2(t)$ on time in the solution of the forward problem \eqref{eq:ODE}
		using the optimal parameters values and optimal forms of the constitutive 
		relation ($\overline{\bbeta},\overline{\alpha},\overline{\omega}$)
		reconstructed by calibrating system \eqref{eq:ODE} using aggregated data $\mathcal{D}_t$ for charge, 
		discharge and OCV regimes, cf.~Section \ref{sec:robust}. The dashed green and the solid blue lines 
		represent the experimental concentrations and the solution of the forward problem \eqref{eq:ODE} 
		using optimal parameters and constitutive relations, respectively.}
	\label{fig:all_c2_trajectory}
\end{figure}

In principle, one uses the inverse problem \eqref{eq:inverse} to train
\eqref{eq:ODE} on individual sequences of data corresponding to
particular C-rates, and for charge and discharge regimes. It is known
however that each of these models can only perform well in the
vicinity of the original C-rate that it has been trained on.  In order
to systematically assess the prediction capability of the calibrated
models, one is required to test the trained models on unseen data from
other cycles. Individual models are therefore trained on each of the
five cycles for charge and discharge regimes. These calibrated models
will then be used to assess the performance of the model on the data
from other cycles by generating performance metrics as cost functional
error
$\JJ_2(\overline{\alpha},\overline{\omega_1},\overline{\omega_2})$.
Additionally, in order to obtain a more robust model that can
generalize well to a range of charge and discharge rates, the model is
trained by using the optimization framework presented in Section
\ref{sec:robust}. The charging regime comprising all cycles
$\mathcal{D}_t^{ch}$ will be used for training a robust model for the
charging regime. A similar calibration procedure will be followed in
the discharge regime by fitting the model to $\mathcal{D}_t^{dch}$.
Also, in another attempt to find a robust model based on a minimal
amount of experimental data, system \eqref{eq:ODE} will be calibrated
based on measurements in the charging cycles C3 and 3C only, i.e.,
$\bigoplus_i \mathcal{D}_i^{ch}, i\in \{\text{C3}, \text{3C}\}$, using
the optimization framework introduced in Section \ref{sec:robust}. A
similar procedure will also be followed in order to calibrate system
\eqref{eq:ODE} based on measurements in the discharge regime, namely,
for $\bigoplus_i\mathcal{D}_i^{dch}, i\in \{\text{C3}, \text{3C}\}$.
The results of this analysis are presented in Figure
\ref{fig:all_cycles}.  Each solid line corresponds to model
\eqref{eq:ODE} trained on a particular individual cycle, with the
dashed line corresponding to the robust model trained on all cycles
and the dotted line corresponding to the robust model trained on the
C3 and 3C cycles only. As can be observed, each trained model performs
best in the vicinity of the training cycle (C-rate), and the
performance deteriorates as we deviate from the C-rate. Also, the
robust model that is trained by fitting to data from all cycles shows
an overall better and more robust performance in comparison to models
that are trained on individual cycles.  In most cycles shown in Figure
\ref{fig:all_cycles}, this robust model (which is trained on all
cycles) outperforms most models on each cycle, Moreover, the robust
model trained by fitting to data from the C3 and 3C cycles (dotted
lines in Figure \ref{fig:all_cycles}) shows an overall good agreement
with the model trained by fitting to data from all cycles (dashed
lines in Figure \ref{fig:all_cycles}), both for the charge and the
discharge regimes. The agreement between two robust models indicates
that a small amount of measurement data may be sufficient to calibrate
our model without sacrificing accuracy, provided the measurement
cycles used correspond to well-separated C-rates.
The results obtained by solving inverse model \eqref{eq:inverse} using
all cycles as training data, for charge ($\mathcal{D}_t^{ch}$) and discharge regimes ($\mathcal{D}_t^{dch}$)
are demonstrated in 
Figures \ref{fig:all_history} and \ref{fig:all_omega}. The evolution of cost functional values and 
the parameter $\alpha$ with iterations of the algorithm are depicted in Figure 
\ref{fig:all_history}, whereas the optimal reconstructed constitutive relations are shown in 
Figure \ref{fig:all_omega}. 
Although the optimal constitutive relations found by fitting forward model \eqref{eq:ODE} to 
individual cycles for charge and discharge regimes are not presented here for brevity, they show 
a similar behaviour to the optimally reconstructed constitutive relations in Figure \ref{fig:all_omega}
Note the magnitude of the function $\omega$ is in both cases of the order of $0.1$, highlighting the
dominating effect of the intercalation/deintercalation relative to plating/stripping.
Also, it is clear from Figures \ref{fig:all_history} and \ref{fig:all_omega}
that the optimal reconstructions of constitutive relations and parameter $\alpha$ are slightly 
different between the charge and discharge regimes. As can be observed, the reconstructed relations $\omega_2$
for the charge and discharge regimes show similar behaviour, however, the relation $\omega_1$
and the parameter $\alpha$ demonstrate different behaviours in the charge and discharge regimes. This is contrary
to what we expect to observe in the system, namely, that the constitutive relations deduced in 
the two regimes should be approximately the same. This could have a few potential reasons.
\begin{enumerate}
	\item The dynamical behaviour of the system for charge and discharge regimes might show some 
	irreversibility. Note that the function $\omega$ is defined as the balance between Li plating and Li intercalation. This implies 
	that the competition between side reaction and intercalation is different between the charge and discharge 
	regimes. At a particular state of the cell, charging might result in an intercalation-plating 
	competition that might be different from the deintercalation-stripping competition when discharging at the 
	same state of the cell. This would indicate that Li metal does not get stripped in exactly the same manner
	as it gets plated. One possibility is that some plated Li loses electrical connectivity with 
	the negative particles, and for this reason, it becomes electrochemically inactive. In other words, not all plated Li
	is recoverable, giving rise to slightly different behaviour of function $\omega$ for stripping 
	in comparison to plating.
	\item The experimental conditions between the charge and discharge regimes might have slightly changed,
	hence, giving rise to different cell behaviour for each regime.
	\item The noise in the experimental data could be a factor that affects the fitting process
	and results in slightly different behaviour between charge and discharge regimes. As the inverse 
	problem tends to be ill-posed, the effect of noise could be significant.
\end{enumerate}
The results of solving the forward problem \eqref{eq:ODE} equipped with the optimally
reconstructed constitutive relations and optimal parameters $\alpha$ and $\bbeta$ by fitting the model to 
all cycles are depicted in Figures \ref{fig:all_c1_trajectory} and \ref{fig:all_c2_trajectory}. 
The optimal relations and parameters used correspond to the dashed line in Figure \ref{fig:all_cycles}, i.e., model that 
is fitted to aggregated data according to Section \ref{sec:robust}.
The results are shown for the time dependence of the reconstructed concentrations for all cycles, along with the 
true experimental concentrations. As can be observed, the concentrations from the 
model follow the overall behaviour of the dynamics of the system, with some minor deviations. 
There can be multiple reasons for this. 
\begin{enumerate}
	\item The noise in the NMR  measurements is one source of 
	inconsistency between the predictions of the mathematical model and the measured concentrations.
	\item The computational framework has one caveat which can potentially limit its performance. The separation of variables 
	($\omega(C_1, C_2) = \omega_1(C_1)\cdot\omega_2(C_2)$) is assumed in the optimal reconstruction formulation. The "true" optimal 
	form of constitutive relation might not be separable as assumed in the computational framework.
	\item This model does not take into account other undesired processes in the 
	cell that might consume some of the interfacial current	density, such as secondary SEI growth. The current density 
	applied to the cell	is entirely consumed by intercalation/deintercalation and plating/stripping processes.
	\item Model \eqref{eq:ODE} is trained on a range of C-rates simultaneously. It is known
	that the dynamics of Li-ion cells highly depend on the C-rate, and different simplified models are developed for 
	describing the dynamics of the cell at different ranges of C-rates, see	\cite{marquis2019asymptotic} and \cite{richardson2020generalised}. Hence, one model could lose its accuracy when trained 
	on a wide range of C-rates.	
	\item The optimization problems of this nature are typically non-convex and may therefore admit multiple local minima. We cannot guarantee that with the gradient-based approach we used the solutions we found are global minimizers.
\end{enumerate}
Thus, due to these reasons, it is unlikely the optimal solutions presented in this section could be further improves.

As can be observed in Figure \ref{fig:all_c2_trajectory}, the experimental concentrations of plated Li demonstrate
a partial recovery (stripping) of plated Li. This implies that some of the plated Li is inactive, and hence 
the calibrated constitutive relation must take into account this phenomenon. As explained before, 
this could be one reason for different behaviour of the constitutive relations between charge and discharge regimes.

\section{Discussion and Conclusions}\label{sec:discussion}

In this study, Li plating was investigated as one of the main degradation mechanisms in Li-ion cells 
using mathematical and computational tools. Physical modeling was employed in order to model the physical 
and chemical processes in the cell. Starting with the DFN model, we employed a variety of techniques, including
asymptotic reduction and averaging, in order to simplify it to an SP model with side reactions, tailored to 
our experimental data. The resulting SP model with side reaction tracks the evolution of two 
lumped concentrations: intercalated Li and plated Li in the cell. Notably, the model has the following properties:
(i) concentrations are averaged over their corresponding spatial domains to eliminate spatial 
dependence from the model, 
(ii) the model takes the form of an ODE system describing the evolution of the 
averaged quantities, circumventing the need to solve for the potential
distribution in the cell, as done in the DFN models, due to the simplifying assumptions of the proposed
framework,
(iii) the model accounts for both relaxation and excitation dynamics in the 
cell, with excitation being the dominant form  of the dynamics in the cell, and 
(iv) the model accounts for both plating and stripping processes in the
cell, allowing for the recovery of some of the plated Li.
These properties make the model a good candidate for online state estimations and 
monitoring of the cells. From the physical modeling perspective, the study by 
Brosa Planella et al.~\cite{planella2023single} bears
the closest resemblance to this work, although it does not account for Li stripping. 
Sahu et al.~\cite{sahu2023continuum} consider more interactions between 
different phases of Li in the cell and develop a more comprehensive mathematical model capable of 
predicting both plating and stripping. Our resulting physical model involves a number of 
physical parameters and a constitutive relation that require calibration using experimental data. 
Inverse modeling and optimization techniques are employed for this 
purpose in order to determine the optimal value of parameters and the optimal form of constitutive relations,
aiming to minimize discrepancies between model outputs and experimental data.
To our knowledge, this study represents the first instance of using inverse modeling to optimally
predict Li plating and stripping in Li-ion cells. 

We note that the negative electrode material utilized in this study is
silicon. This material experiences significant volume variations
during charge/discharge cycles of the cell, a phenomenon linked to its
high charge density. These volume changes may influence the model
performance, as we have not explicitly accounted for this phenomenon.
Nonetheless, the calibrated parameters and constitutive relations of
the model may implicitly account for this effect.

An important consideration is the range of validity of the Li-plating
model. As highlighted by Marquis et al.~\cite{marquis2019asymptotic},
the SP model remains valid up to a C-rate of 1C, beyond which it
begins to diverge from the DFN model. In this study, we have also
developed a variant of the SP model that accounts for Li-plating as a
side reaction. Consequently, it is important to investigate the range
of validity of this model.  As depicted in Figures
\ref{fig:all_c1_trajectory} and \ref{fig:all_c2_trajectory}, the model
trained across a range of cycles demonstrates the ability to predict
intercalation/deintercalation and plating/stripping behaviour in an
overall acceptable manner.  Due to the fact that the mathematical
model is calibrated using data, its fidelity may extend beyond 1C
rate. Figure \ref{fig:all_cycles} suggests that a model calibrated on
a specific C-rate performs well in its proximity, and its performance
gradually deteriorates as the C-rate deviates from the C-rate used for
training. Thus, the range of validity of the model highly depends on
the training process used to calibrate the model, which, in turn, is
determined by the specific application assumed for the model.

The proposed physical modeling and computational framework can also be
extended to differentiate between different phases of Li within the
cell, particularly non-recoverable Li and recoverable Li.  In the
current study, these two phases are not distinguished as the
experimental data for inactive Li is not available. Additionally, this
framework does not account for other degradation mechanisms in the
cell and solely focuses on the Li plating. However, it can be readily
expanded to include other types of degradation mechanisms in the cell.

\section*{Acknowledgments}
The authors thank Jamie Foster for helpful discussions. This research was supported by a Collaborative Research \&  Development grant \# CRD494074-16 from Natural Sciences \&   Engineering Research Council of Canada.


\section*{Conflict of interest}
The authors declare no potential conflict of interests.


\begin{thebibliography}{10}

\bibitem{su2014silicon}
X.~Su, Q.~Wu, J.~Li, X.~Xiao, A.~Lott, W.~Lu, B.~W. Sheldon, and J.~Wu,
  ``Silicon-based nanomaterials for lithium-ion batteries: a review,'' {\em
  Advanced Energy Materials}, vol.~4, no.~1, p.~1300882, 2014.

\bibitem{birkl2017degradation}
C.~R. Birkl, M.~R. Roberts, E.~McTurk, P.~G. Bruce, and D.~A. Howey,
  ``Degradation diagnostics for lithium ion cells,'' {\em Journal of Power
  Sources}, vol.~341, pp.~373--386, 2017.

\bibitem{edge2021lithium}
J.~S. Edge, S.~O’Kane, R.~Prosser, N.~D. Kirkaldy, A.~N. Patel, A.~Hales,
  A.~Ghosh, W.~Ai, J.~Chen, J.~Yang, {\em et~al.}, ``Lithium ion battery
  degradation: what you need to know,'' {\em Physical Chemistry Chemical
  Physics}, vol.~23, no.~14, pp.~8200--8221, 2021.

\bibitem{lin2021lithium}
X.~Lin, K.~Khosravinia, X.~Hu, J.~Li, and W.~Lu, ``Lithium plating mechanism,
  detection, and mitigation in lithium-ion batteries,'' {\em Progress in Energy
  and Combustion Science}, vol.~87, p.~100953, 2021.

\bibitem{zhang2022investigation}
G.~Zhang, X.~Wei, S.~Chen, G.~Han, J.~Zhu, and H.~Dai, ``Investigation the
  degradation mechanisms of lithium-ion batteries under low-temperature
  high-rate cycling,'' {\em ACS Applied Energy Materials}, vol.~5, no.~5,
  pp.~6462--6471, 2022.

\bibitem{santhanagopalan2009analysis}
S.~Santhanagopalan, P.~Ramadass, and J.~Z. Zhang, ``Analysis of internal
  short-circuit in a lithium ion cell,'' {\em Journal of Power Sources},
  vol.~194, no.~1, pp.~550--557, 2009.

\bibitem{fang2022quantifying}
Y.~Fang, A.~J. Smith, R.~W. Lindstr{\"o}m, G.~Lindbergh, and I.~Fur{\'o},
  ``Quantifying lithium lost to plating and formation of the solid-electrolyte
  interphase in graphite and commercial battery components,'' {\em Applied
  Materials Today}, vol.~28, p.~101527, 2022.

\bibitem{bugga2010lithium}
R.~V. Bugga and M.~C. Smart, ``Lithium plating behavior in lithium-ion cells,''
  {\em ECS transactions}, vol.~25, no.~36, p.~241, 2010.

\bibitem{paul2021review}
P.~P. Paul, E.~J. McShane, A.~M. Colclasure, N.~Balsara, D.~E. Brown, C.~Cao,
  B.-R. Chen, P.~R. Chinnam, Y.~Cui, E.~J. Dufek, {\em et~al.}, ``A review of
  existing and emerging methods for lithium detection and characterization in
  li-ion and li-metal batteries,'' {\em Advanced Energy Materials}, vol.~11,
  no.~17, p.~2100372, 2021.

\bibitem{tian2021detecting}
Y.~Tian, C.~Lin, H.~Li, J.~Du, and R.~Xiong, ``Detecting undesired lithium
  plating on anodes for lithium-ion batteries--a review on the in-situ
  methods,'' {\em Applied Energy}, vol.~300, p.~117386, 2021.

\bibitem{sanders2023quantitative}
K.~J. Sanders, A.~A. Ciezki, A.~Berno, I.~C. Halalay, and G.~R. Goward,
  ``Quantitative operando 7li nmr investigations of silicon anode evolution
  during fast charging and extended cycling,'' {\em Journal of the American
  Chemical Society}, vol.~145, no.~39, pp.~21502--21513, 2023.

\bibitem{sethurajan2015accurate}
A.~K. Sethurajan, S.~A. Krachkovskiy, I.~C. Halalay, G.~R. Goward, and
  B.~Protas, ``Accurate characterization of ion transport properties in binary
  symmetric electrolytes using in situ nmr imaging and inverse modeling,'' {\em
  The Journal of Physical Chemistry B}, vol.~119, no.~37, pp.~12238--12248,
  2015.

\bibitem{sethurajan2019dendrites}
A.~K. Sethurajan, J.~M. Foster, G.~Richardson, S.~A. Krachkovskiy, J.~D. Bazak,
  G.~R. Goward, and B.~Protas, ``Incorporating dendrite growth into continuum
  models of electrolytes: Insights from nmr measurements and inverse
  modeling,'' {\em Journal of The Electrochemical Society}, vol.~166, no.~8,
  pp.~A1591--A1602, 2019.

\bibitem{escalante2020discerning}
J.~M. Escalante, W.~Ko, J.~M. Foster, S.~Krachkovskiy, G.~Goward, and
  B.~Protas, ``Discerning models of phase transformations in porous graphite
  electrodes: Insights from inverse modelling based on mri measurements,'' {\em
  Electrochimica Acta}, vol.~349, p.~136290, 2020.

\bibitem{daniels2023learning}
L.~Daniels, S.~Sahu, K.~J. Sanders, G.~R. Goward, J.~M. Foster, and B.~Protas,
  ``Learning optimal forms of constitutive relations characterizing ion
  intercalation from data in mathematical models of lithium-ion batteries,''
  {\em The Journal of Physical Chemistry C}, vol.~127, no.~35,
  pp.~17508--17523, 2023.

\bibitem{newman2021electrochemical}
J.~Newman and N.~P. Balsara, {\em Electrochemical systems}.
\newblock John Wiley \& Sons, 2021.

\bibitem{planella2022continuum}
F.~B. Planella, W.~Ai, A.~Boyce, A.~Ghosh, I.~Korotkin, S.~Sahu, V.~Sulzer,
  R.~Timms, T.~Tranter, M.~Zyskin, {\em et~al.}, ``A continuum of physics-based
  lithium-ion battery models reviewed,'' {\em Progress in Energy}, 2022.

\bibitem{o2022lithium}
S.~E. O'Kane, W.~Ai, G.~Madabattula, D.~Alonso-Alvarez, R.~Timms, V.~Sulzer,
  J.~S. Edge, B.~Wu, G.~J. Offer, and M.~Marinescu, ``Lithium-ion battery
  degradation: how to model it,'' {\em Physical Chemistry Chemical Physics},
  vol.~24, no.~13, pp.~7909--7922, 2022.

\bibitem{guo2016li}
Y.~Guo, R.~B. Smith, Z.~Yu, D.~K. Efetov, J.~Wang, P.~Kim, M.~Z. Bazant, and
  L.~E. Brus, ``Li intercalation into graphite: direct optical imaging and
  cahn--hilliard reaction dynamics,'' {\em The journal of physical chemistry
  letters}, vol.~7, no.~11, pp.~2151--2156, 2016.

\bibitem{fullerdoylenewman1994}
T.~F. Fuller, M.~Doyle, and J.~Newman, ``{Simulation and Optimization of the
  Dual Lithium Ion Insertion Cell},'' {\em J. Electrochem. Soc.}, vol.~141,
  no.~1, pp.~1--10, 1994.

\bibitem{Atlung1979dynamic}
S.~Atlung, K.~West, and T.~Jacobsen, ``Dynamic aspects of solid solution
  cathodes for electrochemical power sources,'' {\em Journal of The
  Electrochemical Society}, vol.~126, no.~8, p.~1311, 1979.

\bibitem{prada2012simplified}
E.~Prada, D.~Di~Domenico, Y.~Creff, J.~Bernard, V.~Sauvant-Moynot, and F.~Huet,
  ``Simplified electrochemical and thermal model of lifepo4-graphite li-ion
  batteries for fast charge applications,'' {\em Journal of The Electrochemical
  Society}, vol.~159, no.~9, p.~A1508, 2012.

\bibitem{planella2023single}
F.~B. Planella and W.~D. Widanage, ``A single particle model with electrolyte
  and side reactions for degradation of lithium-ion batteries,'' {\em Applied
  Mathematical Modelling}, vol.~121, pp.~586--610, 2023.

\bibitem{marquis2019asymptotic}
S.~G. Marquis, V.~Sulzer, R.~Timms, C.~P. Please, and S.~J. Chapman, ``An
  asymptotic derivation of a single particle model with electrolyte,'' {\em
  Journal of The Electrochemical Society}, vol.~166, no.~15, p.~A3693, 2019.

\bibitem{richardson2020generalised}
G.~Richardson, I.~Korotkin, R.~Ranom, M.~Castle, and J.~Foster, ``Generalised
  single particle models for high-rate operation of graded lithium-ion
  electrodes: Systematic derivation and validation,'' {\em Electrochimica
  Acta}, vol.~339, p.~135862, 2020.

\bibitem{sahu2023continuum}
S.~Sahu and J.~M. Foster, ``A continuum model for lithium plating and dendrite
  formation in lithium-ion batteries: Formulation and validation against
  experiment,'' {\em Journal of Energy Storage}, vol.~60, p.~106516, 2023.

\bibitem{dickinson2020butler}
E.~J. Dickinson and A.~J. Wain, ``The butler-volmer equation in electrochemical
  theory: Origins, value, and practical application,'' {\em Journal of
  Electroanalytical Chemistry}, vol.~872, p.~114145, 2020.

\bibitem{arora1999mathematical}
P.~Arora, M.~Doyle, and R.~E. White, ``Mathematical modeling of the lithium
  deposition overcharge reaction in lithium-ion batteries using carbon-based
  negative electrodes,'' {\em Journal of The Electrochemical Society},
  vol.~146, no.~10, p.~3543, 1999.

\bibitem{yang2018look}
X.-G. Yang, S.~Ge, T.~Liu, Y.~Leng, and C.-Y. Wang, ``A look into the voltage
  plateau signal for detection and quantification of lithium plating in
  lithium-ion cells,'' {\em Journal of Power Sources}, vol.~395, pp.~251--261,
  2018.

\bibitem{ovejas2019effects}
V.~Ovejas and A.~Cuadras, ``Effects of cycling on lithium-ion battery
  hysteresis and overvoltage,'' {\em Scientific reports}, vol.~9, no.~1,
  p.~14875, 2019.

\bibitem{nw00}
J.~Nocedal and S.~Wright, {\em Numerical Optimization}.
\newblock Springer, 2002.

\bibitem{bukshtynov20111228}
V.~Bukshtynov, O.~Volkov, and B.~Protas, ``On optimal reconstruction of
  constitutive relations,'' {\em Physica D: Nonlinear Phenomena}, vol.~240,
  no.~16, pp.~1228 -- 1244, 2011.

\bibitem{bukshtynov2013889}
V.~Bukshtynov and B.~Protas, ``Optimal reconstruction of material properties in
  complex multiphysics phenomena,'' {\em Journal of Computational Physics},
  vol.~242, pp.~889 -- 914, 2013.

\bibitem{pnm14}
B.~Protas, B.~R. Noack, and M.~Morzynski, ``An optimal model identification for
  oscillatory dynamics with a stable limit cycle,'' {\em J. Nonlin. Sci.},
  vol.~24, pp.~245--275, 2014.

\bibitem{pftv86}
W.~H. Press, B.~P. Flanner, S.~A. Teukolsky, and W.~T. Vetterling, {\em
  Numerical Recipes: the Art of Scientific Computations}.
\newblock Cambridge University Press, 1986.

\end{thebibliography}

\appendix

\section{Adjoint Sensitivities in Relaxation Dynamics}
\label{sec:relaxation_appendix}
In order to compute components of the gradient vector in
\eqref{eq:gradient_OCV}, adjoint sensitivity analysis is employed
\cite{bukshtynov20111228, bukshtynov2013889, pnm14}. We begin by
computing the directional derivatives
\begin{equation}
\begin{alignedat}{3}
\JJ_1^\prime(\bbeta;\beta_i^\prime) &= \lim_{\epsilon\to 0}\epsilon^{-1} \left[\JJ_1(\bbeta; \beta_i+\epsilon\beta_i^\prime)-\JJ_1(\bbeta)\right] &&= \int_{0}^{T} (\bw \, \br(t;\bbeta))^\top \bC^\prime(\beta_i^\prime) dt,\\
\bC^\prime(\beta_i^\prime) &= \begin{bmatrix}
C_{1}^\prime(\bbeta;\beta_i^\prime)\\
C_{2}^\prime(\bbeta;\beta_i^\prime)
\end{bmatrix},\\
\bw &= \begin{bmatrix} 1&0 \\ 0&w\end{bmatrix},
\end{alignedat}
\label{eq:directional_OCV}
\end{equation}
where $i\in\{1,2,3,4\}$, and $\bC^\prime(\beta_i^\prime)$ is the solution of a system of equations describing 
perturbations of the state variables resulting from the perturbations of each of the parameters.
In order to derive this system, the parameters are perturbed, and the state variables are 
perturbed with respect to each of the parameters in $\bbeta$ as 
\begin{equation}
\begin{alignedat}{1}
\beta_i &= \widehat{\beta}_i + \epsilon \beta_i^\prime,\\
\bC(\bbeta) &= \widehat{\bC}(\widehat{\bbeta}) + \epsilon \bC^\prime(\bbeta;\beta_i^\prime) + \mathcal{O}(\epsilon^2),
\end{alignedat}
\label{eq:pert1}
\end{equation}
where $ i\in\{1,2,3,4\}$, the variables with a hat sign represent the unperturbed version of 
the variables, and the prime sign represents the perturbation. 
Substituting \eqref{eq:pert1} into the system of equations \eqref{eq:ODE}, collecting terms with respect to
different powers of $\epsilon$, at the leading-order we get one system of equations corresponding to the unperturbed 
version of equations, $d\widehat{\bC}/dt = \widehat{\bB} + \widehat{\bA} \widehat{\bC}$. At the order of $\epsilon$,
four systems of equations are obtained corresponding to the perturbation of each of the parameters in the 
vector $\bbeta$. The four systems of equations are obtained as 
\begin{equation}
\begin{alignedat}{1}
\frac{d}{dt}\bC^\prime(\beta_i^\prime) &= \widehat{\bA}\bC^\prime(\beta_i^\prime) + \bI_i \beta_i^\prime \widehat{\bC} + z_i \bI_0, \\
\bI_0 &= \begin{bmatrix}
1\\
0
\end{bmatrix}, \qquad
\bI_1 = \begin{bmatrix}
0&0\\
0&0
\end{bmatrix}, \qquad
\bI_2 = \begin{bmatrix}
1&0\\
0&0
\end{bmatrix}, \qquad
\bI_3 = \begin{bmatrix}
0&\lambda^{-1}\\
0&-1
\end{bmatrix}, \qquad
\bI_4 = \begin{bmatrix}
0&0\\
1&0
\end{bmatrix},
\end{alignedat}
\end{equation}
where $i\in\{1,2,3,4\}$, $z_1 = \beta_1^\prime$, and $z_i = 0$ for $i\in\{2,3,4\}$.
Dotting this equation with the vectors of the adjoint variables 
$\bC_i^\ast(t) = \left[C_1^\ast(t), C_2^\ast(t) \right]^\top$, and 
integrating in time we obtain
\begin{equation}
\begin{alignedat}{1}
\int_{0}^{T} {\bC_i^\ast}^{\top}  \frac{d}{dt}\bC^\prime(\beta_i^\prime) ~dt
- \int_{0}^{T} {\bC_i^\ast}^{\top}  \widehat{\bA}\bC^\prime(\beta_i^\prime) ~dt
- \int_{0}^{T} {\bC_i^\ast}^{\top} \bI_i \beta_i^\prime \widehat{\bC} ~dt
- \int_{0}^{T} {\bC_i^\ast}^{\top} z_i \bI_0 ~dt &= 0.
\end{alignedat}
\end{equation}
Note that four different adjoint vectors are required, each of which correspond to one system of equations
resulting from perturbation of one parameter. Performing integration by parts on the first term, and applying initial 
conditions, we get
\begin{equation}
\begin{alignedat}{1}
- {\bC_i^\ast}^{\top}(T) \bC^\prime(\beta_i^\prime)(T)
+ \int_{0}^{T} \frac{d}{dt}{\bC_i^\ast}^{\top}  \bC^\prime(\beta_i^\prime) ~dt
+ \int_{0}^{T} {\bC_i^\ast}^{\top}  \widehat{\bA}\bC^\prime(\beta_i^\prime) ~dt\\
+ \int_{0}^{T} {\bC_i^\ast}^{\top} \bI_i \beta_i^\prime \widehat{\bC} ~dt
+ \int_{0}^{T} {\bC_i^\ast}^{\top} z_i \bI_0 ~dt &= 0.
\end{alignedat}
\end{equation}
Factoring out $\bC^\prime$ results in
\begin{equation}
\begin{alignedat}{1}
- {\bC_i^\ast}^{\top}(T) \bC^\prime(\beta_i^\prime)(T)
+ \int_{0}^{T} \left[ \frac{d}{dt}{\bC_i^\ast}^{\top} + {\bC_i^\ast}^{\top}  \widehat{\bA}  \right]  \bC^\prime(\beta_i^\prime) ~dt 
&=
- \int_{0}^{T} {\bC_i^\ast}^{\top} \bI_i \beta_i^\prime \widehat{\bC} ~dt
- \int_{0}^{T} {\bC_i^\ast}^{\top} z_i \bI_0 ~dt.
\end{alignedat}
\label{eq:adj10}
\end{equation}
Thus, we define the adjoint system in a judicious manner to provide a convenient expression for the directional 
derivative as 
\begin{equation}
\begin{alignedat}{1}
\frac{d}{dt}{\bC^\ast}^{\top} &=  (\bw \, \br(t;\bbeta))^\top -  {\bC^\ast}^{\top}\widehat{\bA} ,  \\
{\bC^\ast}(T) &= \boldsymbol{0}.
\end{alignedat}
\label{eq:adjoint_OCV}
\end{equation}
Note that different adjoint systems defined for each adjoint vector $\bC_i^\ast, i\in\{1,2,3,4\}$ are identical, and hence, the 
subscript $i$ is removed.
Consequently, with this definition of the adjoint system  \eqref{eq:adj10} reduces to 
\begin{equation}
\begin{alignedat}{1}
\JJ_1^\prime(\bbeta;\beta_i^\prime)
&= 
- \int_{0}^{T} {\bC^\ast}^{\top} \bI_i \beta_i^\prime \widehat{\bC} ~dt
- \int_{0}^{T} {\bC^\ast}^{\top} z_i \bI_0 ~dt.\\
\end{alignedat}
\end{equation}
Note that we can also compute the directional derivative as $\JJ_1^\prime(\bbeta;\beta_i^\prime) = \partial \JJ_1/\partial \beta_i \cdot \beta_i^\prime $. Thus, the gradient of cost functional is obtained as 
\begin{equation}
\begin{alignedat}{1}
\bnab_\bbeta \JJ_1
&= \begin{bmatrix}
- \int_{0}^{T} {\bC^\ast}^{\top} \bI_0 ~dt 
&\quad
- \int_{0}^{T} {\bC^\ast}^{\top} \bI_2 \widehat{\bC} ~ dt 
&\quad
- \int_{0}^{T} {\bC^\ast}^{\top} \bI_3 \widehat{\bC} ~ dt 
&\quad
- \int_{0}^{T} {\bC^\ast}^{\top} \bI_4 \widehat{\bC} ~ dt 
\end{bmatrix}.
\end{alignedat}
\end{equation}

\section{Validation of the Computational Framework}\label{sec:validate}
\subsection{Validation of Gradients}
\begin{figure}[!ht]
	\centering
	\begin{subfigure}[b]{0.45\textwidth}
		\centering
		\includegraphics[width=1\textwidth]{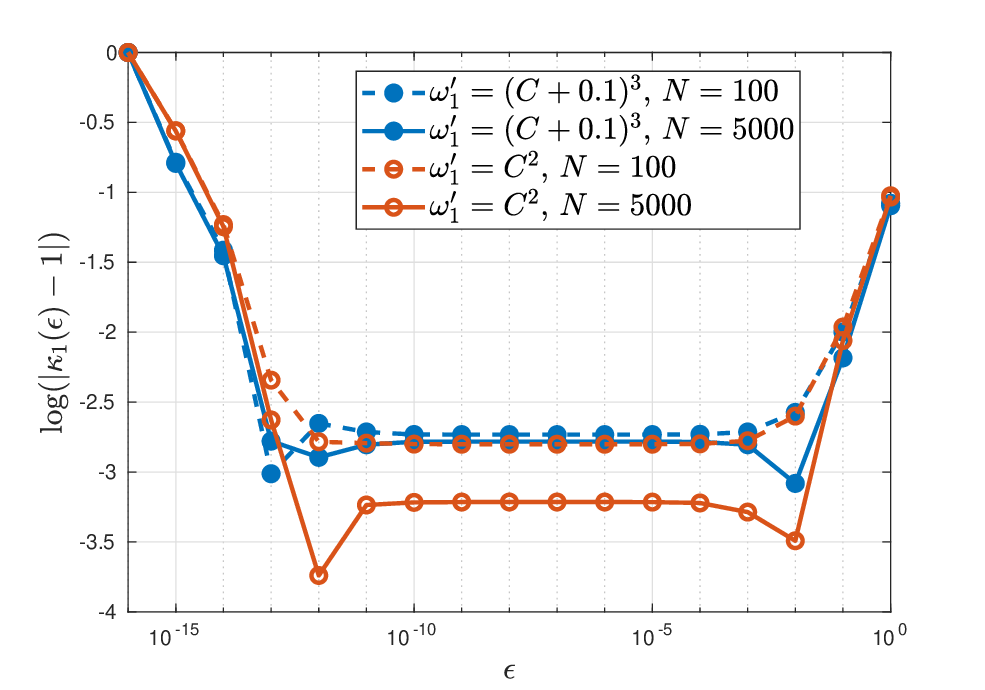}
		\subcaption{}
	\end{subfigure}
	\begin{subfigure}[b]{0.45\textwidth}
		\centering
		\includegraphics[width=1\textwidth]{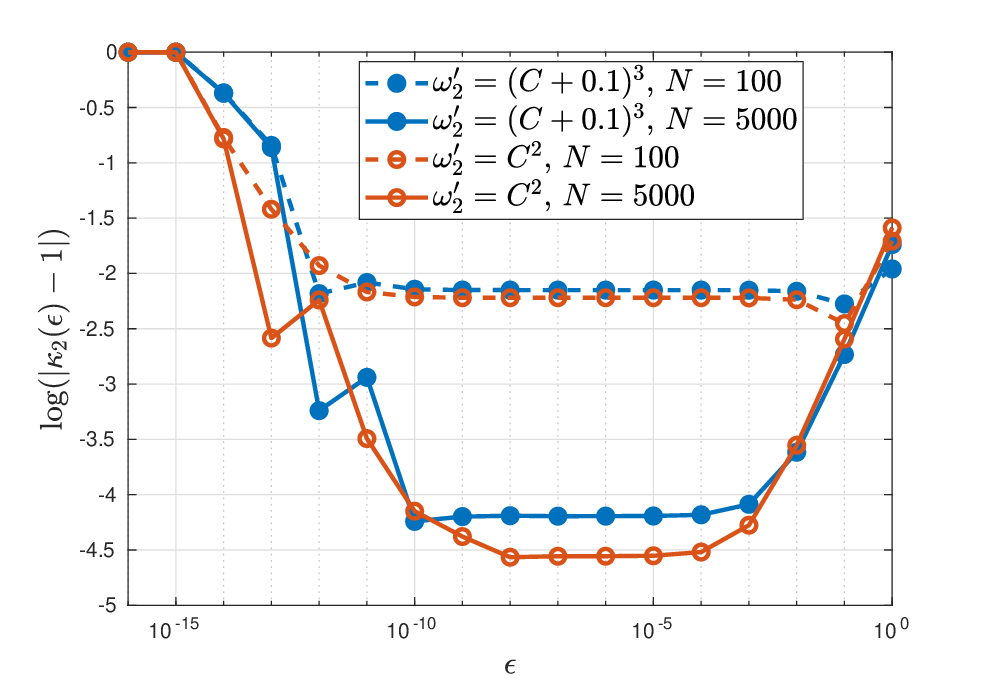}
		\subcaption{}
	\end{subfigure}
	\begin{subfigure}[b]{0.45\textwidth}
		\centering
		\includegraphics[width=1\textwidth]{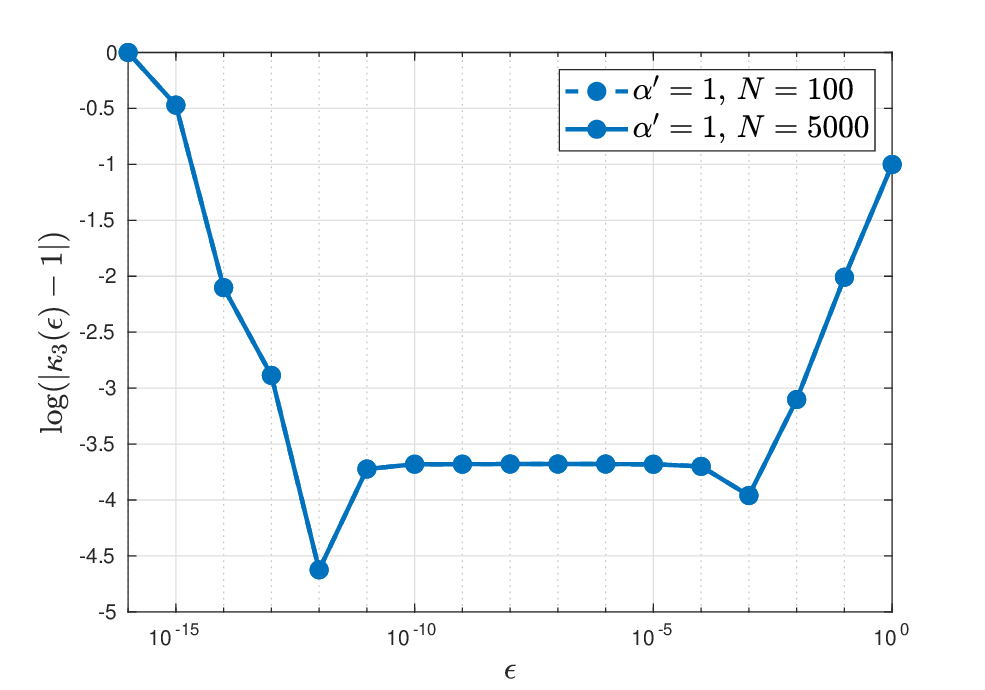}
		\subcaption{}
	\end{subfigure}
	\caption{The behaviour of $\kappa_1(\epsilon)$ (a), $\kappa_2(\epsilon)$ (b), and  
		$\kappa_3(\epsilon)$ (c), over a wide range of $\epsilon$ values, by using 
		$\omega_1^{(0)}=\omega_2^{(0)}=0.7$, $\alpha^{(0)}=3$ as the starting point, and
		using different perturbations of constitutive relations and parameters.		
		Two different discretizations of the interval $\L$ are
		used, namely, $N=100$ (dashed lines) and $N=5000$ (solid lines). 
		Note that discretization of the state interval $\L$ does not affect the quantity 
		$\kappa_3(\epsilon)$, as its partial derivative \eqref{eq:L2} is computed without discretizing the 
		state space $\L$. Note that $\bbeta = [-0.1,-0.1,-0.1,-0.1]$ in this experiment.}
	\label{fig:kappa_1}
\end{figure}

\begin{figure}[!ht]
	\centering
	\begin{subfigure}[b]{0.4\textwidth}
		\centering
		\includegraphics[width=1\textwidth]{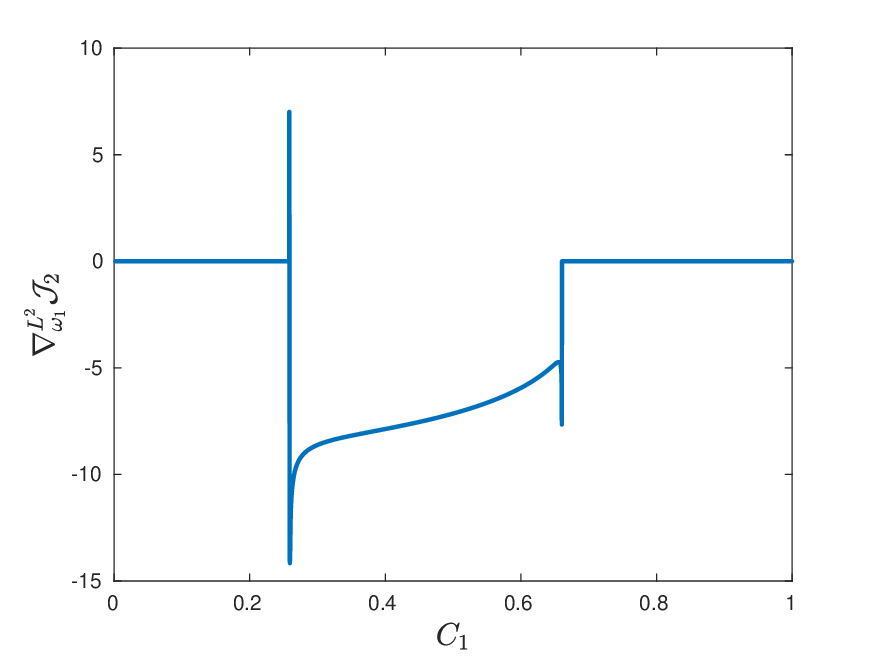}
		\caption{\centering $\bnab_{\omega_1}^{L^2} \JJ_2$}
	\end{subfigure}
	\begin{subfigure}[b]{0.4\textwidth}
		\centering
		\includegraphics[width=1\textwidth]{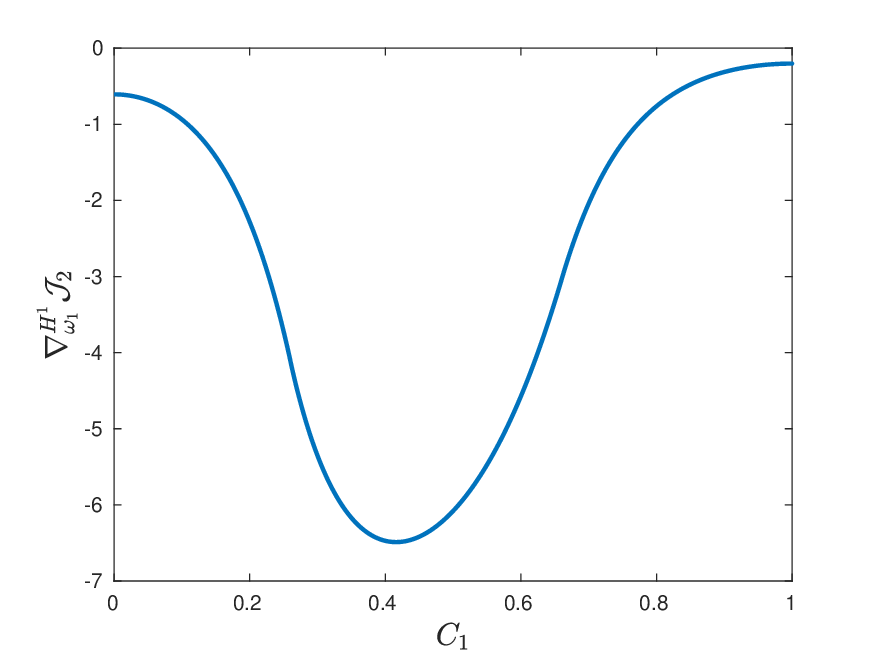}
		\caption{\centering $\bnab_{\omega_1}^{H^1} \JJ_2$}
	\end{subfigure}
	\begin{subfigure}[b]{0.4\textwidth}
		\centering
		\includegraphics[width=1\textwidth]{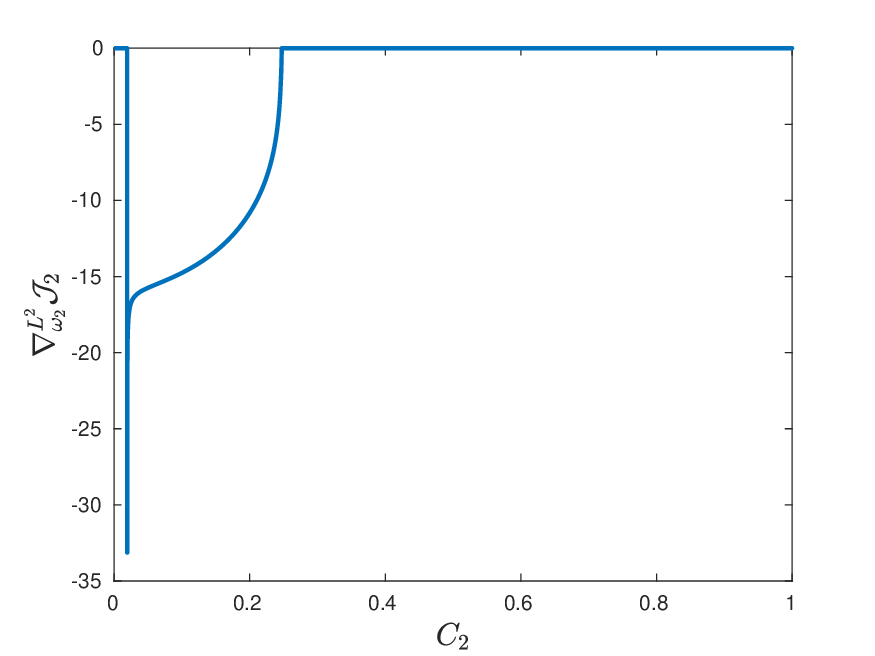}
		\caption{\centering $\bnab_{\omega_2}^{L^2} \JJ_2$}
	\end{subfigure}
	\begin{subfigure}[b]{0.4\textwidth}
		\centering
		\includegraphics[width=1\textwidth]{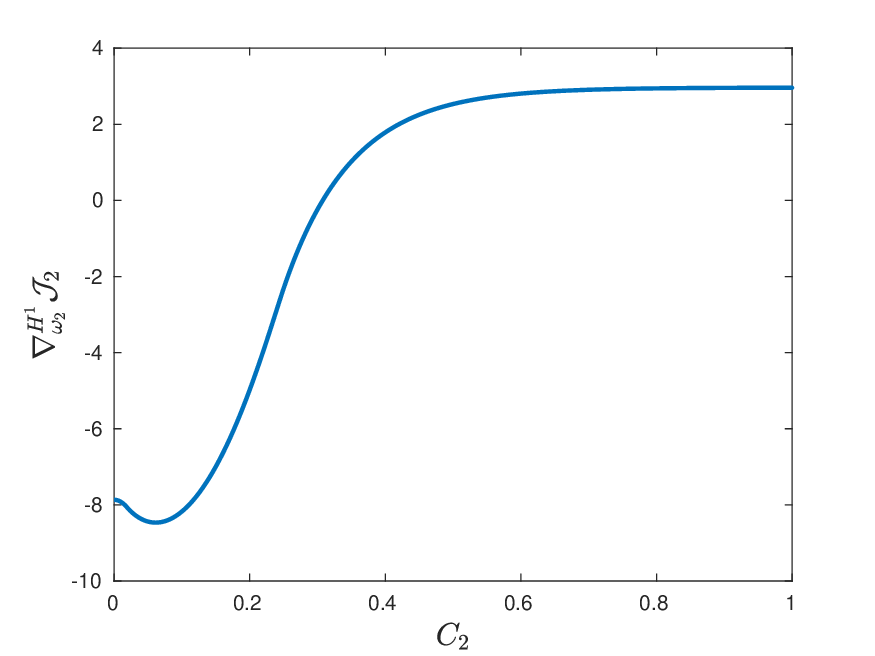}
		\caption{\centering $\bnab_{\omega_2}^{H^1_0} \JJ_2$}
	\end{subfigure}
	\caption{$\bnab_{\omega_1}^{L^2} \JJ_2$ (a), $\bnab_{\omega_2}^{L^2} \JJ_2$ (c), $\bnab_{\omega_1}^{H^1} \JJ_2$ (b), and $\bnab_{\omega_2}^{H^1_0} \JJ_2$ (d) at the first iteration of Algorithm \ref{alg:optimal}. Note the mean 
		of the gradient in (d), as it is reconstructed in $H^1_0$ space.}
	\label{fig:manufactured_gradients}
\end{figure}

To validate the derivation and computation of 
the gradients computed using adjoint analysis we will compare the adjoint-based 
expression for the Gateaux differential, cf.~\eqref{eq:riesz} and \eqref{eq:L2}, 
with a finite-difference approximation of the differential.
In order to determine the consistency of the gradients, we define the ratio of
the directional derivative evaluated as described above for each of the gradients as 
\begin{equation}
\begin{alignedat}{3}
\kappa_1(\epsilon) &= \frac{\epsilon^{-1} \left[\JJ_2(\alpha,\omega_1+\epsilon\omega_1^\prime,\omega_2)-\JJ_2(\alpha,\omega_1,\omega_2)\right]}{\int_{C_1^{\alpha}}^{C_1^{\beta}} \bnab_{\omega_1}^{L^2} \JJ_2 \cdot \omega_1^\prime ds},\\
\kappa_2(\epsilon) &= \frac{\epsilon^{-1} \left[\JJ_2(\alpha,\omega_1,\omega_2+\epsilon\omega_2^\prime)-\JJ_2(\alpha,\omega_1,\omega_2)\right]}{\int_{C_2^{\alpha}}^{C_2^{\beta}} \bnab_{\omega_2}^{L^2} \JJ_2 \cdot \omega_2^\prime ds},\\
\kappa_3(\epsilon) &= \frac{\epsilon^{-1} \left[\JJ_2(\alpha+\epsilon\alpha^\prime,\omega_1,\omega_2)-\JJ_2(\alpha,\omega_1,\omega_2)\right]}{ \frac{\partial\JJ_2} {\partial\alpha} \cdot \alpha^\prime },
\label{eq:kappa}
\end{alignedat}
\end{equation}
which we will refer to as the $\kappa$-test. We note that either functional spaces, $L^2$ or $H^1$, 
could be used to evaluate the expressions in the denominator and for simplicity we choose the $L^2$ 
gradients here. When the gradients are approximated correctly, the quantities $\kappa_1$, $\kappa_2$ 
and $\kappa_3$ should be close to unity for a broad range of 
$\epsilon$ values. However, these quantities deviate from the unity for very small or 
very large values of $\epsilon$ due to round-off and truncation errors, respectively, 
which are well-understood effects. The results of the $\kappa$-test are demonstrated in 
Figure \ref{fig:kappa_1}. In this test, two different discretization of the 
interval $\L$ are used. In Figures \ref{fig:kappa_1}a,b we see that, as expected, when the discretization $N$ 
of the state interval $\L$ is refined, the quantities $\kappa_1(\epsilon)$ and $\kappa_2(\epsilon)$ approach unity for 
a broad range of values of epsilon $\epsilon$. This trend is absent from Figure \ref{fig:kappa_1}c, since 
approximation of the derivative \eqref{eq:L2} does not depend on the discretization of the interval $\L$.
The results of constructing the $L^2$ and $H^1$ gradients of the constitutive relations
in the first iteration of the Algorithm \ref{alg:optimal} are demonstrated in 
Figure \ref{fig:manufactured_gradients}, with $\omega_1^{(0)}=\omega_2^{(0)}=0.7$ 
and $\alpha^{(0)}=3$ as the initial guess and $5000$ grid points in the $\L$ interval. 
Note that $\bbeta = [-0.1,-0.1,-0.1,-0.1]$ in this experiment. As it can be observed 
in Figure \ref{fig:manufactured_gradients}, the $L^2$ gradients are discontinuous and 
vanish outside the identifiability region (the discontinuity occurs on the boundary 
of the identifiability region). However, the $H^1$ gradients behave well outside the 
identifiability region where their behavior is determined by the choice of the boundary 
conditions in \eqref{eq:BVP_1} and \eqref{eq:BVP_2} whereas their smoothness is controlled 
by the parameter $l$ in the definition of the $H^1$ inner product \eqref{eq:Sobolev}. A Neumann
boundary condition, and a smoothing parameter of $l=1$ is used for this experiment.

\subsection{Validation Based on a Manufactured Solution}
\begin{figure}[!ht]
	\centering
	\begin{subfigure}[b]{0.45\textwidth}
		\centering
		\includegraphics[width=1\textwidth]{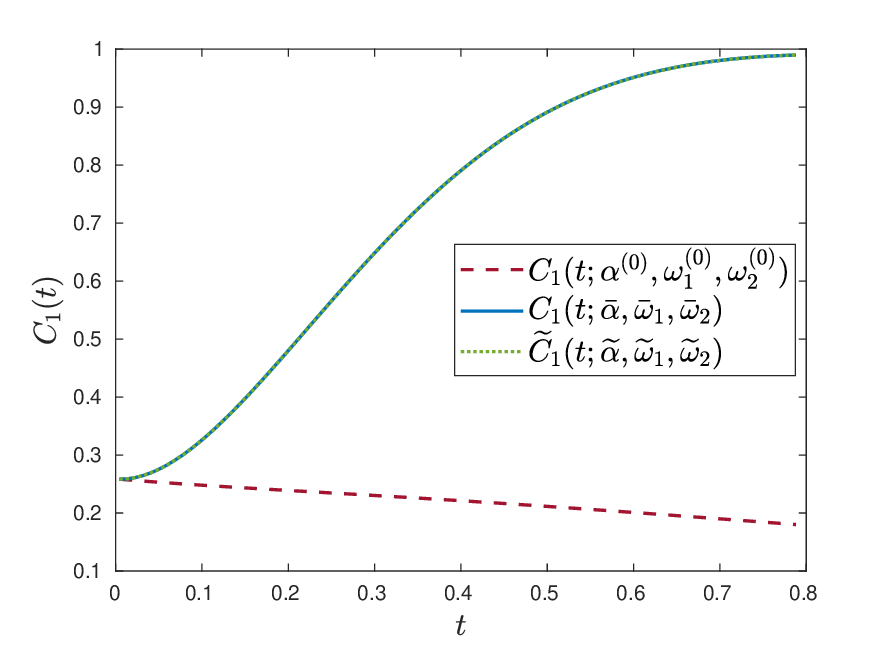}
		\subcaption{}
	\end{subfigure}
	\hfill
	\begin{subfigure}[b]{0.45\textwidth}
		\centering
		\includegraphics[width=1\textwidth]{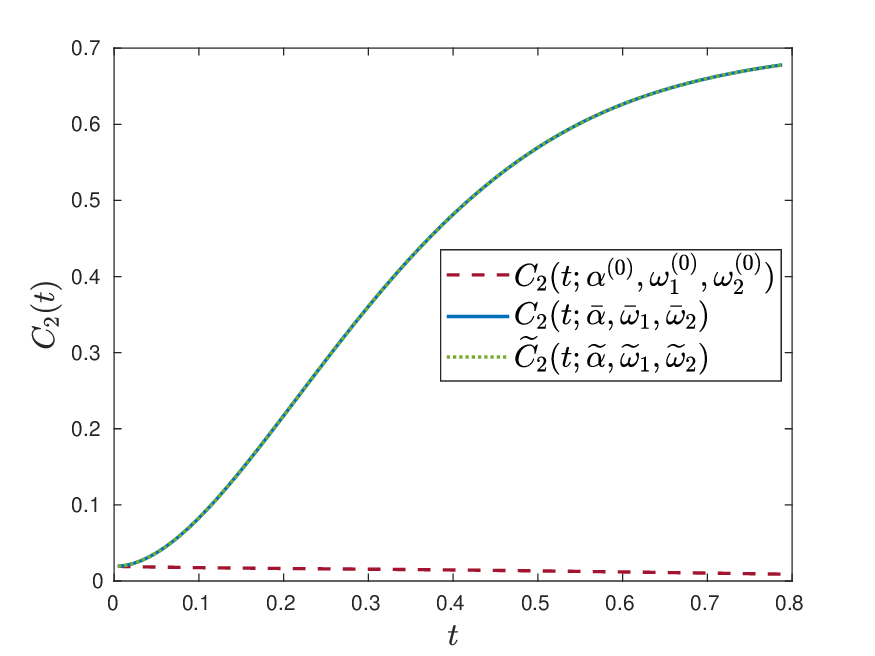}
		\subcaption{}
	\end{subfigure}
	\caption{Time history of concentrations $C_1(t)$ (a) and $C_2(t)$ (b) obtained using the true 
		parameter and constitutive relations (dotted green line), the initial guess of parameter 
		and relations (dashed red line), and the optimal reconstructed parameter and relations 
		(solid blue line).}
	\label{fig:simulated_concentrations}
\end{figure}

\begin{figure}[!ht]
	\centering
	\begin{subfigure}[b]{0.45\textwidth}
		\centering
		\includegraphics[width=1\textwidth]{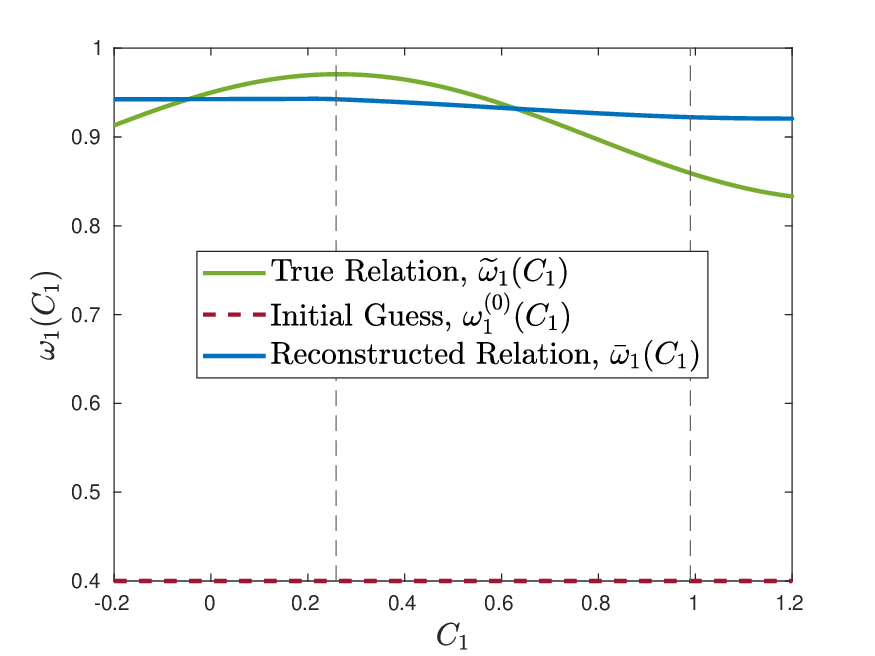}
		\subcaption{}
	\end{subfigure}
	\hfill
	\begin{subfigure}[b]{0.45\textwidth}
		\centering
		\includegraphics[width=1\textwidth]{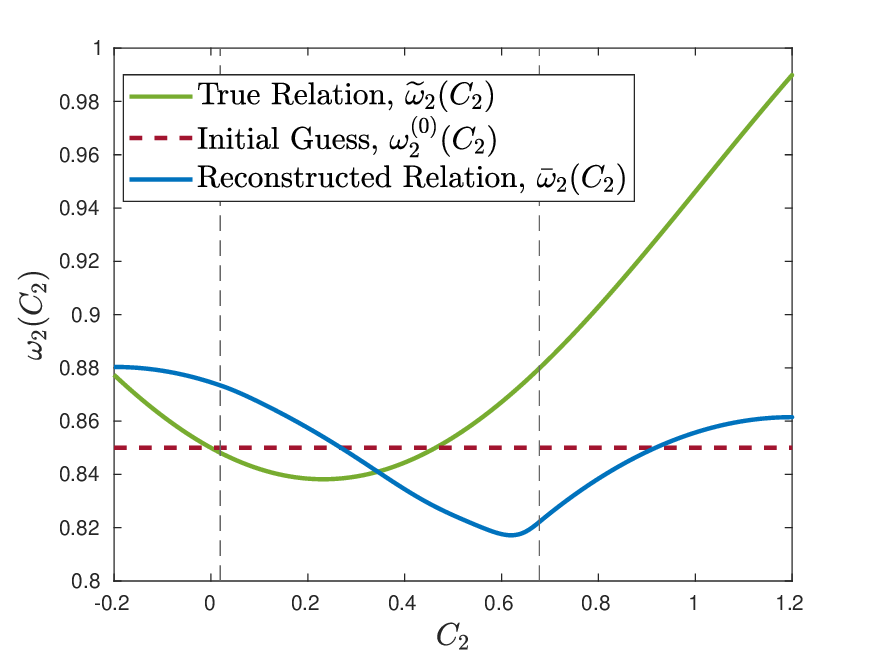}
		\subcaption{}
	\end{subfigure}
	\hfill
	\begin{subfigure}[b]{0.45\textwidth}
		\centering
		\includegraphics[width=1\textwidth]{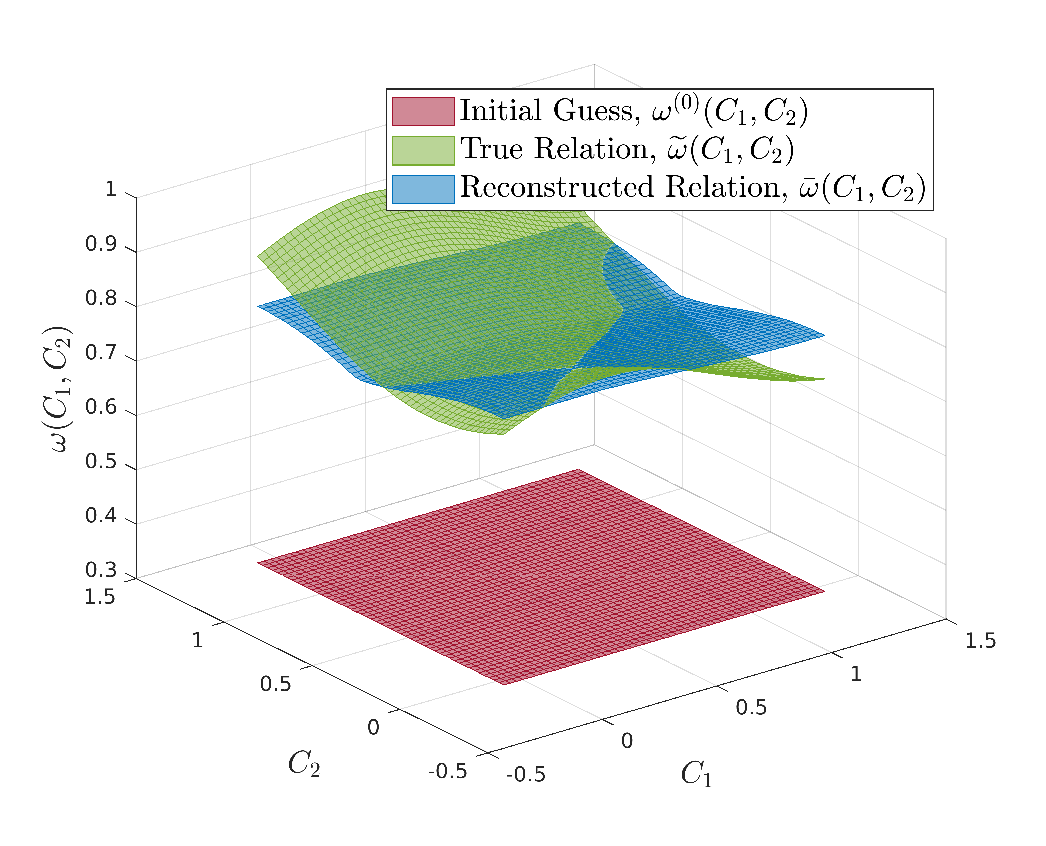}
		\subcaption{}
	\end{subfigure}
	\caption{Constitutive relations (a) $\omega_1(C_1)$, (b)
		$\omega_2(C_2)$, and (c) $\omega(C_1,C_2)$.  Optimally
		reconstructed constitutive relations $\overline{\omega}$
		(blue), along with the initial guess of relations
		$\omega^{(0)}$ (red) and the true relations
		$\widetilde{\omega}$ (green) are shown. The grey vertical
		lines in panels (a) and (b) denote the identifiability
		region for the last iteration of the Algorithm
		\ref{alg:optimal}.}
	\label{fig:manufactured_results}
\end{figure}

In order to validate the computational framework, one can
manufacture synthetic "experimental" data using some assumed forms of the constitutive relations 
and parameter values, and then 
seek to reconstruct them based on the manufactured data using Algorithm \ref{alg:optimal}, 
starting from arbitrary initial guesses. One can assume an arbitrary functional form of the 
factors $\omega_1$ and $\omega_2$ defining the constitutive relation, along with an
arbitrary parameter $\alpha$. Subsequently, synthetic
experimental concentrations can be manufactured based on these assumed functional forms and parameters. 
Finally, the manufactured experimental concentrations can be employed
to reconstruct the "unknown" parameters and relations optimally. By comparing the reconstructed relations 
to their assumed forms, this process allows for the validation of the proposed methodology and ensures its 
effectiveness under controlled conditions. Note that for the purpose of 
computational validation, we only perform the validation on the adjoint analysis of the excitation
dynamics (stage II of Algorithm \ref{alg:optimal}) as it is computationally more
complex. The analogous tests have also been performed for the relaxation dynamics, 
but are omitted here due to the simpler nature of the problem.
Figure \ref{fig:manufactured_results} demonstrates the assumed functional forms
of the factors determining the constitutive relation and the resulting relation. 
Also, the optimal parameter value is chosen as $\alpha=5$ for this experiment.
Based on the forms of the constitutive relations presented in Figure \ref{fig:manufactured_results}, the 
corresponding manufactured concentrations with some arbitrary initial conditions will be generated, 
as shown in Figure \ref{fig:simulated_concentrations}. This data will be used as the "true" data 
for computation of cost functionals in the current section.

\begin{figure}[!ht]
	\centering
	\begin{subfigure}[b]{0.45\textwidth}
		\centering
		\includegraphics[width=1\textwidth]{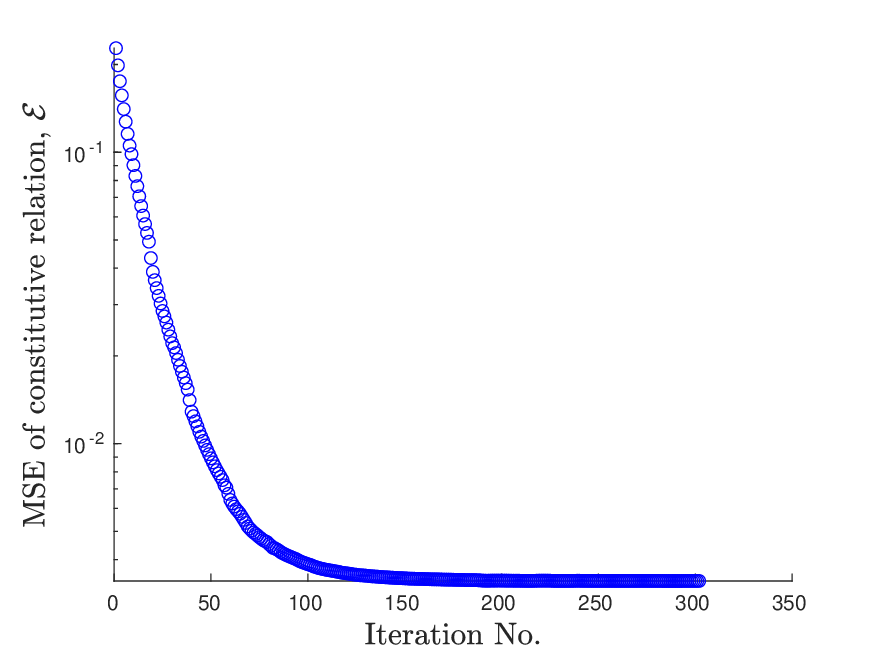}
		\subcaption{}
	\end{subfigure}
	\hfill
	\begin{subfigure}[b]{0.45\textwidth}
		\centering
		\includegraphics[width=1\textwidth]{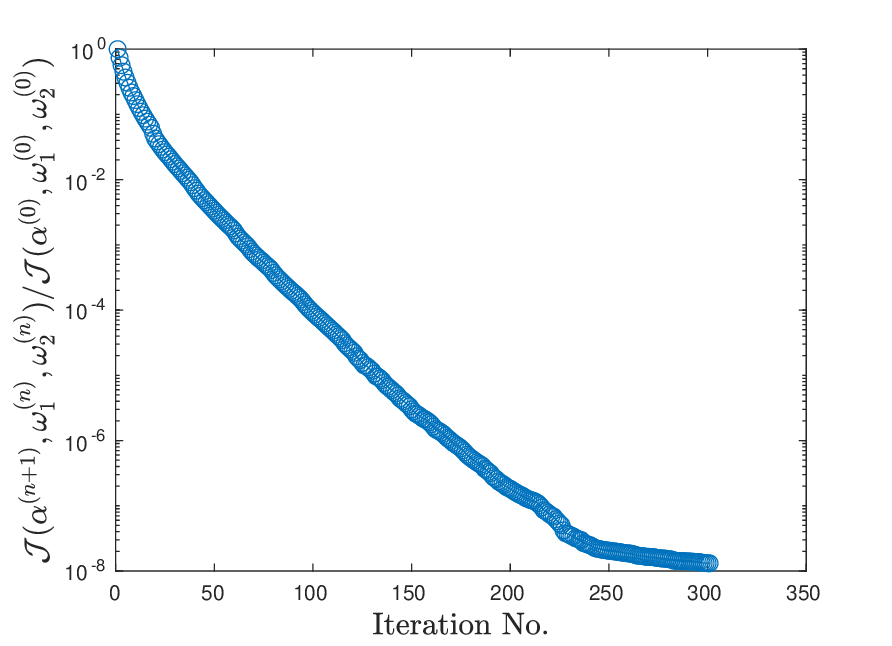}
		\subcaption{}
	\end{subfigure}
	\begin{subfigure}[b]{0.45\textwidth}
		\centering
		\includegraphics[width=1\textwidth]{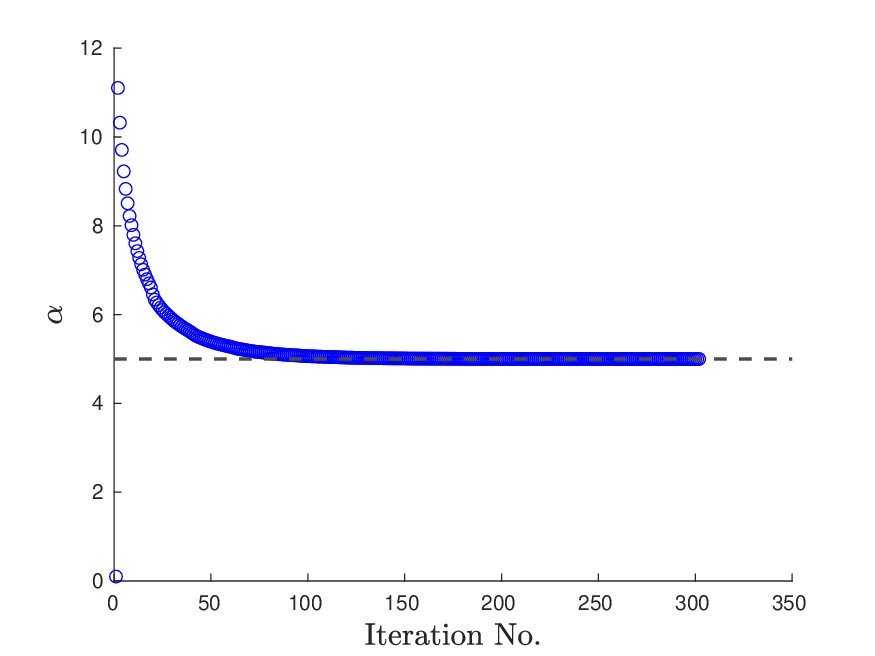}
		\subcaption{}
	\end{subfigure}
	\caption{Performance evaluation of the iterative algorithm
		according to Algorithm \ref{alg:optimal}.  The mean squared
		error between the true and the reconstructed constitutive
		relations (a), the relative decay of cost functional
		normalized with respect to its initial value (b), and the
		evolution of the parameter $\alpha$ (c), all shown as functions
		of iteration count $n$.}
	\label{fig:J_history_manufactured}
\end{figure}

The results of the optimal reconstruction of the constitutive relations are presented in Figure \ref{fig:manufactured_results},
along with their "true" forms. In this experiment the initial guesses of the algorithm are chosen as 
$\omega_1^{(0)} = 0.4$, $\omega_2^{(0)} = 0.85$ and $\alpha^{(0)} = 0.1$. 
Algorithm \ref{alg:optimal} is terminated when the relative decrease of the objective 
functional between two consecutive iterations becomes smaller than a prescribed tolerance 
($TOL = 10^{-6}$) or the maximum number of iterations ($N=300$) has been exceeded.
Note that the mean squared error for the omega relation between the true and the reconstructed one is 
defined as 
\begin{equation}
\begin{alignedat}{1}
\mathcal{E}(\omega) &= \frac{1}{(C_1^b-C_1^a)(C_2^b-C_2^a)}\int_{C_1^{a}}^{C_1^{b}} \int_{C_2^{a}}^{C_2^{b}} \left[\omega(C_1,C_2)-\widetilde{\omega}\right]^2 dC_2 dC_1.
\label{eq:MSE_omega}
\end{alignedat}
\end{equation}
The performance of the algorithm is illustrated in Figure \ref{fig:J_history_manufactured}, 
in which the mean squared error of the reconstruction of $\omega$ with iterations, 
the relative decay of cost functional with respect to its initial value, and the evolution 
of the parameter $\alpha$ with iterations are plotted.
As can be observed, the parameter $\alpha$ is approaching to its true value, $\alpha=5$.

The time histories of concentrations corresponding to the true constitutive relations $\widetilde{C}_1(t;\widetilde{\alpha},\widetilde{\omega}_1,\widetilde{\omega}_2)$ and $\widetilde{C}_2(t;\widetilde{\alpha},\widetilde{\omega}_1,\widetilde{\omega}_2)$, 
the time evolution of concentrations corresponding to the initial guess of parameter and relations $C_1(t;\alpha^{(0)},\omega_1^{(0)},\omega_2^{(0)})$ and 
$C_2(t;\alpha^{(0)},\omega_1^{(0)},\omega_2^{(0)})$, 
and the time evolution of concentrations corresponding to the optimal reconstructed relations 
$C_1(t;\overline{\alpha},\overline{\omega}_1,\overline{\omega}_2)$ and 
$C_2(t;\overline{\alpha},\overline{\omega}_1,\overline{\omega}_2)$
are shown in Figure \ref{fig:simulated_concentrations}.
As can be observed, the model \eqref{eq:ODE} equipped with the optimally reconstructed constitutive 
relations and parameters can very well predict the time evolution of concentrations. 
The optimal reconstruction of constitutive relation is demonstrated in Figure 
\ref{fig:manufactured_results}. As can be observed, there are slight differences 
between the true and the reconstructed relations, however, the time evolution of 
concentrations matches the true data very accurately, cf.~Figure \ref{fig:simulated_concentrations}. 
This provides information about the degree of sensitivity of the concentrations to 
the form of constitutive relations. Comparing the initial guess for constitutive relation $\omega^{(0)}$ 
to its optimal reconstruction $\overline{\omega}$, it is clear that there is a significant improvement. 
The small differences between the true and the reconstructed relations have two main 
reasons. First, the sensitivity of the concentrations to the constitutive relations is small, meaning 
that small perturbations in constitutive relation will not have significant impact on the 
results. This is a measure of the ill-posedness of the inverse problem \eqref{eq:inverse}.
Second, the constitutive relations are extended beyond the identifiability region
based on some boundary conditions that might not be completely correct. For this reason, 
the deviation between the reconstructed function and the true one beyond the identifiability 
region becomes larger.
This concludes the validation of the computational framework. It is shown that the computational
framework is capable of reconstructing constitutive relations to minimize the mismatch
between experimental and predicted concentrations.

\end{document}